\documentclass[letterpaper,twocolumn,english,reprint, aps, prb, showpacs]{revtex4-1}
\usepackage[T1]{fontenc}
\usepackage[latin9]{inputenc}
\usepackage{babel}
\usepackage{array}
\usepackage{multirow}
\usepackage{amsmath}
\usepackage{amssymb}
\usepackage{graphicx}
\usepackage{subscript}
\usepackage[unicode=true]
 {hyperref}

\makeatletter

\pdfpageheight\paperheight
\pdfpagewidth\paperwidth

\providecommand{\tabularnewline}{\\}

\addto\captionsenglish{\renewcommand{\figurename}{FIG.}}
\addto\captionsenglish{\renewcommand{\tablename}{TABLE}}

\makeatother

\begin{document}

\preprint{APS/123-QED}

\title{Hard wall edge confinement in 2D topological insulators and the energy
of the Dirac~Point}

\author{P. C. Klipstein}
\email{philip_k@scd.co.il}

\affiliation{Semiconductor Devices, P.O. Box 2250, Haifa 31021, Israel }

\date{\today}
\begin{abstract}
In 2D topological insulators (TIs) based on semiconductor quantum
wells such as HgTe/CdTe or InAs/GaSb/AlSb, spin polarized edge states
have been predicted with a massless Dirac like dispersion. In a hard
wall treatment based on the $4\times4$ BHZ Hamiltonian and open boundary
conditions (OBCs), the wave function is weakly confined near the edge,
with which it makes no contact. In contrast, standard boundary conditions
for the wave function and its derivative (SBCs) lead to strong confinement
with a peak amplitude at the edge. Unfortunately, weak confinement
exhibits unphysical behavior related to a spurious gap solution that
is included in the OBC wave function. This is confirmed by the gap
solutions of the parent multiband Hamiltonian from which the smaller
Hamiltonian is derived, which exhibit physical behavior and do not
satisfy OBCs. Unlike OBCs or other approaches based on phenomenological
boundary conditions, SBCs treat the wall explicitly. Using a basis
of empty crystal free electron states for the vacuum with the same
symmetry as the TI states, it is shown that a large wall band gap
overlapping that of the TI can only be achieved by including a thin
passivation layer. For passivation materials such as silicon dioxide
where the mid gap energy is nearly degenerate with that of the TI,
the Dirac point is very close to mid gap and virtually independent
of the TI band asymmetry. The treatment also demonstrates that a significant
shift of the dispersion may be introduced by interface band mixing.
The shift is largest at the Dirac point and decreases monotonically
with edge state wave vector, vanishing when the edge states merge
with the bulk band edges. 
\end{abstract}



\maketitle

\section{\label{sec:1-INTRODUCTION}INTRODUCTION\protect \\
 }

Surface states occur in many areas of condensed matter physics, often
when some material parameter changes sign across a boundary. Two classic
examples are surface plasmons and surface optical phonons, where a
sign reversal of the dielectric function leads to edge confinement
with an amplitude that decays away from the edge, both into the material
and into the surroundings.\citep{YuCardona1996,Ritchie1957,Pitarke2007,Fuchs1965,Sood1985}
More recently, electronic edge states have been discovered in topological
insulators (TIs), where a change in sign of the band gap parameter
reverses the ordering of even and odd parity crystal periodic basis
functions.\citep{Qi2010} These edge states also have the unique property
that their direction of motion depends on the electron spin, leading
to unusual phenomena such as dissipationless ballistic transport and
the quantum spin Hall effect.\citep{Bernevig2006,Konig2008} Their
experimental observation, however, remains a challenge and in some
cases more advanced surface passivation techniques may be needed,
in order to eliminate parallel conduction paths through trivial edge
states due, for example, to unsatisfied dangling bonds.\citep{Nichelel2016}

A popular treatment for the spin polarized edge states in TIs is based
on the $\mathbf{k}\cdot\mathbf{p}$ theory, where useful results can
be obtained with the simple four band BHZ Hamiltonian based on $2\times2$
blocks for each spin direction.\citep{Bernevig2006} Each $2\times2$
block is associated with a Chern number whose value on each side of
the boundary can be related to the number of spin-polarized edge states.
At a hard wall, the nature of the edge state confinement can depend
strongly on the boundary conditions used. Standard boundary conditions
for the wave function and its derivative (SBCs) and other related
approaches, can lead to strong confinement, where the wave function
has a peak amplitude at the edge, similar to the classic surface states
described above.\cite{Klipstein2015,*KlipErratum2016,Tkachov2010,Tkachov2013,Medhi2012}
On the other hand, open boundary conditions (OBCs) lead to weak confinement,
where the amplitude is zero at the edge and only reaches a peak typically
10-100~Å away.\citep{Konig2008,Zhou2008,Essert2015}While deep quantum
well (QW) states also have negligible amplitude at the edge of the
well, the bound state vanishes when one of the barriers is removed
and this should not be confused with true edge confinement. Of these
methods, the OBC approach is by far the most popular because it apparently
avoids any explicit treatment of the wall.\citep{Qi2006,Konig2007,Dai2008,Konig2008,Liu2008,Zhou2008,Linder2009,Liu2010,Lu2010,Shan2010,Sonin2010,Murakami2011,Shen2011,Wada2011,Michetti2012a,Michetti2012b,Takagaki2012,CanoCortes2013,Hohenadler2013,Sengupta2013,Wang2014,Xu2014,Essert2015,Enaldiev2015,Durnev2016,Entin2017,Candido2018,Gioia2018,Durnev2019,Gioia2019,Bottcher2020,Durnev2020}
Notwithstanding the greater simplicity of OBCs, this author has previously
argued that the spin polarized edge states in TIs cannot have zero
amplitude at the edge, and in this respect they are no different from
their classic plasmon or phonon counterparts. The weakly confined
wave function in the OBC treatment arises from a mathematically correct
but physically spurious solution of the four band BHZ Hamiltonian
for each spin direction, which not only gives an unphysical wave function
but also leads to other unphysical properties. The aim of the present
work is to provide further evidence for this point of view, based
on an eight band $\mathbf{k}\cdot\mathbf{p}$ treatment. It is shown
that edge states are not possible in the eight band model using OBCs,
while the four band and eight band results correspond very well when
SBCs are used. In contrast to some alternatives to OBCs, which are
phenomenological in nature and do not consider the wall explicitly,\citep{Medhi2012,Enaldiev2015,Entin2017}
SBCs provide a realistic physical picture on both sides of the boundary.
They can also give a physical edge dispersion whose Dirac point is
close to mid gap, even when band structure asymmetry would cause the
Dirac point to be near a bulk band edge in the four band OBC treatment. 

In most cubic semiconductors, one of the crystal periodic basis functions
of the BHZ Hamiltonian is based on antibonding \textit{s}-orbitals,
while the other is based on bonding \textit{p}-orbitals, corresponding
to the conduction and valence bands, respectively. In two dimensional
TIs, usually based on semiconductor QWs such as HgTe/CdTe or InAs/GaSb/AlSb,
their order can be reversed by increasing the QW width. Spin polarized
edge states are then predicted according to the change in Chern number
at the boundary with the wall or other semiconductor material.\citep{Kane2013}
For a TI with symmetrical bands, the $2\times2$ spin up Hamiltonian
can be written: $H_{2\times2\uparrow}=A\left(\sigma_{x}k_{x}-\sigma_{y}k_{y}\right)+\sigma_{z}\left(M+\mathrm{\mathbf{k\cdot\mathbf{\mathit{B}}k}}\right)$
and the Chern number is given by $N_{C}=-\frac{1}{2}\left[\mathrm{sgn}\left(M\right)-\mathrm{sgn}\left(B\right)\right]$,
where $2M$ is the semiconductor band gap, $B^{-1}$ is related to
the band effective mass, and $\mathbf{k}=\left(k_{x},k_{y}\right)$
is the in-plane wave-vector.\citep{Shen2011} The symmetrical operator
ordering used in the quadratic term has been justified in Ref.~\onlinecite{Klipstein2010}.
For a boundary between a TI material with $M<0$ , $B>0$ and a wall
with $M$- and $B$-parameter values, $M_{0}>0$ , $B_{0}>0$, the
change in Chern number is $\Delta N_{C}=1$, supporting a single spin-polarized
edge state. If the sign of $B_{0}$ is reversed, $\Delta N_{C}=2$
and two edge states are predicted. Examples of both types will be
discussed in this work.
\begin{figure}
\begin{tabular}{cc}
\includegraphics[scale=0.31]{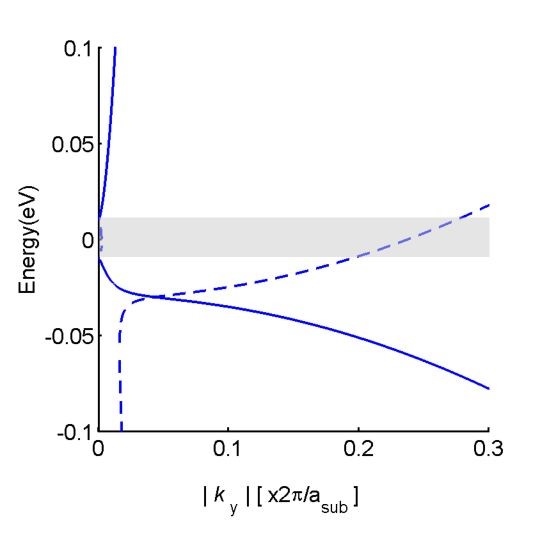} & \includegraphics[scale=0.31]{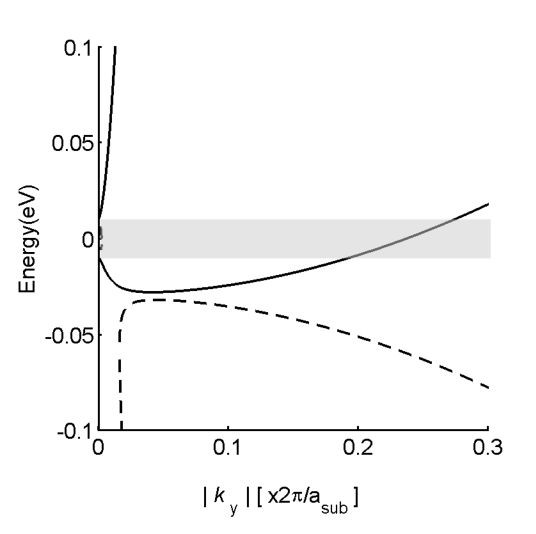}\tabularnewline
(a)\label{1(a)} & (b)\label{1(b)}\tabularnewline
\multicolumn{2}{c}{\includegraphics[scale=0.31]{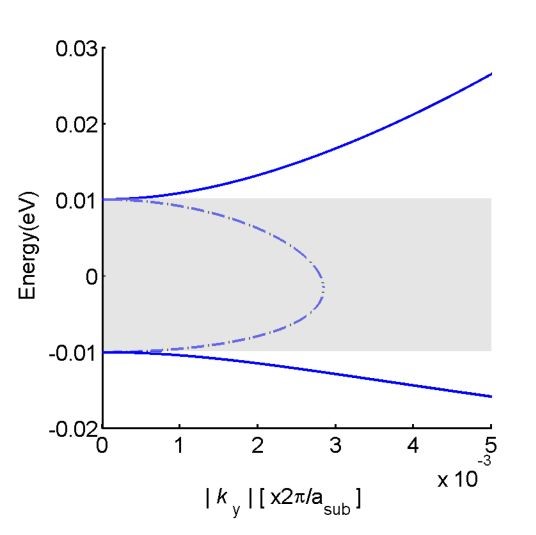}}\tabularnewline
\multicolumn{2}{c}{(c)\label{1(c)}}\tabularnewline
\end{tabular}

\caption{{\footnotesize{}\label{Figure 1} (Color online) Dispersions of the
2 band spin up Hamiltonian in the $y$-direction with $M=-0.01\,\mathrm{eV}$,
 $A=4\,\mathrm{eV\,\mathring{A}}$, $B=200\,\mathrm{eV}\,\mathrm{\mathring{A}^{2}}$
and (a) $D=199.5\,\mathrm{eV\,\mathring{A}}^{2}$ (blue curves), (b)
$D=200.5\,\mathrm{eV\,\mathring{A}}^{2}$ (black curves), and (c)
both $D$-values (blue and black superimposed). Note the shorter range
of wave vector in (c). Extended (evanescent) states are depicted as
solid (dashed) lines ($a_{\mathrm{sub}}=6.0954\,\mathrm{\mathring{A}})$.}}
\end{figure}

Band asymmetry can be incorporated into the 2 x 2 spin up Hamiltonian
by adding an additional term proportional to the identity matrix,
$I_{2\times2}$, yielding $H'_{2\times2\uparrow}=H_{2\times2\uparrow}+I_{2\times2}\mathbf{k\cdot\mathit{D}k}$,
and parameter values can be found, usually empirically, which provide
a good description of the small wave vector states near the band edges
of most semiconductor materials. There are also solutions in the band
gap energy range, $E<|M|$. Considering the dispersion in the $y$-direction
($k_{x}=0$), the \textquotedblleft middle states\textquotedblright{}
have a small imaginary wave vector, $k_{y}=i\sigma_{m}\left(E\right)$,
and describe tunneling, for example when the semiconductor is used
as a thin barrier material. The \textquotedblleft wing states\textquotedblright{}
, on the other hand, have a wave vector, $k_{y}=i\sigma_{w}\left(E\right)$,
that is imaginary or real, depending on whether $D<B$ or $D>B$,
respectively. When the difference between $B$ and $D$ is small,
the decay parameter at zero energy is given by the simple formula:
$\sigma_{w}\left(0\right)=\pm\sqrt{\left(2MB+A^{2}\right)/2B\left(B-D\right)}$,
showing that its magnitude depends on the difference, and that the
state is only evanescent when \textit{D} < \textit{B}. This is demonstrated
in Fig.~\ref{Figure 1}, which compares two TI Hamiltonians with
$M=-0.01\,\mathrm{eV}$, $A=4\,\mathrm{eV\,\mathring{A}}$, and $B=200\,\mathrm{eV\mathring{A}^{2}}$,
and where extended and evanescent states are depicted as solid and
dashed lines, respectively. Wave vectors are plotted in units of $\frac{2\pi}{a_{\mathrm{sub}}}$,
where $a_{\mathrm{sub}}=6.0954\,\mathrm{\mathring{A}}$ (the value
of the GaSb cubic lattice parameter). The blue plot in Fig.~1(a)
is for $D=199.5\,\mathrm{eV\,\mathring{A}^{2}}$, while the black
plot in Fig.~1(b) is for $D=200.5\,\mathrm{eV\,\mathring{A}^{2}}$.
The same color scheme is used in Fig.~1(c), where the results for
both $D$-values are superimposed in the small wave vector region.
The switch between large imaginary and real wing wave vectors of the
same magnitude shows clearly in the shaded band gap region of Figs.~1(a)
and 1(b), occurring when the crossing of extended and evanescent
states below the band gap changes to an anti-crossing. In contrast,
the middle states with small imaginary wave vector in Fig.~1(c)
are indistinguishable for the two cases and totally insensitive to
the increase in $D$. The spurious nature of the wing states was first
discussed by White and Sham\citep{White1981}, and later by Schuurmans
and t'Hooft \citep{Schuurmans1985}. Both the extended and evanescent
solutions do not correspond to any plausible dispersion and are obviously
unphysical.\footnote{A physical dispersion must be evanescent in the band gap, starting
and finishing at a band edge. } As discussed in these same references, and demonstrated explicitly
in Sec.~\ref{sec:2-MULTI-BAND}, the wing solutions arise because
the number of basis states is too small. The effect of remote states
which have been omitted from the Hamiltonian is incorporated in the
$k-$quadratic terms (proportional to $B$ and $D$) using perturbation
theory, and this always leads to a spurious solution. Moreover, since
the range of real or imaginary wave vectors must then be limited to
magnitudes, $k_{\mathrm{max}}<\sqrt{\frac{|M|}{|B|+|D|}}$, where
the perturbation terms are less than the typical band energy, the
spurious solution is usually found to lie well beyond this limit (see
Sec.~\ref{sec:3-TWO-BAND}). For the example in Fig.~\ref{Figure 1},
the valid range corresponds to $k_{\mathrm{max}}<0.005\times\frac{2\pi}{a_{\mathrm{sub}}}$,
which is the range plotted for the middle solution in Fig.~1(c).
The valid range thus includes the middle solution but is about 50
times smaller than the band gap wave vectors of the wing solutions
in Figs.~1(a) and 1(b).

The behavior shown in Fig.~\ref{Figure 1} is quite general, as can
be seen from the simple formula given above for $\sigma_{w}\left(0\right)$,
which always gives a switch been real and imaginary wing wave vectors
when \textit{D} = \textit{B}, regardless of the sign of \textit{M},
or the magnitude of \textit{B} and \textit{D}. When \textit{B} and
\textit{D} are small, the wing solution can lie far outside the Brillouin
zone, while the middle solution, $\sigma_{m}\left(0\right)\simeq M/A$,
is very insensitive to their values. The eigenvectors for the $\sigma_{m}$
and $\sigma_{w}$ solutions at a given energy are usually different.
However, in a TI material with \textit{M} < 0 and \textit{D} < \textit{B},
it turns out that they are identical at energy $E_{\mathrm{OBC}}^{\mathrm{D}}=-M\frac{D}{B}$
(see Sec.~\ref{sec:3-TWO-BAND}), which is just below the conduction
band edge in the example of Fig.~1(a).\citep{Zhou2008} This has
led to the common but unfortunate practice of adopting OBC boundary
conditions, where both solutions are treated completely seriously
and combined into a single wave function with an envelope of the form,
$\psi=e^{-\sigma_{m}\left(E_{\mathrm{OBC}}^{\mathrm{D}}\right)y}-e^{-\sigma_{w}\left(E_{\mathrm{OBC}}^{\mathrm{D}}\right)y}$,
so that the amplitude is zero at the sample boundary, which is assumed
to be a hard wall at $y=0$. Although this would be correct if $H'_{2\times2\uparrow}$
yielded only physical solutions, the inclusion of the spurious solution
in the wave function leads to several problematic results, which have
been discussed previously and which will be elaborated below. In order
to clarify the matter, the results of the two band spin up model are
compared in Sec.~\ref{sec:2-MULTI-BAND}, with a multiband spin up
model where the remote states are included explicitly. When at least
four bands are included, both small and large imaginary wave vector
solutions still appear, but in this case the latter is a physical
tunneling state containing large eigenvector amplitudes from the additional
bands. On the other hand, the small imaginary wave vector solution
has almost zero amplitudes from these bands (as for the two band,
spin up model), so it is now impossible to combine the two solutions
and satisfy OBCs. Since the two band spin up model is simply a perturbation
approximation of the multiband spin up model, this confirms that something
is wrong with the OBC approach.

The problem with OBCs has been highlighted previously by the author
and an alternative approach was proposed for the two band spin up
model, based on SBC boundary conditions.\citep{Klipstein2015,Klipstein2018}
Although SBCs treatments have been reported by other workers, they
are generally based on a soft wall and evolve into OBCs when the wall
potential increases, because the wave function still includes the
spurious solution\citep{Michetti2012b,Durnev2020}. The SBC approach
proposed by the author is for a hard wall and does not include the
spurious and physical solutions in the same wave function. It was
shown in previous work that two exponential edge solutions can be
found when $B_{0}$ (and $D_{0}$) is (are) negative, consistent with
$\Delta N_{C}=2$ when $D_{0}=0$. One of these involves only $\sigma_{m}$
solutions in the TI and wall, and the other only $\sigma_{w}$ solutions.
The former yields a simple edge state dispersion in the useful limit,
$M_{0}\rightarrow\infty$ , and $B_{0},D_{0}\rightarrow0$, close
to $E_{2\times2\uparrow}^{\mathrm{SBC}}\simeq-Ak_{x}+Dk_{x}^{2}$,
which merges smoothly with the bulk band edges, while the latter shows
unphysical merging behavior and is rejected as spurious. It will be
shown in Sec.~\ref{sec:2-MULTI-BAND} that these results are entirely
consistent with the SBC multiband spin up model, which also puts the
Dirac point close to mid gap. When $B_{0}$ (and $D_{0}$) is (are)
positive a single, non-exponential solution exists, again consistent
with both models.

In previous work based on the two band spin up model, the edge dispersion
$E_{2\times2\uparrow}^{\mathrm{SBC}}(k_{x})$ was obtained by solving
a characteristic equation numerically, for both strongly hybridized
HgTe/CdTe and weakly hybridized InAs/GaSb/AlSb QWs.\citep{Klipstein2018}
In Sec.~\ref{sec:3-TWO-BAND} of this work, an analytical solution
is obtained for the dispersion, which provides a simple expression
for the energy of the Dirac point. The dependence of the Dirac point
on the wall hybridization parameters is then compared for the two
and four band spin up models in Sec.~\ref{sec:4-MULTI-BAND-A0}.
Although the wall hybridization parameters are found to have a relatively
small effect, a suitable model of the wall region is lacking. Therefore,
in Sec.~\ref{sec:5-WALL}, an approach is proposed for a consistent
$\mathbf{k}\cdot\mathbf{p}$ treatment. This approach yields a wall
hybridization parameter in the two band spin up model that is fairly
similar to that in the TI material, and much smaller than the free
electron Dirac value, which was previously suggested as a possibility.
It also highlights the importance of an edge passivation material
and introduces interface band mixing terms which may cause a significant
additional shift of the Dirac point. In Sec.~\ref{sec:6-CONCLUSION},
conclusions are summarized. Even though this work is focused primarily
on the physical nature of the wave function confinement and its influence
on the energy of the Dirac point, a multiband spin up treatment of
the full edge dispersion is presented in Appendix A, where it is
shown to be quite consistent with the SBC two band spin up approach.

\section{\label{sec:2-MULTI-BAND}MULTIBAND MODEL OF DIRAC POINT}

\subsection{$\mathbf{k\cdot p}$ Hamiltonian}

In 1955, Luttinger and Kohn derived the $\mathbf{k}\cdot\mathbf{p}$
theory in Fourier space and showed how it could be used to treat the
potential of an impurity atom in a bulk crystal.\citep{LuttKohn1955}
This was extended by Volkov and Takhtamirov in the 1990s, who used
the same approach to treat semiconductor superlattices.\citep{Volkov1997,Takhtamirov1997,Takhtamirov1999}
When transformed into real space and with a few basic assumptions,
their Hamiltonian can be written:\citep{Klipstein2010}

\begin{equation}
\begin{array}[b]{ccc}
E\widetilde{F}_{n}\left(\mathbf{r}\right)= &  & \left(-\frac{\hbar^{2}}{2m_{0}}\nabla^{2}+E_{n}\right)\widetilde{F}_{n}\left(\mathbf{r}\right)\\
 &  & -\underset{n'}{\sum}{\displaystyle \frac{i\hbar}{m_{0}}\mathbf{p}_{nn'}\cdot\mathbf{\nabla}\widetilde{F}_{n'}\left(\mathbf{r}\right)}\\
 &  & +\underset{n'}{\sum}H_{nn'}^{\mathrm{mod}}\left(y\right)\widetilde{F}_{n'}\left(\mathbf{r}\right)
\end{array}\label{eq:1}
\end{equation}
where $E_{n}$ is the band edge of the zone center basis state $|n\rangle$,
$\widetilde{F}_{n}\left(\mathbf{r}\right)$ is the envelope function
which only contains Fourier components in the first Brillouin zone,
and $\mathbf{p}_{nn'}=\langle n|-i\hbar\mathbf{\nabla}|n'\rangle$
is a momentum matrix element between crystal periodic basis states
$|n\rangle$ and $|n'\rangle$. For an infinite bulk material, $\widetilde{F}_{n}\left(\mathbf{r}\right)$
is a plane wave. The term $H_{nn'}^{\mathrm{mod}}\left(y\right)$
represents additional terms introduced when the crystal potential
is modulated along the \textit{y}-direction, for instance at a boundary
between two different materials.\citep{Volkov1997,Takhtamirov1997,Takhtamirov1999,Klipstein2010}
This term is not included in the present bulk treatment, but is discussed
further in Sec.~\ref{sec:5-WALL}. 

Eq.~\eqref{eq:1} can be written in matrix form with elements, $H_{nn'}$,
and for a bulk material with $H_{nn'}^{\mathrm{mod}}=0$, it is essentially
an exact description of the crystal if enough basis states are included.
For example, in their seminal work of 1966, Cardona and Pollak were
able to model the whole Brillouin zones of silicon and germanium with
15 zone center basis states.\citep{Cardona1966} However, in cases
where only a small local region of the Brillouin zone is of interest,
it is possible to eliminate a large number of the remote states using
perturbation theory, leaving only those states in the local energy
range. The Bir and Pikus expression for the local Hamiltonian, can
be written up to second order as\citep{BirPikus1974}: 

\begin{equation}
\begin{array}[b]{ccc}
\bar{H}_{mm'}= &  & H_{mm'}-\frac{1}{2}\underset{s}{\sum}H'_{ms}H'_{sm'}\\
 &  & \times\left(\dfrac{1}{E_{s}-E_{m}}+\dfrac{1}{E_{s}-E_{m'}}\right)+\ldots
\end{array}\label{eq:2}
\end{equation}
in which the local states are $|m\rangle$, $|m'\rangle$, etc., the
remote states are $|s\rangle$, $|s'\rangle$, etc., and $H'_{ms}=$
$\mathbf{k}\cdot\langle m\mid\frac{\hbar}{m_{0}}\mathbf{p}\mid s\rangle$
is an off-diagonal matrix element of $H_{nn'}$ where $\mathbf{k}=-i\mathbf{\nabla}$.
This reduces the size of the Hamiltonian, but introduces additional
\textit{k}-quadratic matrix elements into the Hamiltonian of local
states, $\bar{H}_{mm'}$. 

The model Hamiltonian for a 2D quantum well, $H'_{2\times2\uparrow}$,
discussed in Sec.~\ref{sec:1-INTRODUCTION}, is thus derived from
a larger number of basis states, using Eq.~\eqref{eq:2}. As a simple
example, consider the following unperturbed $4\times4$ spin up Hamiltonian
based on Eq.~\eqref{eq:1}, with a basis, $\mid n\rangle$:

\begin{equation}
\begin{array}[b]{ccc}
H'_{4x4\uparrow}\left(\mathbf{k}\right)= &  & I_{4\times4}D'k^{2}+\\
\\
 &  & \left[\begin{array}{cccc}
\Delta_{1} & 0 & iQ_{1}k_{-} & 0\\
0 & M & Ak_{+} & -iQ_{2}k_{+}\\
-iQ_{1}k_{+} & Ak_{-} & -M & 0\\
0 & iQ_{2}k_{-} & 0 & -\Delta_{2}
\end{array}\right]
\end{array}\label{eq:3}
\end{equation}
where $k_{\pm}=k_{x}\pm ik_{y}$, $D'=\tfrac{\hbar^{2}}{2m_{0}}=3.8\,\mathrm{eV\,\mathring{A}^{2}}$,
and $n=1,2,...,4$. The states $|2\rangle$ and $|3\rangle$ are anti-bonding
\textit{s}- and bonding \textit{p}-states discussed in Sec.~\ref{sec:1-INTRODUCTION}.
These interact with remote states $|4\rangle$ and $|1\rangle$, at
energies, $-\Delta_{2}$ and $\Delta_{1}$, respectively.\footnote{If $k_{+}$ and $k_{-}$ are interchanged in the \textit{Q}-terms
of Eq.~\eqref{eq:3}, corresponding to a change in symmetry of the
remote states, there is no change to the energy of the Dirac point
based on the present multiband model, and the edge state dispersions
are essentially the same as those derived in Appendix A.} 

Because $D'$ is quite small, the quadratic term in the first line
of Eq.~\eqref{eq:3} will be ignored. Moreover, its inclusion would
lead to non-exponential hard wall edge state solutions which are not
considered in this work. The effect of including this term will be
discussed further at the end of Sec.~\ref{sec:5-WALL}. Using the
perturbation expression in Eq.~\eqref{eq:2}, $H'_{4\times4\uparrow}\left(\mathbf{k}\right)$
is then reduced to $H'_{2\times2\uparrow}\left(\mathbf{k}\right)$
where:

\begin{subequations}
\begin{equation}
\begin{array}{ccc}
B= & \dfrac{1}{2}\left(\dfrac{Q_{2}^{2}}{\Delta_{2}+M}+\dfrac{Q_{1}^{2}}{\Delta_{1}+M}\right)\end{array}\label{eq:4a}
\end{equation}
\begin{equation}
D=\dfrac{1}{2}\left(\dfrac{Q_{2}^{2}}{\Delta_{2}+M}-\dfrac{Q_{1}^{2}}{\Delta_{1}+M}\right)\label{eq:4b}
\end{equation}
\label{Eq. 4}
\end{subequations}
If more bands are included in the multiband Hamiltonian of Eq.~\eqref{eq:3}
that interact with bands $|2\rangle$ and $|3\rangle$, there will
simply be an additional term for each band in the expressions for
\textit{B} and \textit{D}. This procedure is analogous to that used
to derive the BHZ Hamiltonian in the supplementary material of Ref.~\onlinecite{Bernevig2006},
where \textit{B} and \textit{D} were calculated from the Luttinger
parameters. The Luttinger parameters depend on interactions with remote
states, which can be expressed in a similar form to Eq.~\eqref{Eq. 4}.\citep{Lawaetz1971}

Eq.~\eqref{Eq. 4} shows that the size of the reciprocal mass terms,
\textit{B} and \textit{D}, in $H'_{2\times2\uparrow}\left(\mathbf{k}\right)$
is determined by the interaction with the remote states, $|1\rangle$
and $|4\rangle$ that have been eliminated. If these states are removed
to infinity, i.e. $\Delta_{1},\Delta_{2}\rightarrow\infty$, the reciprocal
mass terms are reduced to zero. This highlights a first problem with
OBCs. As discussed in Sec.~\ref{sec:1-INTRODUCTION}, the Dirac point
of the edge states in the OBC model occurs at $E_{\mathrm{OBC}}^{\mathrm{D}}=-M\frac{D}{B}$.
Thus for a given ratio, $\frac{D}{B}$, the shift of the Dirac point
from mid gap remains fixed, even when \textit{D} and \textit{B} become
vanishingly small. In fact, the whole dispersion remains fixed (see
Sec.~\ref{sec:3-TWO-BAND}). This kind of behavior is not physical,
because remote states which are far away in energy cause the same
shift as when they are much closer. 

\begin{figure}
\begin{tabular}{c}
\includegraphics[scale=0.35]{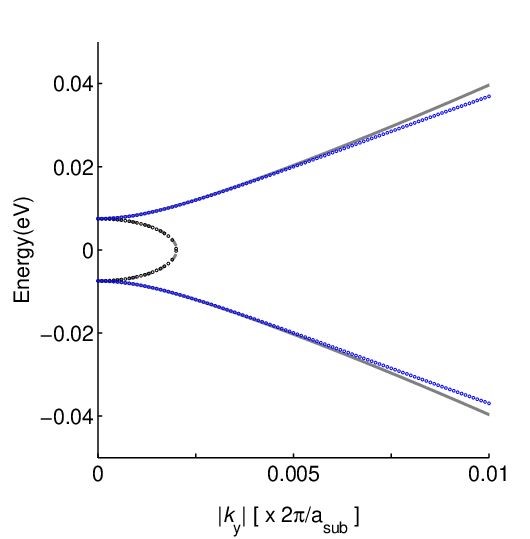}\tabularnewline
(a)\label{2(a)}\smallskip{}
\tabularnewline
\includegraphics[scale=0.35]{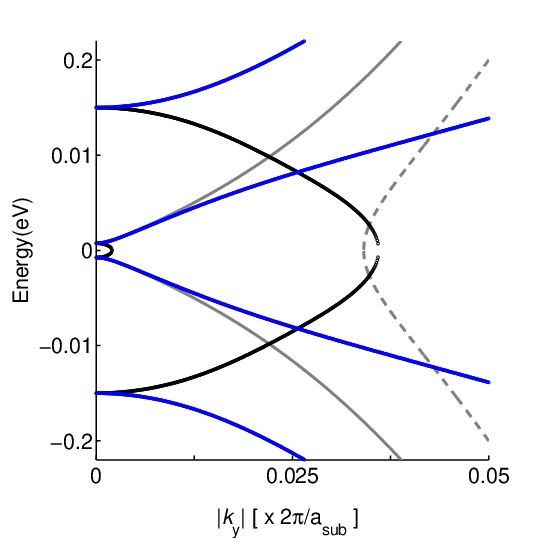}\tabularnewline
(b)\label{2(b)}\tabularnewline
\end{tabular}

\caption{{\footnotesize{}\label{Figure 2} (Color online) Dispersions in the
}\textit{\footnotesize{}y}{\footnotesize{}-direction with $M=-0.0075eV$,
$A=Q_{1}$= $Q_{2}=3.83\,\mathrm{eV}\mathrm{\mathring{A}}$, and $\Delta_{1}=\Delta_{2}=0.15\:\mathrm{eV}$
for the 4 band spin up Hamiltonian over (a) a short, and (b) a wide,
range of wave vector, onto which are superimposed the equivalent 2
band spin up results with $B=102.9\,\mathrm{eV\,\mathring{A}}^{2}$
and }\textit{\footnotesize{}D}{\footnotesize{} = 0. Extended (evanescent)
states are depicted as small blue (black) circles and as solid (dashed)
gray lines for the 4 and 2 band spin up Hamiltonians, respectively.}}
\end{figure}

\subsection{Failure of OBC's}

Fig.~\ref{Figure 2} shows the extended and evanescent states, in
blue and black, respectively, calculated from $H'_{4\times4\uparrow}\left(\mathbf{k}\right)$
with $k_{x}$= 0, for the case of a symmetric inverted band gap where
$M=-0.0075eV$, $A=Q_{1}$= $Q_{2}=Q=3.83\,\mathrm{eV}\mathrm{\mathring{A}}$
and $\Delta_{1}=\Delta_{2}=\Delta=0.15\:\mathrm{eV=}$ $-20M$. In
addition to physical tunneling states connecting bands $|2\rangle$
and $|3\rangle$, as in the two band spin up model discussed in Sec.~\ref{sec:1-INTRODUCTION},
there are now physical tunneling states also connecting bands $|1\rangle$
and $|4\rangle$. Superimposed in gray are the solutions of $H'_{2\times2\uparrow}$
with the same \textit{M} and \textit{A} values, and with $B=102.9\,\mathrm{eV\,\mathring{A}}^{2}$,
$D=0$, calculated from the \textit{M, Q} and $\Delta$-values using
Eq.~\eqref{Eq. 4}. It can be seen in Fig.~2(a) that the two dispersions
correspond very well for the middle states and for the conduction
and valence band edges out to $k_{y}=0.007\times\frac{2\pi}{a_{\mathrm{sub}}}$,
which is consistent with the valid range of wave vectors for the two
band spin up model, $|k_{y}|<k_{\mathrm{max}}$, discussed in Sec.~\ref{sec:1-INTRODUCTION}.
However Fig.~2(b) shows that the two dispersions are completely
different in the vicinity of the wing states. This highlights the
spurious nature of the wing dispersion of $H'_{2\times2\uparrow}\left(\mathbf{k}\right)$,
even when \textit{D} = 0. The two models only agree near imaginary
wave vector, $k_{y}=i\sigma_{w}\left(0\right)$, demonstrating that
the spurious $2\times2$ branch is a phantom like dispersion associated
with bands $|1\rangle$ and $|4\rangle$  that have been eliminated.
It cannot merge with these band edges as in the multiband model, because
they are no longer there, so adopts a meaningless trajectory towards
the Brillouin zone boundary.

The zero energy eigenvectors of the middle and wing solutions of the
two band spin up model are both $\frac{1}{\sqrt{2}}\left[1,-1\right]$,
which enables their combination into an OBC wave function at the Dirac
point, as discussed in Sec.~\ref{sec:1-INTRODUCTION}. Such combination
is unfortunate, because a proper description of the wing solution
should contain a significant amplitude from the absent states, $|1\rangle$
and $|4\rangle$. This can be seen in the zero energy eigenvectors
of the multiband spin up model, which are given in the first two columns
of Table~\ref{tab:Table 2} for the parameters used in Fig.~\ref{Figure 2},
namely $\left[i0.037,\,0.706,\,-0.706,\,i0.037\right]$ and $\left[i0.486,\,0.513,\,-0.513,\,i0.486\right]$.
These eigenvectors contain contributions from the relevant bands and
represent the different physical nature of the two states correctly.
However, in consequence, they can no longer be combined into a wave
function which satisfies OBCs, showing that there is a fundamental
problem with the OBC approach.

\subsection{Edge states using SBC's}

In the rest of this work the hard wall SBC approach, described in
Sec.~\ref{sec:1-INTRODUCTION} and introduced by the author in an
earlier four band treatment,\citep{Klipstein2015,Klipstein2016,Klipstein2018}
is discussed within the context of the multiband model. The wall region
is treated explicitly, because the edge state wave functions decay
on both sides of the boundary. Since the eigenvectors of the middle
and wing solutions are different in the multiband model, it is necessary
to ensure continuity of the two vector components independently. This
is entirely consistent with the separation of the middle and wing
solutions in the previously reported four band treatment, and leads
to similar results. In contrast to the OBC treatment, the Dirac point
always remains very close to mid gap, even when the semiconductor
band dispersions become asymmetric. It is assumed that the wall has
a large fundamental band gap, $2M_{0}$, that is symmetrically disposed
about that of the semiconductor. How this is realized in practice,
and the consequences of when this is not the case are discussed in
Sec.~\ref{sec:5-WALL}. It turns out that there are only exponential
edge state solutions when the wall parameters have a certain relationship
with those in the semiconductor. Since the number of eigenvalues is
conserved when some system parameter is varied, solutions exist for
other combinations of wall and semiconductor parameters but they are
no longer exponential. An exponential solution exists, however, when
$M_{0}>>|M|$ and the wall band gap is effectively infinite, so this
is the limit that is used in both models. It can be reached by a trajectory
in parameter space that involves only exponential solutions (as in
this work) or otherwise. At the same time, a band gap that is truly
infinite results in an unphysically rapid decay of the wavefunction
in the wall region. Typical band gap values for the wall based on
real physical systems are discussed in Sec.~\ref{sec:5-WALL}.

Two specific cases are now treated for the multiband spin up model.
The simplest case is for a symmetric semiconductor band structure.
Next, band asymmetry is included. In both treatments, the same values
are used for the electron-hole hybridization parameters in the semiconductor
and the wall, and it is shown that this always puts the Dirac point
exactly at mid gap, even when the band structure is asymmetric. The
multiband spin up treatment where these parameters are different is
reserved to Sec.~\ref{sec:4-MULTI-BAND-A0}, after the corresponding
results for the two band spin up model are derived in Sec.~\ref{sec:3-TWO-BAND}. 

\begin{table}
\caption{{\footnotesize{}\label{tab:Table1}Zero energy decay parameters and
corresponding eigenvectors for the 4 band spin up Hamiltonian with
a symmetric band structure, where $\Delta=\Delta_{1}=\Delta_{2}$
and $Q=Q_{1}=Q_{2}$. The functions }\textit{\footnotesize{}S}{\footnotesize{},
}\textit{\footnotesize{}R}{\footnotesize{} and }\textit{\footnotesize{}N}{\footnotesize{}
are defined in Eq.~\eqref{Eq. 5}.}}

\begin{ruledtabular}
\begin{tabular*}{8cm}{@{\extracolsep{\fill}}ll}
\noalign{\vskip\doublerulesep}
Decay Parameter & Eigenvector\tabularnewline[\doublerulesep]
\noalign{\vskip\doublerulesep}
\hline 
\noalign{\vskip\doublerulesep}
\multirow{4}{*}{$\sigma_{\pm}=\theta\mathrm{sgn}\left(\Delta\right)\dfrac{\Delta}{Q^{2}}\sqrt{\dfrac{S_{\pm}}{2}}$} & $a=\theta\mathrm{sgn}\left(\Delta\right)\dfrac{i}{N}\dfrac{R}{A}\sqrt{\dfrac{S_{\pm}}{2}}$\tabularnewline[\doublerulesep]
\noalign{\vskip\doublerulesep}
\noalign{\vskip\doublerulesep}
 & $b=\theta\mathrm{sgn}\left(\Delta\right)\dfrac{Q}{N}\sqrt{\dfrac{2}{S_{\pm}}}$\tabularnewline[\doublerulesep]
\noalign{\vskip\doublerulesep}
\noalign{\vskip\doublerulesep}
 & $c=-\dfrac{Q}{N}\dfrac{R}{A}$\tabularnewline[\doublerulesep]
\noalign{\vskip\doublerulesep}
\noalign{\vskip\doublerulesep}
 & $d=\dfrac{i}{N}$\tabularnewline[\doublerulesep]
\noalign{\vskip\doublerulesep}
\end{tabular*}
\end{ruledtabular}

\end{table}

Table~\ref{tab:Table1} gives expressions for the pair of zero energy
decay parameters, $\sigma_{\pm}$, in the \textit{y}-direction and
their corresponding eigenvectors, $[a,b,c,d]_{\pm}$, of the symmetric
multiband spin up model, expressed in terms of the following quantities:

\begin{subequations}
\begin{equation}
S_{\pm}=A^{2}+2\dfrac{M}{\Delta}Q^{2}\pm\mathrm{sgn}\left(\Delta\right)A\sqrt{A^{2}+4\dfrac{M}{\Delta}Q^{2}}\label{eq:5a}
\end{equation}
\begin{equation}
R=\left[1-2\dfrac{M}{\Delta}\dfrac{Q^{2}}{S_{\pm}}\right]\label{eq:5b}
\end{equation}
\begin{equation}
N=\left[\sqrt{\left(Q^{2}+\dfrac{S_{\pm}}{2}\right)\left(\dfrac{R^{2}}{A^{2}}+\dfrac{2}{S_{\pm}}\right)}\right]\label{eq:5c}
\end{equation}

\label{Eq. 5}
\end{subequations}

\begin{table}[b]
\caption{{\footnotesize{}\label{tab:Table 2}Zero energy decay parameters and
eigenvectors for the symmetric 4 band spin up Hamiltonian at $k_{x}=0$,
with two types of wall in which $\Delta_{0}$ has opposite signs.
Middle (small type face) and wing (large type face) solutions are
listed below the material parameters. The equivalent 2 band spin up
results are given in the last four rows. The solutions are calculated
for $M=-0.0075\,\mathrm{eV}$ and $M_{0}=200\,\mathrm{eV}$, but the
middle or wing eigenvectors in the wall and semiconductor are equal
for any $M_{0}$ when $\Delta_{0}<0$ and $\Delta/M=\Delta_{0}/M_{0}$.
When $\Delta_{0}>0,$ they are similar (dissimilar) for the middle
(wing) solution, implying that only a single non-exponential middle
solution exists. }}

\begin{ruledtabular}
\begin{tabular}{ccccccc}
{\small{}Parameter} & \multicolumn{2}{c}{{\small{}Semicond.}} & \multicolumn{2}{c}{{\small{}Wall (-ve $\Delta_{0}$)}} & \multicolumn{2}{c}{{\small{}Wall (+ve $\Delta_{0}$)}}\tabularnewline
\hline 
\multicolumn{7}{c}{{\small{}4 band}}\tabularnewline
\textit{\small{}$A,A_{0}$}{\small{} (eV Å)} & \multicolumn{2}{c}{{\small{}3.83}} & \multicolumn{2}{c}{{\small{}3.83}} & \multicolumn{2}{c}{{\small{}3.83}}\tabularnewline
\textit{\small{}$Q,Q_{0}$}{\small{} (eV Å)} & \multicolumn{2}{c}{{\small{}3.83}} & \multicolumn{2}{c}{{\small{}3.83}} & \multicolumn{2}{c}{{\small{}3.83}}\tabularnewline
{\small{}$\Delta/M,\Delta_{0}/M_{0}$} & \multicolumn{2}{c}{{\small{}-20}} & \multicolumn{2}{c}{{\small{}-20}} & \multicolumn{2}{c}{{\small{}50}}\tabularnewline
\noalign{\vskip\doublerulesep}
$\sigma_{\pm},\sigma_{0\pm}\left(\times\frac{2\pi}{a_{sub}}\right)$ & {\tiny{}0.0020} & 0.0360 & {\tiny{}-53.48} & -959.7 & {\tiny{}-49.68} & -2582.6\tabularnewline[\doublerulesep]
\noalign{\vskip\doublerulesep}
\textit{\small{}a} & \textit{\tiny{}i}{\tiny{}0.037} & \textit{i}0.486 & \textit{\tiny{}i}{\tiny{}0.037} & \textit{i}0.486 & \textit{\tiny{}-i}{\tiny{}0.014} & \textit{i}0.505\tabularnewline
\textit{\small{}b} & {\tiny{}0.706} & 0.513 & {\tiny{}0.706} & 0.513 & {\tiny{}0.707} & 0.495\tabularnewline
\textit{\small{}c} & {\tiny{}-0.706} & -0.513 & {\tiny{}-0.706} & -0.513 & {\tiny{}-0.707} & 0.495\tabularnewline
\textit{\small{}d} & \textit{\tiny{}i}{\tiny{}0.037} & \textit{i}0.486 & \textit{\tiny{}i}{\tiny{}0.037} & \textit{i}0.486 & \textit{\tiny{}-i}{\tiny{}0.014} & \textit{-i}0.505\tabularnewline
\hline 
\noalign{\vskip\doublerulesep}
\multicolumn{7}{c}{{\small{}2 band}}\tabularnewline[\doublerulesep]
\noalign{\vskip\doublerulesep}
{\small{}$B,B_{0}(\mathrm{eV\mathring{A}^{2}})$} & \multicolumn{2}{c}{{\small{}102.9}} & \multicolumn{2}{c}{{\small{}-0.00386}} & \multicolumn{2}{c}{{\small{}0.00144}}\tabularnewline
{\small{}$\sigma_{\pm},\sigma_{0\pm}\left(\times\frac{2\pi}{a_{sub}}\right)$} & {\tiny{}0.0020} & 0.0341 & {\tiny{}-53.65} & -908.9 & {\tiny{}-49.7} & -2633.3\tabularnewline
\textit{\small{}a} & {\tiny{}0.707} & 0.707 & {\tiny{}0.707} & 0.707 & {\tiny{}0.707} & 0.707\tabularnewline
\textit{\small{}b} & {\tiny{}-0.707} & -0.707 & {\tiny{}-0.707} & -0.707 & {\tiny{}-0.707} & 0.707\tabularnewline
\end{tabular}
\end{ruledtabular}

\end{table}
The parameter, $\theta$, has a value of +1 for the semiconductor
and -1 for the wall. In contrast to the two band spin up model, where
boundary conditions exist for both the wave function and its derivative,
the only boundary condition in the multiband spin up model is the
continuity of the wave function, because $H'_{4\times4\uparrow}\left(\mathbf{k}\right)$
is linear in $k_{y}$. For an edge at \textit{y} = 0, this means that
edge solutions must have the same eigenvectors on each side of the
boundary. Since the zero energy eigenvectors in Table~\ref{tab:Table1}
only depend on the ratio of the two band gap parameters, \textit{M}
and $\Delta$, it is always possible to match the solutions with small
or large decay parameters, independently, for a given \textit{A} and
\textit{Q,} provided the band gap parameters in the semiconductor
and wall have the same ratio. Examples are shown in Table~\ref{tab:Table 2}
for the same parameters used in Fig.~\ref{Figure 2}, with $\frac{\Delta}{M}=\frac{\Delta_{0}}{M_{0}}=-20$,
where $|2M_{0}|$ and $|2\Delta_{0}|$ are the band gaps between the
inner and outer bands, respectively, in the wall material. Note that
$\Delta_{0}$ is negative, because \textit{M} and $M_{0}$ have opposite
signs. In the Table, the eigenvectors for the small decay parameter
(large decay parameter) on each side of the boundary are written with
a small (large) typeface, so that the correspondence between the eigenvectors
in each group can be seen clearly. Inserting $\frac{\Delta}{M}=\frac{\Delta_{0}}{M_{0}}$
into Eq.~\eqref{Eq. 4} gives $B_{0}=B\frac{M}{M_{0}}<0$ for the
two band spin up model. It was shown in previous work that for this
value of $B_{0}$ there are two exponential edge states corresponding
to the same matching of solutions with either a small or a large decay
parameter, respectively.\citep{Klipstein2015,Klipstein2018} Both
models thus obey the same condition for the two exponential edge solutions.
Moreover, $B_{0}\rightarrow0$ or $\Delta_{0}\rightarrow-\infty$
as $M_{0}\rightarrow\infty$, both of which are realistic descriptions
of a hard wall.

\begin{figure}
\includegraphics[viewport=120bp 0bp 860bp 520bp,clip,scale=0.33]{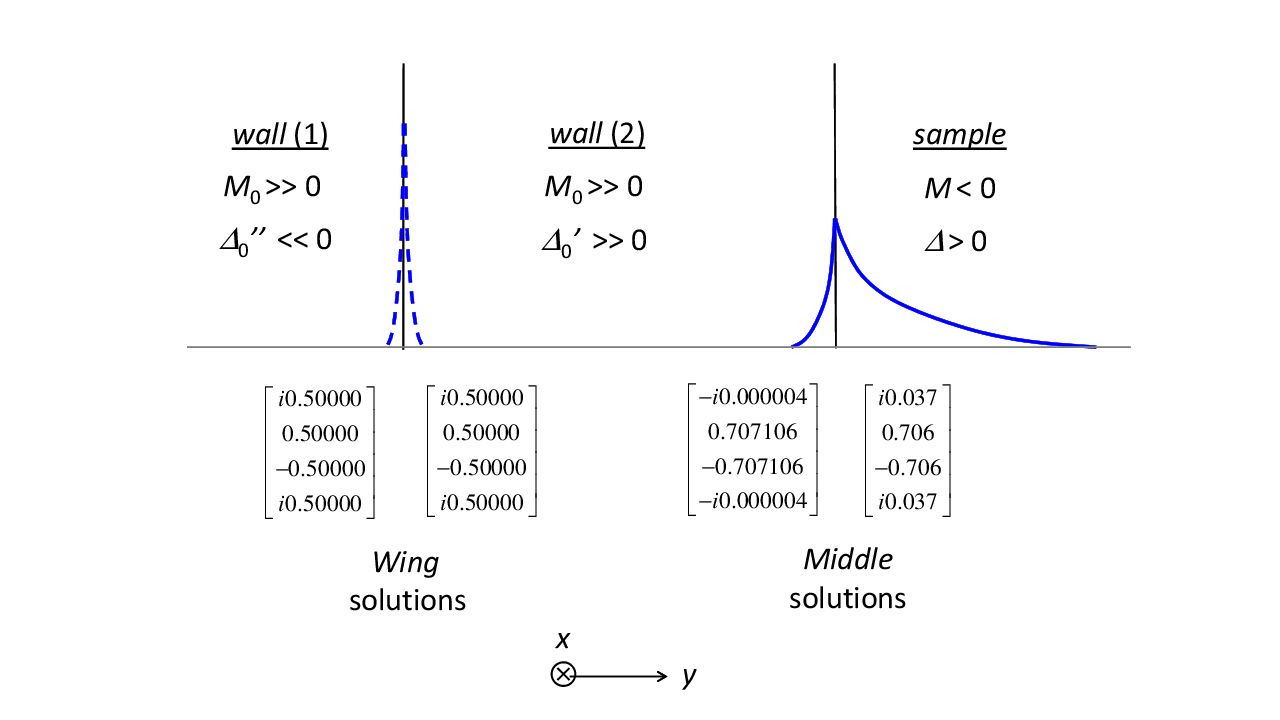}

\caption{{\footnotesize{}\label{Figure 3}(Color online) Schematic depiction
of Dirac point edge state wave functions at a compound, double layer
wall, based on the middle (solid) and wing (dashed) solutions of the
symmetric 4 band spin up Hamiltonian. The inner band gap parameter
of the wall, $M_{0}>>M$, is constant and positive, while the outer
band gap parameter changes sign in each of the two wall layers, with
values $\Delta'_{0}>>0$ ($\Delta''_{0}<<0$) closest (furthest) from
the semiconductor. Similar zero energy eigenvectors on each side of
an interface are listed below the diagram for the semiconductor and
wall parameters listed in Table~\ref{tab:Table 2}, except that $|\Delta'_{0}|=|\Delta''_{0}|>4\times10^{7}$eV.}}
\end{figure}
Positive $\Delta_{0}$ solutions (corresponding to positive $B_{0}$
in the two band spin up model) are shown in the last column of Table~\ref{tab:Table 2},
where there is only fairly close agreement between eigenvectors for
the small decay parameters on each side of the boundary. The values
for \textit{b}, \textit{c} on each side are almost identical while
\textit{a}, \textit{d} although of opposite sign, are very small.
Presumably a small deformation of the wave function might lead to
a perfect match. This can only be checked using numerical methods,
since the deformed wave function will no longer be exponential. In
contrast, the solutions with large decay parameters have completely
different eigenvectors in the semiconductor and the wall, with roughly
equal magnitudes for all four components, but where the ratios \textit{b}/\textit{c}
and \textit{a}/\textit{d} have opposite signs on each side of the
boundary. The solutions with large decay parameters are thus unlikely
to satisfy wave function continuity under any circumstances. These
results appear to be consistent with the single edge state predicted
from the Chern numbers in the two band spin up model. Although there
is only one non-exponential solution in this model when $B_{0}$ is
positive, it must converge to the same physical solution as for negative
$B_{0}$ when the wall band gap is infinite in each case and $|B_{0}|\rightarrow0.$\citep{Klipstein2018}
The wing solutions for negative $B_{0}$ or $\Delta_{0}$ are thus
rejected as unphysical. This point is also demonstrated by the diagram
in Fig.~\ref{Figure 3}, where a wall layer with positive $\Delta_{0}$
($=\Delta'_{0}$) is sandwiched between the semiconductor and another
wall layer with negative $\Delta_{0}$ ($=\Delta_{0}''$). As these
parameters become very large, $|\Delta'_{0}|=|\Delta''_{0}|>4\times10^{7}$eV,
only the eigenvectors of the wing solutions (listed below the dashed
wave function) are equal at the interface between the two wall materials,
while only the eigenvectors of the middle solutions (listed below
the solid wave function) are close in value at the interface with
the semiconductor. If the central layer thickness is expanded to infinity,
then only the single edge state based on the small decay parameters
of the middle solutions remains at the semiconductor boundary, while
if the central layer thickness is reduced to zero, there are two edge
states localized at an interface between the semiconductor and a wall
with negative $\Delta_{0}$, which is the situation already discussed
above. This provides additional support for the conclusion that a
wall with negative $\Delta_{0}$ supports both the physical and spurious
edge states.\footnote{Even though the wing solution of the four band spin up Hamiltonian
is not necessarily spurious, the reason it gives an unphysical edge
state is discussed at the end of Appendix A. } Fig.~\ref{Figure 3} is analogous to the equivalent treatment for
the two band spin up case, in Fig.~3 of Ref.~\onlinecite{Klipstein2018}.

Exponential solutions can also be found for the zero energy edge states
when $\Delta_{1}\neq\Delta_{2}$, and/or $Q_{1}\neq Q_{2}$, and the
band structure is asymmetric. In the multiband spin up model, the
ratio of the outer to inner band energies for a wall with hybridization
parameters $A_{0}$, $Q_{10}$ and $Q_{20}$, that correspond to a
particular eigenvector {[}\textit{a}, \textit{b}, \textit{c}, \textit{d}{]}
at energy \textit{E}, are given as a function of $k_{x}$ by Eq.~\eqref{Eq. A2}
in Appendix A. Setting $k_{x}=0$, gives:

\begin{subequations}
\begin{equation}
\dfrac{\Delta_{10}}{M_{0}}=\dfrac{Q_{10}}{Q_{20}}\left\{ \dfrac{b}{d}\right\} \left\{ \dfrac{c}{a}\right\} \dfrac{1-\dfrac{E}{M_{0}}}{1+i\dfrac{A_{0}}{Q_{20}}\left\{ \dfrac{c}{d}\right\} }+\dfrac{E}{M_{0}}\label{eq:6a}
\end{equation}
\begin{equation}
\dfrac{\Delta_{20}}{M_{0}}=-\left\{ \dfrac{b}{d}\right\} ^{2}\dfrac{1-\dfrac{E}{M_{0}}}{1+i\dfrac{A_{0}}{Q_{20}}\left\{ \dfrac{c}{d}\right\} }-\dfrac{E}{M_{0}}\label{eq:6b}
\end{equation}

\label{Eq: 6}
\end{subequations}
When the hybridization parameters on each side of the boundary are
equal, namely $Q_{1}=Q_{10}$, $Q_{2}=Q_{20}$ and $A=A_{0}$, the
expressions for the eigenvector components in the right hand column
of Table~\ref{tab:Table 4} of Appendix A can be used in Eq.~\eqref{Eq: 6}
to show that the zero energy eigenvectors in the semiconductor and
wall are equal, provided $\frac{\Delta_{10}}{M_{0}}=\frac{\Delta_{1}}{M}$
and $\frac{\Delta_{20}}{M_{0}}=\frac{\Delta_{2}}{M}$, consistent
with the symmetric case discussed above. Therefore, as in the symmetric
case, solutions with either small or large decay parameters, can be
matched independently, when the outer band parameters in the wall,
$\Delta_{01}$ and $\Delta_{02}$, are negative. Note that in the
two band spin up model, the semiconductor band asymmetry is defined
by the \textit{B-} and \textit{D}-parameters. Eq.~\eqref{Eq. 4}
shows that a given asymmetry can be achieved by various combinations
of the semiconductor \textit{Q}- and $\Delta$-parameters in the four
band spin up model. In all these cases, the Dirac point does not move
from gap center when the hybridization parameters in the semiconductor
are equal to those in the wall, in complete contrast to the large
shift observed in the OBC model, but in good agreement with the two
band spin up SBC model. Both two band spin up models are discussed
in the next Section, where the effect of unequal hybridization parameters
in the wall and semiconductor is now considered ($A_{0}\neq A$). 

\section{\label{sec:3-TWO-BAND}FOUR BAND MODEL USING SBC'\protect\lowercase{S}}

It has been shown previously that in the band gap region of the four
band BHZ Hamiltonian, the two exponential decay parameters for each
$2\times2$ spin block can be expressed in terms of the wave vector
parallel to the edge, $k_{x}$, and energy, \textit{E}, as:

\begin{equation}
\sigma(k_{x},E)=\sqrt{k_{x}^{2}+F\pm\sqrt{F^{2}-G}}\label{eq:7}
\end{equation}
where $F=\frac{A^{2}+2\left(MB+ED\right)}{2B_{+}B_{-}}$, $G=\frac{M^{2}-E^{2}}{B_{+}B_{-}}$
and $B_{\pm}=B\pm D$.\cite{Klipstein2015,*KlipErratum2016,Zhou2008}At
$k_{x}=0$, these are just the decay parameters of the middle and
wing solutions discussed in Sec.~\ref{sec:1-INTRODUCTION}, i.e.
$\sigma_{m,w}\left(E\right)=\sigma_{\pm}\left(k_{x}=0,E\right)=\sqrt{F\pm F^{2}-G}.$
The spin up eigenvectors corresponding to each solution\citep{Klipstein2015}
can be set equal to give the following characteristic equation for
the OBC edge state energy: 

\begin{equation}
\dfrac{M+E+B_{-}\left(k_{x}^{2}-\sigma_{-}^{2}\right)}{A\left(k_{x}+\sigma_{-}\right)}=-\dfrac{A\left(k_{x}-\sigma_{+}\right)}{M-E+B_{+}\left(k_{x}^{2}-\sigma_{+}^{2}\right)}\label{eq:8}
\end{equation}

Making use of the transformation, $E\rightarrow-E$, $D\rightarrow-D$,
$k_{x}\rightarrow-k_{x}$ and solving Eq.~\ref{eq:8} for the energy
yields the well-known OBC dispersion formula for spin up:\citep{Zhou2008,Wada2011}

\begin{equation}
E_{\mathrm{2\times2\uparrow}}^{\mathrm{OBC}}=-Ak_{x}\sqrt{1-\dfrac{D^{2}}{B^{2}}}-M\dfrac{D}{B}\label{eq:9}
\end{equation}

As discussed above for the Dirac point, the OBC solution is unphysical
because it only depends on the ratio of reciprocal mass parameters,
and not on their size.

The characteristic equation for both SBC edge solutions, physical
and spurious, is derived from the boundary conditions for the wave
function and its derivative with negative $B_{0},D_{0}\rightarrow0$
in the limit $M_{0}\rightarrow\infty$. It has been reported previously
by the author and has the form:\citep{Klipstein2018}

\begin{equation}
\dfrac{M+E+B_{-}\left(k_{x}^{2}-\sigma_{\pm}^{2}\right)}{A\left(k_{x}+\sigma_{\pm}\right)}=\dfrac{\sigma_{\pm}D-\sqrt{\sigma_{\pm}^{2}D^{2}+A_{0}^{2}}}{A_{0}}\label{eq:10}
\end{equation}

This equation can be solved for the energy in an analogous way to
the OBC case, yielding a dispersion:

\begin{equation}
E_{2\times2\uparrow}^{\mathrm{SBC}}=-Ak_{x}\sqrt{1+\dfrac{\sigma^{2}D^{2}}{A_{0}^{2}}}+Dk_{x}^{2}+D\sigma^{2}\left(\dfrac{A}{A_{0}}-1\right)\label{eq:11}
\end{equation}
where $\sigma$ is the smaller decay parameter in the semiconductor
corresponding to the physical solution. Strong hybridization is assumed,
as in the previous Section, where the decay parameter is real (the
weak case is discussed below). As demonstrated in Ref. \onlinecite{Klipstein2015},
$\sigma=\sigma_{-}$ ($\sigma=\sigma_{+}$) when \textit{D} <\textit{
B} (\textit{D} > \textit{B}). When \textit{D} > \textit{B}, the $\sigma_{-}$
solution in Eq.~\eqref{eq:7} is imaginary and corresponds to the
spurious gap solution with a large real wave vector discussed in Sec.~\ref{sec:1-INTRODUCTION}.
Note that for spin down, the sign of the first term reverses in both
Eqs.~\eqref{eq:9} and \eqref{eq:11}. 

For \textit{D} < \textit{B}, a numerical solution of Eq.~\eqref{eq:10}
was used previously to demonstrate that the $\sigma_{+}$ solution
has a larger phase velocity than the $\sigma_{-}$ solution and does
not merge smoothly with the bulk band edges (see for example Fig.~2(a)
in Ref.~\onlinecite{Klipstein2015,*KlipErratum2016}). Since this
decay parameter is equal to the wing solution when $k_{x}=0$, this
was taken as further evidence of the spurious nature of the edge state
based on the larger decay parameter. On the other hand, the physical
solution based on the smaller decay parameter does merge smoothly
with the bulk band edges, at which point $\sigma\rightarrow0$. If
the wave vector at which the bulk and edge states merge is $k_{x}^{m}$,
then Eq.~\eqref{eq:11} shows that the merging energy is $E_{2\times2\uparrow}^{\mathrm{SBC}}\overset{\lim\sigma\rightarrow0}{\longrightarrow}-Ak_{x}^{m}+D\left(k_{x}^{m}\right)^{2}.$
Comparing this energy with the bulk dispersion, given in Eq. (6) of
Ref.~\onlinecite{Klipstein2015,*KlipErratum2016}, yields $k_{x}^{m}=\pm\sqrt{\frac{-M}{B}}$
. Note that substituting the merging energy and $k_{x}=k_{x}^{m}$
into Eq.~\eqref{eq:7} yields $\sigma=0$, as required. 

Eq.~\eqref{eq:11} confirms the result of the multiband spin up model
discussed at the end of the previous Section, namely that when the
hybridization parameters in the semiconductor and wall are equal,
the Dirac point is at mid gap, regardless of the degree of band asymmetry.
On the other hand, it was previously pointed out that it is unlikely
that these parameters are exactly equal.\citep{Klipstein2018} This
is confirmed in Sec.~\ref{sec:5-WALL} where a hybridization parameter
is estimated for the wall that is slightly larger than the value typically
used for strongly hybridized TIs such as HgTe/CdTe QWs. Eq.~\eqref{eq:11}
then puts the Dirac point at $E_{\mathrm{2\times2\uparrow}}^{\mathrm{D}}=D\left(\frac{A}{A_{0}}-1\right)\sigma_{\mathrm{D}}^{2}$,
which is slightly below mid gap when $A_{0}>A$. The physical zone
center decay parameter, $\sigma_{\mathrm{D}}$, can be evaluated by
solving Eq.~\eqref{eq:7} with $k_{x}=0$ and $E=E_{\mathrm{2\times2\uparrow}}^{\mathrm{D}}$.
In the limit $A_{0}>>A$, it turns out to be independent of \textit{D},
and is then given by: 

\begin{equation}
\sigma_{\mathrm{D}}=\dfrac{A}{2B}-\sqrt{\dfrac{A^{2}}{4B^{2}}+\dfrac{M}{B}}\label{eq:12}
\end{equation}
which is the same expression as for the \textit{D} = 0 case, given
in equation (3b) of Ref.~\onlinecite{Klipstein2015,*KlipErratum2016}.
Since the Dirac point is always close to mid gap where the decay parameter
varies slowly with energy (see Fig.~1(c)), Eq.~\eqref{eq:12} is
a good approximation for any value of $A_{0}\geq A$. 

For equal hybridization parameters in the TI and wall, the band gap
ratios, $\frac{\Delta_{i0}}{M_{0}}=\frac{\Delta_{i}}{M}$, at the
end of Sec.~\ref{sec:2-MULTI-BAND} can be substituted into Eq.~\ref{Eq. 4}
to show that $B_{0}=\frac{M}{M_{0}}B$ and $D_{0}=\frac{M}{M_{0}}D$.
Substituting these relations into the expression for $k_{\mathrm{max}}$
in Sec.~\ref{sec:1-INTRODUCTION} yields the valid range of wave
vectors in the wall, $k_{\mathrm{max},0}=\frac{M_{0}}{|M|}k_{\mathrm{max}}$.
In a wall where $M_{0}$ is effectively infinite and the physical
decay parameter is comparable to the size of the Brillouin zone, its
value can be estimated from $\sigma\left(0,0\right)$ in Eq.~\eqref{eq:7}
in the limit of vanishing $B_{0},D_{0}$, giving $\sigma_{0-}\simeq-\frac{M_{0}}{A}$.
While it is clear that the SBC wave function in the TI is consistent
with perturbation theory, to show that this is also the case in the
wall it is required that $|\sigma_{0-}|<k_{\mathrm{max,0}}$, i.e.
$\frac{M_{0}}{A}<\frac{M_{0}}{|M|}\sqrt{\frac{|M|}{B+D}}$. This expression
can be rearranged to give $\frac{A^{2}}{B^{2}}>\frac{|M|}{B}$, which
is always true in strongly hybridized TIs, where $\sigma_{\mathrm{D}}$
in Eq.~\eqref{eq:12} is real. In addition, since the wing decay
parameter in the TI is greater than or equal to $\frac{A}{B}$,\citep{White1981,Klipstein2018}
this inequality also proves that the wing solution is outside the
valid range, as already pointed out in Sec.~\ref{sec:1-INTRODUCTION}.
While it is sometimes argued that the inclusion of the wing solution
in the OBC wave function is benign,\citep{Gioia2018} especially when
it decays over several lattice spacings,\citep{Gioia2019,Durnev2020}
this shows that it cannot be included without violating perturbation
theory. Further anomalies related to OBCs are discussed in Appendix B.

Before comparing the Dirac point predicted above for $A_{0}\neq A$
with the multiband spin up result, a few more properties of the newly
derived dispersion relation in Eq.~\eqref{eq:11} are discussed,
all of which have been confirmed previously from a direct solution
of Eq.~\eqref{eq:10}.\citep{Klipstein2015,KlipErratum2016,Klipstein2018}
First, in strongly hybridized TIs such as HgTe/CdTe, it was noted
above that the edge state velocity for the spurious solution with
$D<B$ is larger than for the physical solution. This is confirmed
by replacing the decay parameter in Eq.~\eqref{eq:11} with $\sigma_{+}>>\sigma_{-}$,
which leads to a larger square root term and hence a larger velocity.
Second, in weakly hybridized InAs/GaSb/AlSb, the decay parameters
near $k_{x}=0$ are complex conjugates. As discussed previously, the
physical and spurious wave functions are obtained from symmetric and
antisymmetric combinations of these two solutions.\citep{Klipstein2016}
However, it was noted in Ref.~\onlinecite{Klipstein2018} that the
characteristic equation in this regime does not have an exact solution
when $D\neq0$, although the error is small. This is now understood
from Eq.~\eqref{eq:11}, where the energy, $E_{\mathrm{2\times2\uparrow}}^{\mathrm{SBC}}$,
is no longer real when $\sigma_{\pm}^{2}D^{2}/A_{0}^{2}$ is complex.
The absence of an exact solution when \textit{D} is finite shows that
the edge state wave function is no longer purely exponential, and
numerical methods must be used to obtain a precise solution. Note
also, that for the exponential case with $D=0$, SBCs give $B_{0}=\frac{M}{M_{0}}B$
and $\mathrm{|Re}(\sigma_{0})|=\frac{A}{2|B_{0}|}$ when $A_{0}=A$.\citep{Klipstein2015}
The condition for a complex TI decay parameter is $\frac{A^{2}}{4B^{2}}<\frac{|M|}{B}$,
yielding $\mathrm{|Re}(\sigma_{0})|<k_{\mathrm{max,0}}$ and confirming,
also for weak hybridization, that the SBC wave function in a wall
with effectively infinite $M_{0}$ is still consistent with perturbation
theory.

\section{\label{sec:4-MULTI-BAND-A0}SENSITIVITY TO WALL HYBRIDIZATION}

In this Section the energies of the Dirac point are compared for the
two band and multiband spin up models, when one or more of the electron-hole
hybridization parameters in the semiconductor are unequal to those
in the wall. As shown in the previous Section, when $A_{0}>A$, the
two band spin up model predicts a negative Dirac point energy of $E_{\mathrm{2\times2\uparrow}}^{\mathrm{D}}=D\left(\frac{A}{A_{0}}-1\right)\sigma_{\mathrm{D}}^{2}$,
where $\sigma_{\mathrm{D}}$ is given to a very good approximation
by Eq.~\eqref{eq:12}.

Since the Dirac point is no longer at $E=0$, it is not a simple matter
to find an analytical solution for its energy in the multiband spin
up model. Instead a numerical solution can be performed based on Eq.~\eqref{Eq: 6}.
By inserting the eigenvector for the middle solution of the semiconductor
at energy \textit{E} into Eq.~\eqref{Eq: 6}, and using the predicted
ratio of the outer to inner band energies in the wall to calculate
its middle solution at the same energy, the energy of the Dirac point,
$E_{4\times4\uparrow}^{\mathrm{D}}$, is found when the wall and semiconductor
eigenvectors are equal. The results are shown in Table \ref{tab:Table 3}
for typical semiconductor parameters and two different values of the
wall hybridization parameter, $A_{0}$. The effect of changing the
secondary wall hybridization parameters, $Q_{10}$ and $Q_{20}$ is
also studied. The calculation was performed such that each of the
eigenvector components in the semiconductor and wall agree to better
than 0.001\%. Results for the two band spin up model are also shown
in the lower part of the Table, for comparison. 
\begin{table}
\caption{{\footnotesize{}\label{tab:Table 3}The eigenvectors of the 4 band
spin up middle solutions in three types of wall with negative $\Delta_{i0}$
and $A_{0}\protect\neq A$, at the energy of the Dirac point, $E_{4\times4\uparrow}^{\mathrm{D}}$,
for the semiconductor parameters defined in the first column. The
solutions are calculated for $M=-0.0075\,\mathrm{eV}$ and $M_{0}=200\,\mathrm{eV}$,
but the eigenvectors and Dirac point are essentially unchanged for
any $M_{0}>1\,\mathrm{eV}$. The equivalent 2 band results are given
in the lower rows. The ratio of the Dirac point energies calculated
by the two methods is shown in the last row. The listed outer band
energies of the wall were calculated using Eq.~\eqref{Eq: 6}. For
other values, the solutions are not exponential.}}

\begin{ruledtabular}
\begin{tabular}{cccc||c||c||c||cc||c||c||c||c||c||c||c}
{\small{}Parameter} & Semicond. & {\small{}Wall (1)} & \multicolumn{5}{c}{{\small{}Wall (2)}} & \multicolumn{8}{c}{{\small{}Wall (3)}}\tabularnewline
\hline 
\multicolumn{16}{c}{4 band}\tabularnewline
\textit{\small{}$A,A_{0}$}{\small{} (eV Å)} & {\small{}3.83} & {\small{}11.0} & \multicolumn{5}{c}{{\small{}11.0}} & \multicolumn{8}{c}{{\small{}1973.5}}\tabularnewline
{\small{}$Q_{1},Q_{10}$ (eV Å)} & {\small{}2.0} & {\small{}2.0} & \multicolumn{5}{c}{{\small{}11.0}} & \multicolumn{8}{c}{{\small{}2.0}}\tabularnewline
{\small{}$Q_{2},Q_{20}$ (eV Å)} & {\small{}2.0} & {\small{}2.0} & \multicolumn{5}{c}{{\small{}11.0}} & \multicolumn{8}{c}{{\small{}2.0}}\tabularnewline
{\small{}$\Delta_{1}/M,\Delta_{10}/M_{0}$} & {\small{}-20} & {\small{}-6.723} & \multicolumn{5}{c}{{\small{}-40.18}} & \multicolumn{8}{c}{{\small{}-0.037}}\tabularnewline
{\small{}$\Delta_{2}/M,\Delta_{20}/M_{0}$} & {\small{}-3.2} & {\small{}-1.066} & \multicolumn{5}{c}{{\small{}-6.52}} & \multicolumn{8}{c}{{\small{}-0.0058}}\tabularnewline
\textit{\small{}a} &  & \textit{\small{}i}{\small{}0.0195} & \multicolumn{5}{c}{\textit{\small{}i}{\small{}0.0200}} & \multicolumn{8}{c}{\textit{\small{}i}{\small{}0.0193}}\tabularnewline
\textit{\small{}b} &  & {\small{}0.6971} & \multicolumn{5}{c}{{\small{}0.6769}} & \multicolumn{8}{c}{{\small{}0.7016}}\tabularnewline
\textit{\small{}c} &  & {\small{}-0.7064} & \multicolumn{5}{c}{{\small{}-0.7268}} & \multicolumn{8}{c}{{\small{}-0.7017}}\tabularnewline
\textit{\small{}d} &  & \textit{\small{}i}{\small{}0.1211} & \multicolumn{5}{c}{\textit{\small{}i}{\small{}0.1149}} & \multicolumn{8}{c}{\textit{\small{}i}{\small{}0.1225}}\tabularnewline
{\small{}$E_{4\times4\uparrow}^{\mathrm{D}}$(eV)} &  & {\small{}-0.00020} & \multicolumn{5}{c}{{\small{}0.00027}} & \multicolumn{8}{c}{{\small{}-0.00030}}\tabularnewline
\hline 
\multicolumn{16}{c}{2 band}\tabularnewline
{\small{}$B,B_{0}(\mathrm{eV\mathring{A}^{2}})$} & {\small{}135.3} & {\small{}-0.153} & \multicolumn{5}{c}{{\small{}-0.063}} & \multicolumn{8}{c}{{\small{}0.0204}}\tabularnewline
{\small{}$D,D_{0}(\mathrm{eV\mathring{A}^{2}})$} & {\small{}107.2} & {\small{}-0.149} & \multicolumn{5}{c}{{\small{}-0.047}} & \multicolumn{8}{c}{{\small{}-0.00032}}\tabularnewline
{\small{}$E_{\mathrm{2\times2\uparrow}}^{\mathrm{D}}$(eV)} &  & {\small{}-0.00031} & \multicolumn{5}{c}{{\small{}-0.00031}} & \multicolumn{8}{c}{{\small{}-0.00048}}\tabularnewline
\hline 
 &  &  & \multicolumn{5}{c}{} & \multicolumn{8}{c}{}\tabularnewline
{\small{}$E_{4\times4\uparrow}^{\mathrm{D}}/E_{\mathrm{2\times2\uparrow}}^{\mathrm{D}}$} &  & {\small{}0.629} & \multicolumn{5}{c}{{\small{}-0.856}} & \multicolumn{8}{c}{{\small{}0.633}}\tabularnewline
\end{tabular}
\end{ruledtabular}

\end{table}

In the four band spin up model there is a dependence on the secondary
hybridization parameters and almost perfect correspondence exists
with the two band spin up model when these parameters are very small
(not shown in Table~\ref{tab:Table 3}). For the parameter values
shown in the Table, the correspondence between models is reasonable,
with agreement for the energy of the Dirac point to better than 1.0
meV or just a few percent of the TI band gap. Note that the Dirac
point even has a small positive energy when all hybridization parameters
in the wall are equal. It can be concluded, however, that the shift
of the Dirac point from mid gap is extremely small in both models,
and it is suggested that in the absence of an accurate knowledge for
the values of $Q_{10}$ and $Q_{20}$, the two band model can be taken
to provide a reasonable estimate. 

The large wall hybridization parameter of 1973.5\,eV\,Å in the right
hand column of Table~\ref{tab:Table 3} corresponds to $A_{0}=\hbar c$
in the relativistic Dirac equation. It was previously suggested by
the author that this value might be used for a vacuum wall.\citep{Klipstein2018}
In the next Section, however, it is argued that the relativistic value
cannot be justified and an alternative picture is presented. This
predicts a smaller value for the hybridization parameter closer to
that in the semiconductor and with the same order of magnitude as
the value of 11\,eV\,Å used here.

\section{\label{sec:5-WALL}DISCUSSION OF THE WALL REGION}

\subsection{$\mathbf{k\cdot p}$ treatment of the interface}

Although the 2D Dirac Hamiltonian for the electron is similar in form
to the BHZ Hamiltonian, a treatment is required for the wall that
is conceptually consistent within a non-relativistic $\mathbf{k}\mathbf{\cdot p}$
framework. In this Section an attempt is made to address the problem.

The final term in Eq.~\eqref{eq:1}, was not required for the preceding
treatment of the evanescent band gap states which are essentially
properties of the bulk material. However, this term is required when
considering properties related specifically to the sample edge. If
we ignore derivative of a delta-function terms, and consider only
a single boundary at $y=0$, then $H_{nn'}^{\mathrm{mod}}\left(y\right)$
is given to a reasonable approximation by the following expression:
\begin{equation}
H_{nn'}^{\mathrm{mod}}\left(y\right)=\delta U_{nn'}\widetilde{G}\left(y\right)-D_{0,nn'}\widetilde{\delta}\left(y\right)\label{eq:13}
\end{equation}
where $G(y)$ is a step function as shown in the middle panel of Fig.~4(a)
and $\delta(y)$ is a Dirac delta-function, and $\widetilde{G}(y)$
and $\widetilde{\delta}(y)$ are the same functions with Fourier components
limited to the first Brillouin zone.\citep{Klipstein2010} While the
first pair of functions are mathematically abrupt, the second pair
change over a distance of about one monolayer, or $a'=\frac{a_{s\mathrm{ub}}}{2}$.
The matrix element, $\delta U_{nn'}$, is defined as $\langle n|\delta U|n'\rangle$
where $\delta U=U_{B}-U_{A}$ and $U_{A}$ and $U_{B}$ are the microscopic
crystal potentials for materials \textit{A} and \textit{B} on each
side of the interface. Material \textit{A} is treated as a reference
crystal and $\delta U$ is thus the perturbation that transforms material
\textit{A} into material \textit{B}. $E_{n}$ in Eq.~\eqref{eq:1}
is then a local band edge in the reference crystal, and the perturbation
term, $\delta U_{nn'}\widetilde{G}\left(y\right)$, in Eq.~\eqref{eq:13}
can be used with the Pikus-Bir formula in Eq.~\eqref{eq:2} to generate
the band edge positions of the local states in the other material.
Together with the interface terms, $D_{0,nn'}$, which are discussed
in detail below, it is possible to describe the local band edges and
basis states in materials \textit{A} or \textit{B}, and their evolution
on passing from one material to the other.\citep{Takhtamirov1997,Takhtamirov1999}
\begin{figure*}
\begin{tabular}{cc}
\includegraphics[viewport=100bp 0bp 750bp 540bp,clip,scale=0.42]{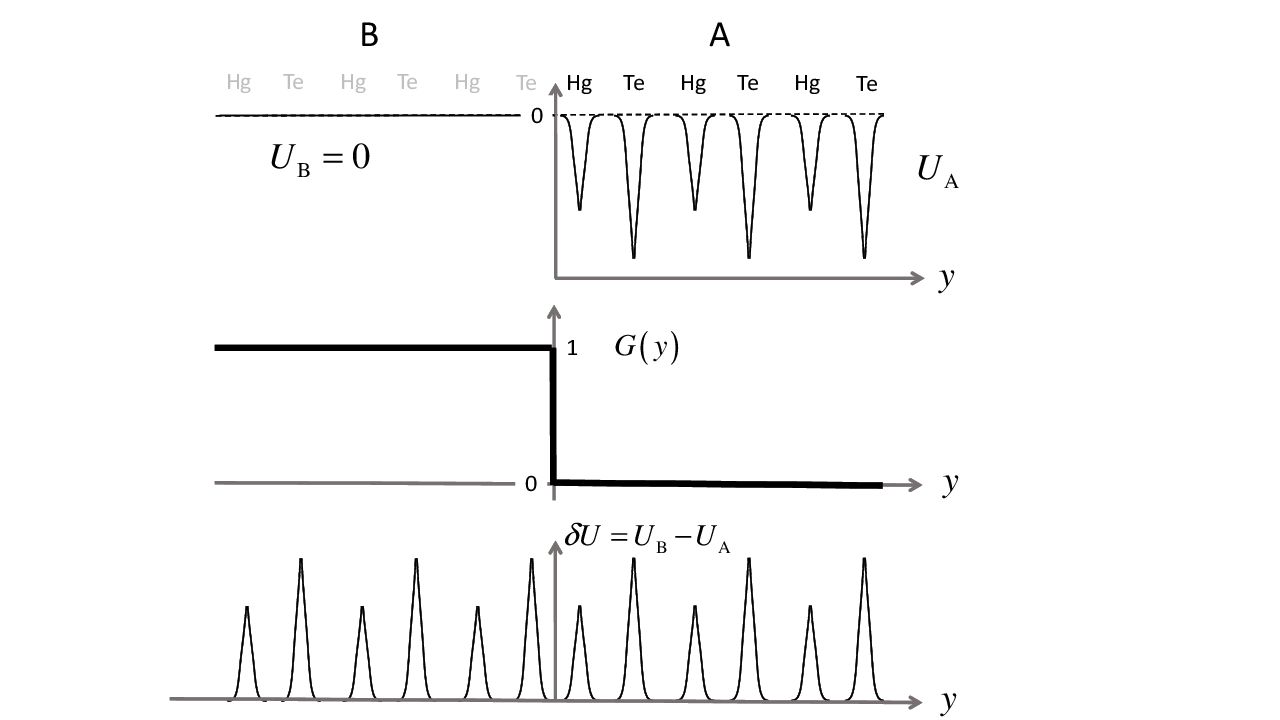} & \includegraphics[viewport=250bp 0bp 650bp 540bp,clip,scale=0.42]{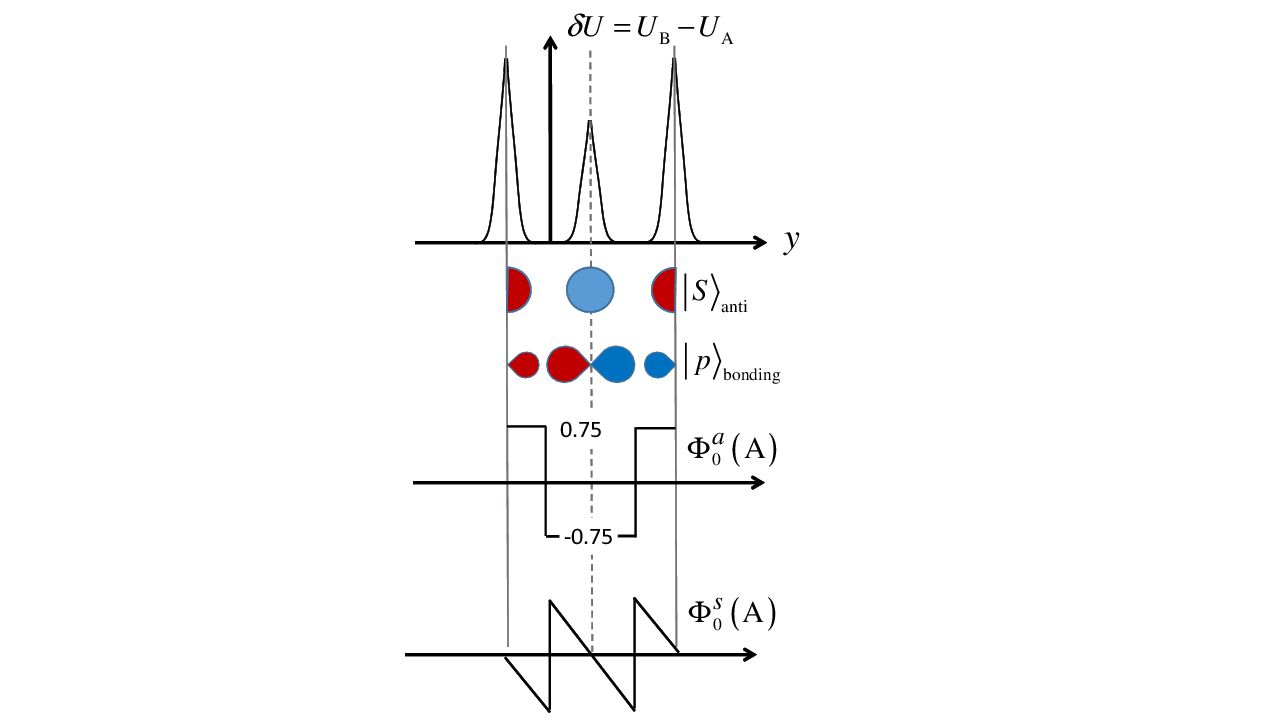}\tabularnewline
(a)\label{4(a)} & (b)\label{4(b)}\tabularnewline
\end{tabular}\caption{{\footnotesize{}\label{Figure 4}(Color online) (a) Schematic microscopic
potential of Mercury Telluride near the edge of a HgTe/CdTe two dimensional
QW. The upper panel in (a) shows the microscopic crystal potential,
$U_{A}$, in the QW, and $U_{B}=0$ is the potential in the wall,
where $U_{B}=U_{A}+G\left(y\right)\delta U$. The step function, $G\left(y\right)$,
is shown in the middle panel and the perturbing potential, $\delta U=U_{B}-U_{A}$,
is depicted in the lower panel. (b) Components $\Phi_{0}^{a}\left(y\right)$
and $\Phi_{0}^{s}\left(y\right)$ of the crystal periodic interface
function, $\Phi_{0}$, are shown over a distance of one monolayer
(width $a'$) for a mathematically abrupt interface, together with
the }\textit{\footnotesize{}s}{\footnotesize{}-antibonding and }\textit{\footnotesize{}p}{\footnotesize{}-bonding
crystal periodic functions, and the perturbation, $\delta U$. These
functions are used to calculate the interface potentials, $D_{0,nn'}=\langle n|\Phi_{0}\delta U|n'\rangle$.}}
\end{figure*}

Fig.~4(a) illustrates how materials \textit{A} and \textit{B} are
defined in the present treatment. Material \textit{A} represents the
TI semiconductor, which is depicted schematically along the \textit{y}-direction
in Fig.~\ref{Figure 4} for mercury telluride in the central plane
of a HgTe/CdTe QW grown in the \textit{z}-direction. If $U_{A}$ is
the microscopic crystal potential in the QW, $U_{B}=0$ represents
the potential in the wall which is treated as the vacuum or \textquotedblleft empty
crystal\textquotedblright{} with the same lattice parameter as material
\textit{A}. Thus $\delta U=-U_{A}$, which is depicted schematically
in the lower panel of Fig.~4(a). Local antibonding\textit{ s}- and
bonding \textit{p}-like band edge states in the semiconductor should
thus evolve into empty crystal states of the same symmetry in the
vacuum. 

\subsection{Four band empty crystal}

In the empty crystal, antibonding\textit{ s}- and bonding \textit{p}-like
states with a given spin can be constructed directly from degenerate
free electron states with wave vectors, $\frac{\pi}{a'}\left[\pm2,0,0\right]$,
$\frac{\pi}{a'}\left[0,\pm2,0\right]$ and $\frac{\pi}{a'}\left[0,0,\pm2\right]$,
normalized to an effective unit cell volume, $a'^{3}$, as follows:\citep{YuCardona1996,Cardona1966} 

\begin{subequations}
\begin{equation}
u_{s}=i\sqrt{\dfrac{2}{3a'^{3}}}\left[\cos\left(\dfrac{2\pi x}{a'}\right)+\cos\left(\dfrac{2\pi y}{a'}\right)+\cos\left(\dfrac{2\pi z}{a'}\right)\right]\label{eq:14a}
\end{equation}
\begin{equation}
u_{p}=\sqrt{\dfrac{1}{a'^{3}}}\left[\sin\left(\dfrac{2\pi x}{a'}\right)+i\sin\left(\dfrac{2\pi y}{a'}\right)\right]\label{eq:14b}
\end{equation}

\label{Eq. 14}
\end{subequations}

The hybridization parameter in the wall is thus:

\begin{equation}
A_{0}=-\dfrac{\hbar^{2}}{m_{0}}\langle u_{p}|\dfrac{\partial}{\partial y}|u_{s}\rangle\label{eq:15}
\end{equation}

which yields: 

\begin{subequations}
\begin{equation}
A_{0}=\hbar v_{w}\label{eq:16a}
\end{equation}
\begin{equation}
v_{w}=\dfrac{h}{\sqrt{6}m_{0}a'}\label{eq:16b}
\end{equation}
\end{subequations}
Taking $a'$= 3Å gives a value of $v_{w}=9.90\times10^{5}$~m/s.
Thus, the hybridization parameter for the wall calculated from Eq.~\eqref{eq:16b}
is 6.5 eV\,Å, which is much smaller than the free electron Dirac
value of 1973.5 eV\,Å.

\subsection{The need for passivation}

Unfortunately, a vacuum wall based on the empty crystal wave functions
defined in Eq.~\eqref{Eq. 14} does not obey a Dirac-like Hamiltonian,
because these wave functions are degenerate with an energy that is
much higher than the energies of the band gap states in the semiconductor.
Instead the wall behaves for spin up as:

\begin{equation}
H_{2\times2\uparrow}^{\mathrm{wall}}=A_{0}\left(\sigma_{x}k_{x}-\sigma_{y}k_{y}\right)+\sigma_{z}M_{0}+I_{2\times2}M_{1}\label{eq:17}
\end{equation}
with $M_{0}=0$ and $M_{1}=\hbar^{2}\left(2\pi/a'\right)^{2}/2m_{0}=16.7\,\mathrm{eV}$.
Although topological edge states are predicted from the change in
Chern number or $\mathbb{Z}_{2}$ index when the semiconductor band
gap inverts,\citep{Bernevig2006,Shen2011,Enaldiev2015} this wall
Hamiltonian does not ensure that the Dirac point lies in the band
gap of the semiconductor. For example, the SBC wave function has no
exponential solution in the wall with a real energy at $k_{x}=0$.

Treatments that realize a massless Dirac like edge dispersion, including
those based on OBCs\citep{Michetti2012b} and the SBC treatment presented
in this work, generally require a wall with a large band gap that
overlaps that of the semiconductor. Since this does not occur with
a vacuum, a passivation material is required, which should also have
band edge states with opposite parity, ideally with \textit{s}- and
\textit{p}-like symmetry, so that it behaves for spin up like $H_{2\times2\uparrow}^{\mathrm{wall}}$
with a large value of $M_{0}$, and with $M_{1}\simeq0$ (additional
quadratic terms, $\sigma_{z}\mathbf{k}\cdot B_{0}\mathbf{k}+I_{2\times2}\mathbf{k\cdot}D_{0}\mathbf{k}$,
can also be included). A good candidate appears to be silicon dioxide
(SiO\textsubscript{2}), which is the material deposited on the delineated
bar-shaped sample in at least two cases where the observation of one
dimensional edge states has been reported\citep{Konig2008,KnezParkin2014}.
Harrison\citep{Harrison1980} discusses the band structure of this
material which has a band gap of approximately 9 eV. The upper valence
band is composed of states which have bonding \textit{p}-like symmetry,
and in the cubic $\beta$-cristobalite phase, they mimic the $\left(X\pm iY\right)/\sqrt{2}$
basis of the TI Hamiltonian. The conduction band is composed of an
antibonding state with \textit{s}-like symmetry on both the silicon
and oxygen sites. The lattice parameter of the cubic unit cell is
about 7.1Å, but a tetragonal version exists with \textit{a} and \textit{c}-parameters
of 5.0 and 6.9 Å\citep{Coh2008}. The cubic lattice parameters for
HgTe/CdTe or InAs/GaSb/AlSb QWs, are 6.5 Å and 6.1 Å, respectively,
which are fairly similar to these values. It is reasonable to suppose
that in the ideal case the first few atomic layers of the passivation
layer will grow in registration with the TI semiconductor lattice
before the SiO\textsubscript{2} structure adopts a more complex lower
energy phase. A SiO\textsubscript{2} band gap of $\sim9$ eV corresponds
to $M_{0}=4.5\,\mathrm{eV}$. In addition, the difference between
the electron affinities of SiO\textsubscript{2} and HgTe or InAs
is about $4.3\pm0.3\,\mathrm{eV}$,\citep{Tash2016,Voitsekhovskii2010,Madelung1982}
so $M_{0}>>M_{1}$. Since the band edge states have the same \textit{s}-
and \textit{p}-like symmetries as the empty crystal states in Eq.~\eqref{Eq. 14},
the value of $A_{0}=6.5\mathrm{eV\,\mathring{A}}$ calculated above
should give a correct order of magnitude. The wave function decay
parameter in the SiO\textsubscript{2} is thus approximately $\frac{M_{0}}{A_{0}}\simeq0.7\mathrm{\mathring{A}}^{-1}$.
This corresponds to a decay length of the order of one effective lattice
parameter, $a'$, which is reasonably consistent with a $\mathrm{\mathbf{k\cdot p}}$
wave function containing only Fourier components in the first Brillouin
zone. It also shows that only a few SiO\textsubscript{2} monolayers
are required before the wave function has fully decayed, consistent
with the structural assumptions made above. 

If a passivation layer such as silicon dioxide is not used intentionally,
it is still quite possible that a thin native oxide with a large band
gap can form at an untreated wall after sample deliniation, e.g. mercury
oxide and/or tellurium oxide for an HgTe/CdTe QW sample. Even here,
the conduction and valence bands may be composed of the \textit{s}-and
\textit{p}-like orbitals of the oxygen and semiconductor atoms, so
it may still be possible to define a wall Hamiltonian with $M_{0}>>M_{1},M$.
An additional complication is the possible formation of trivial edge
states, whose presence may well depend on the choice of surface treatment
and passivation material. For example, bar samples have been fabricated
from InAs/GaSb/AlSb QWs by different groups, using silicon oxide or
silicon nitride \citep{Knez2014,KnezParkin2014}, and aluminium oxide
or hafnium oxide \citep{Nichelel2016}. In the second case there is
evidence of edge conduction in the normal phase, suggesting that trivial
edge states are present and calling into question whether this is
generally the case, or dependent on which passivation material is
used. 

\subsection{Interface band mixing}

The final term of Eq.~\eqref{eq:13} contains interface potentials,
$D_{0,nn'}$, which arise because the orbitals on the boundary layer
of atoms experience a different microscopic potential to their immediate
neighbors. The interface potentials are evaluated as: $D_{0,nn'}=\langle n|\Phi_{0}\delta U|n'\rangle$,
where the crystal periodic interface function, $\Phi_{0}\left(y\right)=\Phi_{0}^{a}\left(y\right)+\Phi_{0}^{s}\left(y\right)$,
is the sum of two components, the first of which has even parity and
the other, odd parity, with respect to the boundary atomic plane.\citep{Klipstein2010}
In the example in Fig.~\ref{Figure 4} with an interface to the vacuum
at $y=0$, this is the plane of mercury atoms closest to the origin.
The functions, $\Phi_{0}^{s/a}\left(y\right)$, are depicted over
one unit cell in the lower part of Fig.~4(b), for a mathematically
abrupt interface. For a more realistic interface with a finite width
they become significantly dampened, as shown in Fig. 3 of Ref. \onlinecite{Klipstein2010}
for an interface with a width of 0.8Å. Also shown in Fig.~4(b) are
the crystal periodic perturbing potential, $\delta U$, as in Fig.~4(a),
and the antibonding \textit{s}- and bonding \textit{p}-like crystal
periodic functions of the reference crystal, which are depicted blue
when positive and red when negative. The example in Fig.~\ref{Figure 4}
is for a vacuum interface, but the perturbing potential, $\delta U=U_{B}-U_{A}$,
can easily be modified to represent an interface with a passivation
material, such as silicon dioxide whose microscopic potential is then
represented by $U_{B}$. This material is assumed to be periodic for
the first one or two monolayers as discussed above, and so can be
treated here with perfect periodicity because we are only interested
in the interface region over the width of the delta function, $\widetilde{\delta}\left(y\right)$,
which is about one monolayer. Inspection of the symmetries in Fig.~4(b)
shows that there will be four finite contributions to $D_{0.nn'}$,
namely $D_{0.\mathrm{SS}}$, $D_{0.\mathrm{XX}}$ and $D_{0.\mathrm{YY}}$,
which have a finite matrix element with $\Phi_{0}^{a}\left(y\right)$,
and $D_{0.\mathrm{SY}}$, which has a finite matrix element with $\Phi_{0}^{s}\left(y\right)$.
The spin up edge state wave function is the product of a hybridized
crystal periodic wave function:\citep{Klipstein2015} $|+\rangle=\frac{1}{\sqrt{2}}|iS\rangle\uparrow-\frac{1}{2}\left(X+iY\right)\uparrow$
and an envelope function, $\psi_{\uparrow\sigma(k_{x})}\left(y\right)$,
so the energy shift of the Dirac point due to the interface band mixing
is approximately: 

\begin{equation}
\begin{array}{ccc}
\delta E & = & \eta|\psi_{\uparrow\sigma\left(0\right)}\left(0\right)|^{2}\\
\\
 &  & \times\left[\dfrac{2D_{0,\mathrm{SS}}+D_{0,\mathrm{XX}}+D_{0,\mathrm{YY}}}{4}-\dfrac{D_{0,\mathrm{SY}}}{\sqrt{2}}\right]\\
 & = & \eta|\psi_{\uparrow\sigma\left(0\right)}\left(0\right)|^{2}\overline{D}_{0}
\end{array}\label{eq:18}
\end{equation}
in which $\sigma\left(k_{x}\right)$ is the decay parameter of the
physical edge state in the semiconductor. The dependence on the squared
amplitude of the envelope function at $y=0$ is due to the delta function
in Eq.~\eqref{eq:13}, where $\widetilde{\delta}\left(y\right)$
has been replaced by $\delta\left(y\right)$. Since the wave function
in the wall decays at the about the same rate as $\widetilde{\delta}\left(y\right)$,
this will lead to an overestimate, so a correction factor, $\eta$,
has been introduced into Eq.~\eqref{eq:18} where $\eta\simeq0.5$.
The value of $\overline{D}_{0}$ depends on the magnitudes and signs
of the four contributions in the square bracket, which can only be
estimated using microscopic calculations. Estimates for the interface
band mixing potentials at a superlattice interface vary widely, and
are typically in the range $0.1-2\,\mathrm{eV}\textrm{Å}$.\cite{Foreman1998,Livneh2012,*LivErratum2014}
Noting that $|\psi_{\uparrow\sigma(0)}\left(0\right)|^{2}=2\sigma_{\mathrm{D}}$
where $\sigma_{\mathrm{D}}$ may be estimated using Eq.~\eqref{eq:12},
and assuming a relatively large value of $\overline{D}_{0}=2\,\mathrm{eV}\textrm{Å}$,
the Dirac point is predicted to shift by $\delta E$ = 0.004~eV when
$B=135.3\,\mathrm{eV\mathring{A}^{2}}$ and $D=107.2\,\mathrm{eV\mathring{A}^{2}}$
(as in Table~\ref{tab:Table 3}). An important point to note is that
the shift of the edge state dispersion is largest at the Dirac point
and decreases with increasing edge state wave vector, because it is
proportional to $2\sigma\left(k_{x}\right)$, which vanishes at the
merging points with the bulk band structure. Thus interface band mixing
may shift the Dirac point and distort the edge state dispersion, but
the merging points will remain fixed. 

\subsection{Effect of the $D'$ term\label{subsec:D primed term}}

Finally, the significance of the quadratic term proportional to $D'$
in the first line of the multiband spin up Hamiltonian of Eq.~\eqref{eq:3}
must be considered. In the present work this term has been ignored
because it is quite small. For example, its inclusion causes a virtually
imperceptible shift of the middle states in Fig.~2(a), of only $-17\,\mu\mathrm{eV}$
at mid gap reducing to zero at the band edges. Moreover its matrix
element with the edge state wave functions diverges due to a discontinuity
in the first derivative of the wave function at the boundary, and
to a contribution in the wall which tends to infinity with increasing
wall potential, $M_{0}$. Assuming that the energy of the Dirac point
varies smoothly with increasing $D'$, this shows that a non-zero
value of $D'$ must lead to a non-exponential wave function, with
a continuous first derivative and with a more linear mode of decay
in the wall. This is consistent with the two band spin up model, where
inclusion of the quadratic term in Eq.~\eqref{eq:3} results in the
addition of $D'=\tfrac{\hbar^{2}}{2m_{0}}$ to the expression for
\textit{D} in Eq.~\eqref{eq:4b}. As for the multiband case, this
has negligible effect on the dispersion of the middle states in the
semiconductor. In the wall, an exponential solution requires that
the band asymmetry parameter varies as $D_{0}=\frac{\sigma}{\sigma_{0}}D$,
vanishing when $M_{0}\rightarrow\infty$.\citep{Klipstein2015} If
instead $D_{0}\rightarrow D'$ and does not vanish, the solution will
again be non-exponential. There will thus be a shift in the edge state
energy in both models, compared with the exponential solutions calculated
for $D'=0$. Based on the small value of $D'$, and in the absence
of an exact numerical solution, this shift is assumed to be small. 

\section{\label{sec:6-CONCLUSION}CONCLUSION}

The four band BHZ Hamiltonian provides a simple but realistic description
of the band edge states in 2D TIs such as HgTe/CdTe and InAs/GaSb/AlSb.
However, the validation of hard wall boundary conditions appropriate
to these materials has remained elusive. The most popular choice is
OBCs which avoid any explicit treatment for the wall, while other
boundary conditions tend to be phenomenological in nature, so the
connection with the microscopic structure of the wall remains unclear.
In this work, OBCs have been ruled out, because they fail when remote
states are included explicitly in the semiconductor Hamiltonian. At
the same time, the other boundary conditions show that a wide variety
of edge state dispersions are possible, depending on the values of
the phenomenological parameters. Therefore a different approach has
been adopted here, based on SBCs which address more directly both
the microscopic properties of the wall, and the conditions that can
lead to a Dirac point in the semiconductor band gap. 

The use of OBCs leads to contactless edge confinement, and a Dirac
point with an unphysical dependence on the TI band parameters. Other
unphysical results have also been reported when these boundary conditions
are used,\citep{Klipstein2018} and further aspects are discussed
in Appendix B. Although mathematically correct, it is known that
one of the two gap solutions for a given spin direction is spurious.
Unfortunately, this solution must be combined with the physical gap
solution in order to satisfy OBCs, and such combination is only possible
because both solutions have the same eigenvector. The spurious solution
is related to diagonal \textit{k}-quadratic terms which are introduced
into the BHZ Hamiltonian when remote states are eliminated using perturbation
theory. If the remote states are not eliminated but included in a
larger, multiband Hamiltonian, the eigenvectors of the two gap solutions
are no longer equal and an OBC solution is impossible. This is because
the spurious solution is replaced by a physical solution with large
eigenvector components from the remote states. 

An alternative approach suggested previously by the author is to use
SBCs. These boundary conditions match both the wave function and the
product of its derivative and a reciprocal mass term at the interface
between the semiconductor and the wall, and are obtained by integrating
the BHZ Hamiltonian across the boundary region. Unlike other SBC treatments
which have been applied to a soft wall and which evolve into OBCs
when the wall becomes hard, the present SBC approach is for a hard
wall and results in wave function confinement with a large amplitude
at the edge. This type of confinement is typical of other classic
surface phenomena, such as surface plasmons and phonons. The SBC edge
state wave function is constructed from just the physical gap solutions
on each side of the boundary. It has been verified in the present
work by comparison with a multiband solution, with which it is in
complete agreement. In both cases, the Dirac point has a dependence
on the TI band parameters that is physically justifiable. 

One of the challenges of the SBC approach is to establish the strength
of the electron-hole hybridization in the wall. When the semiconductor
and wall have the same mid gap energies but different hybridization
parameters, the Dirac point is slightly shifted from mid gap. In this
work the wall was considered initially as an empty crystal, with a
free electron basis that matches the symmetry of the antibonding \textit{s}-states
and bonding \textit{p}-states in the semiconductor. This establishes
an estimate for the wall hybridization parameter that behaves like
the free electron Dirac value, $A_{0}=\hbar c$, but with the speed
of light, \textit{c,} replaced by the velocity, $v_{w}=\frac{h}{\sqrt{6}m_{0}a'}$
in which $a'$ is the monolayer thickness in the semiconductor. The
wall hybridization parameter is then about 6.5 eV Å, which is comparable
to the semiconductor value. The empty crystal Hamiltonian represents
a hard vacuum like wall, but unfortunately there is no splitting between
the electron and hole states, which are also much higher in energy
than those in the semiconductor. In order to ensure edge states with
a Dirac point that lies in the semiconductor band gap, it appears
that a thin passivation layer is required, which may be the native
oxide of the material, or an externally deposited dielectric material
such as silicon dioxide, for which successful observations of the
quantum spin Hall effect have been reported. Silicon dioxide indeed
has bands of the correct \textit{s}- and \textit{p}-like symmetry
separated by a large band gap that overlaps that of the semiconductor
fairly symmetrically. Assuming a pseudomorphic cubic phase immediately
next to the semiconductor, its hybridization parameter should have
a similar magnitude to that estimated for the \textit{s}- and \textit{p}-states
of the empty crystal. The nature of the passivation layer may also
be important to avoid the presence of trivial edge states. 

Another challenge in the SBC approach is that the mode of decay of
the edge state wave function is generally non-exponential. Nevertheless,
it is possible to find useful exponential solutions when the wall
is effectively infinite, $M_{0}>>|M|$, and the edge state decay parameter
is real as in HgTe/CdTe QWs. Although this has been the main focus
of the present work, it was also confirmed that weakly hybridized
systems with a complex decay parameter, such as InAs/GaSb/AlSb, exhibit
non-exponential behavior when the band structure is asymmetric. Even
for weakly hybridized systems with a symmetric band structure, SBCs
have been shown previously to exhibit non-exponential behavior in
narrow samples, except at certain characteristic widths.\citep{Klipstein2018}

The shift of the Dirac point from mid gap is generally small when
SBCs are used with wall passivation. However, several additional factors
could affect its position, including the small quadratic $D'$ term
left out of the multiband Hamiltonian (see Sec.~\ref{sec:5-WALL}),
the difference in the mid gap energies of the semiconductor and passivation
materials ($M_{1}\neq0$ in Eq.~\eqref{eq:17}), and the effect of
interface band mixing. The first two factors should not be too significant,
especially if a passivation material is used with $M_{1}<<M_{0}$,
but the third could be important, producing a shift which is greatest
at the Dirac point and which steadily decreases to zero at the wave
vectors where the edge state dispersion merges with the bulk band
edges. Microscopic calculations of the interface band mixing potentials
are therefore needed to establish the size of this effect. 

In summary, the present work has demonstrated that OBCs must be replaced
by boundary conditions that allow stronger edge confinement with a
large amplitude of the wave function at the sample edge, and that
this results in a modified edge state dispersion. In addition, a passivation
layer is often present, either intentionally or due to the formation
of a native oxide, and it was found that this may even be an essential
way of ensuring that the Dirac point lies in the semiconductor band
gap. Although scattering from edge imperfections is suppressed by
time reversal symmetry, a practical consequence of the present hard
wall SBC approach is that stronger wave function confinement, combined
with significant disorder in the passivation layer, could both lead
to a lower threshold for the breakdown of dissipationless transport
than previously thought.\citep{Pezo2020} 

\href{}{ }
\begin{acknowledgments}
The author acknowledges useful correspondence with Dr. M. V. Durnev.
\end{acknowledgments}

\appendix

\section{\label{Appendix A}Full edge state dispersion in the multiband model}

The full spin up edge state dispersion, $E(k_{x})$ is calculated
by finding solutions for $H'_{4\times4\uparrow}\left(\mathbf{k}\right)$
with the same eigenvectors on both sides of the boundary. Assuming
$D'=0,$ and writing $k_{y}=i\sigma$, Eq.~\eqref{eq:3} can be solved
for the decay parameter, $\sigma$, and eigenvector, {[}\textit{a},
\textit{b}, \textit{c}, \textit{d}{]} in the semiconductor in terms
of the energy, \textit{E}, and wave vector, $k_{x},$ to yield the
relations in Table~\ref{tab:Table 4}, where:

\begin{table*}
\caption{{\footnotesize{}\label{tab:Table 4}Decay parameters in the }\textit{\footnotesize{}y}{\footnotesize{}-direction
at energy, }\textit{\footnotesize{}E}{\footnotesize{}, and edge wave
vector, $k_{x}$, and the corresponding eigenvectors, for the 4 band
spin up Hamiltonian with an asymmetric band structure. The functions,
$Z_{\pm}$, $\Delta'_{1}$, $\Delta'_{2}$, and }\textit{\footnotesize{}N}{\footnotesize{}
are defined in Eq.~\eqref{Eq. A1}.}}

\begin{ruledtabular}
\begin{tabular*}{8cm}{@{\extracolsep{\fill}}ll}
\noalign{\vskip\doublerulesep}
Decay Parameter & Eigenvector\tabularnewline[\doublerulesep]
\noalign{\vskip\doublerulesep}
\hline 
\noalign{\vskip\doublerulesep}
\multirow{4}{*}{$\sigma_{\pm}=\theta\sqrt{k_{x}^{2}+\dfrac{\Delta'_{1}\Delta'_{2}A^{2}+Z_{+}\pm\sqrt{\Delta'_{1}\Delta'_{2}A^{2}\left\{ \Delta'_{1}\Delta'_{2}A^{2}+2Z_{+}\right\} +Z_{-}^{2}}}{2Q_{1}^{2}Q_{2}^{2}}}$} & $a=i\dfrac{Q_{1}\left(k_{x}+\sigma\right)}{AN\Delta'_{1}}\left(\dfrac{Q_{2}^{2}\left(k_{x}^{2}-\sigma^{2}\right)+\left(M-E\right)\Delta'_{2}}{Q_{2}\left(k_{x}^{2}-\sigma^{2}\right)}\right)$\tabularnewline[\doublerulesep]
\noalign{\vskip\doublerulesep}
\noalign{\vskip\doublerulesep}
 & $b=\dfrac{1}{N}\dfrac{\Delta'_{2}}{Q_{2}\left(k_{x}+\sigma\right)}$\tabularnewline[\doublerulesep]
\noalign{\vskip\doublerulesep}
\noalign{\vskip\doublerulesep}
 & $c=-\dfrac{1}{AN}\left(\dfrac{Q_{2}^{2}\left(k_{x}^{2}-\sigma^{2}\right)+\left(M-E\right)\Delta'_{2}}{Q_{2}\left(k_{x}^{2}-\sigma^{2}\right)}\right)$\tabularnewline[\doublerulesep]
\noalign{\vskip\doublerulesep}
\noalign{\vskip\doublerulesep}
 & $d=\dfrac{i}{N}$\tabularnewline[\doublerulesep]
\noalign{\vskip\doublerulesep}
\end{tabular*}
\end{ruledtabular}

\end{table*}

\begin{subequations}
\begin{equation}
Z_{\pm}=\left(M+E\right)\Delta'_{1}Q_{2}^{2}\pm\left(M-E\right)\Delta'_{2}Q_{1}^{2}\label{eq:A1a}
\end{equation}
\begin{equation}
N=\sqrt{T_{1}T_{3}^{2}+T_{2}}\label{eq:A1b}
\end{equation}
\begin{equation}
T_{1}=1+\left(\dfrac{Q_{1}\left(k_{x}+\sigma\right)}{\Delta'_{1}}\right)^{2}\label{eq:A1c}
\end{equation}
\begin{equation}
T_{2}=1+\left(\dfrac{\Delta'_{2}}{Q_{2}\left(k_{x}+\sigma\right)}\right)^{2}\label{eq:A1d}
\end{equation}
\begin{equation}
T_{3}=\dfrac{Q_{2}^{2}\left(k_{x}^{2}-\sigma^{2}\right)+\left(M-E\right)\Delta'_{2}}{AQ_{2}\left(k_{x}^{2}-\sigma^{2}\right)}\label{eq:A1e}
\end{equation}
\begin{equation}
\Delta'_{1}=\Delta_{1}-E\label{eq:A1f}
\end{equation}
\begin{equation}
\Delta'_{2}=\Delta_{2}+E\label{eq:A1g}
\end{equation}

\label{Eq. A1}
\end{subequations}
and $\theta=1$ in the semiconductor. In the wall $\theta=-1$ and
the corresponding parameters are $A_{0}$, $M_{0}$, $Q_{10}$, $Q_{20}$,
$\Delta_{10}$, $\Delta_{20}$, and $\sigma_{0}$. The relations in
Table~\ref{tab:Table 4} can be expressed in terms of these parameters
and rearranged to yield the band gap ratios in the wall that correspond
to a specific eigenvector:

\begin{subequations}
\begin{equation}
\dfrac{\Delta_{10}}{M_{0}}=\dfrac{Q_{10}}{Q_{20}}\left\{ \dfrac{c}{a}\right\} \left[-i\dfrac{2Q_{20}k_{x}}{M_{0}}+\dfrac{\left(1-\dfrac{E}{M_{0}}\right)\left\{ \dfrac{b}{d}\right\} }{1+i\dfrac{A_{0}}{Q_{20}}\left\{ \dfrac{c}{d}\right\} }\right]+\dfrac{E}{M_{0}}\label{eq:A2a}
\end{equation}
\begin{equation}
\dfrac{\Delta_{20}}{M_{0}}=i\dfrac{2Q_{20}k_{x}}{M_{0}}\left\{ \dfrac{b}{d}\right\} -\dfrac{\left(1-\dfrac{E}{M_{0}}\right)\left\{ \dfrac{b}{d}\right\} ^{2}}{1+i\dfrac{A_{0}}{Q_{20}}\left\{ \dfrac{c}{d}\right\} }-\dfrac{E}{M_{0}}\label{eq:A2b}
\end{equation}

\label{Eq. A2}
\end{subequations}

It is clear that these ratios become independent of the wall band
gap, $2M_{0}$, when $M_{0}>>|M|$, since \textit{E} has the same
order of magnitude as the semiconductor band gap parameter, \textit{M},
so $|\Delta_{10}|,|\Delta_{20}|\rightarrow\infty$ as $M_{0}\rightarrow\infty$.
For given values of \textit{E} and $k_{x}$, the semiconductor eigenvector
components are calculated according to the relations in Table~\ref{tab:Table 4}
and inserted into Eq.~\eqref{Eq. A2}. This yields the wall band
gaps for a given set of wall hybridization parameters, and $M_{0}$.
These band gaps should be substituted back into the relations in Table~\ref{tab:Table 4}
and the process repeated at different energies until the wall eigenvector
comes out to be the same as the semiconductor eigenvector, giving
a self-consistent solution at the chosen wave vector. The values of
the decay parameters in the semiconductor and wall can also be found
for each point on the dispersion curve, using the formula in the left
hand column of Table~\ref{tab:Table 4}.

As discussed in Sec.~\ref{sec:2-MULTI-BAND}, it appears to be a
typical feature of SBC edge states that an exponential solution only
exists when specific wall parameters have a certain ratio with the
wall band gap. In the two band case, the parameters in question are
$D_{0}$ and $B_{0}$, where a useful exponential solution can be
found in the infinite wall limit when $D_{0},B_{0}\rightarrow0$.
For other ratios of the wall parameters, a numerical approach must
be used. The multiband spin up model exhibits similar features, where
an exponential solution exists for the wall band gaps given by Eq.~\eqref{Eq. A2},
giving a useful exponential solution in the limit: $|\Delta_{10}|,|\Delta_{20}|\rightarrow\infty$
as $M_{0}\rightarrow\infty$. Practically, this \textquotedblleft hard
wall\textquotedblright{} limit corresponds to $M_{0}>>|M|$, which
is already obeyed very well for $M_{0}>1\,\mathrm{eV}$ in the examples
discussed below.\footnote{This is also true for solutions derived with $k_{+}$ and $k_{-}$
interchanged in the \textit{Q}-terms of Eq.~\eqref{eq:3}, corresponding
to a change in symmetry of the remote states. Note that the interchange
leads to similar but modified expressions for the eigenvectors in
Table~\ref{tab:Table 4} and the band gap ratios in Eq.~\eqref{Eq. A2}.} 

\begin{figure*}
\begin{tabular}{ccc}
\includegraphics[viewport=150bp 50bp 550bp 490bp,clip,scale=0.38]{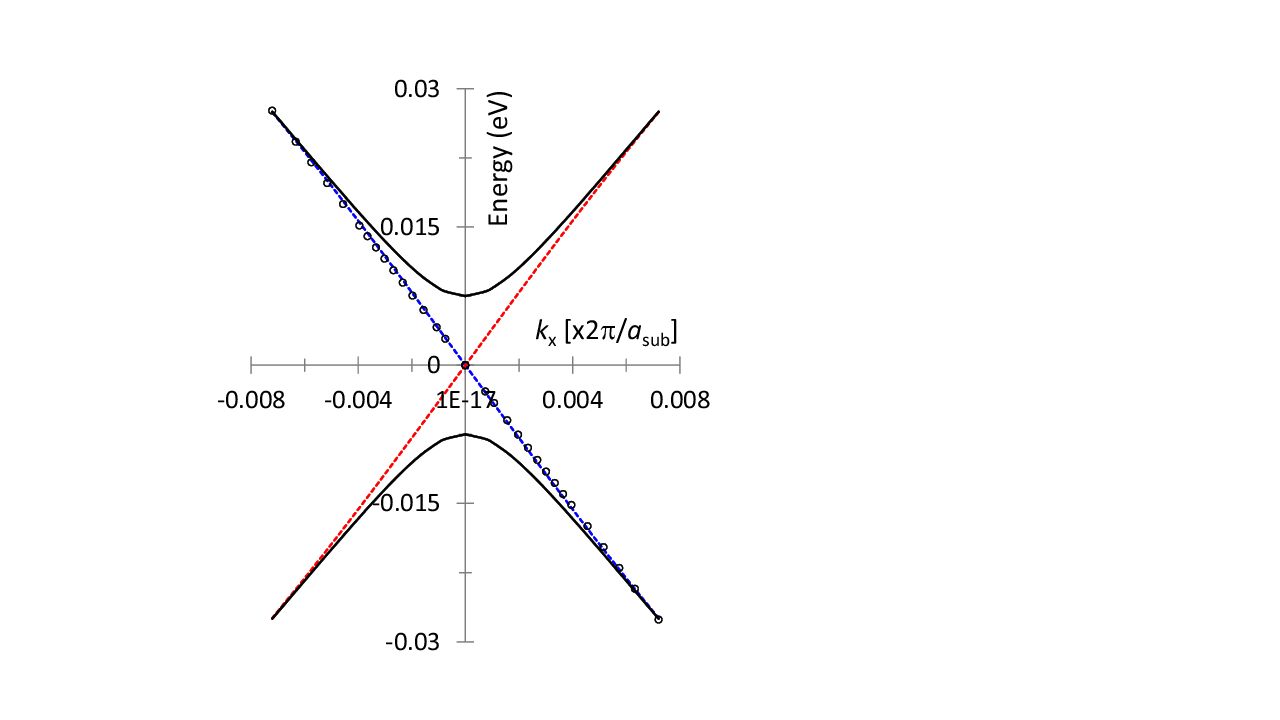} & \includegraphics[viewport=150bp 50bp 535bp 490bp,clip,scale=0.38]{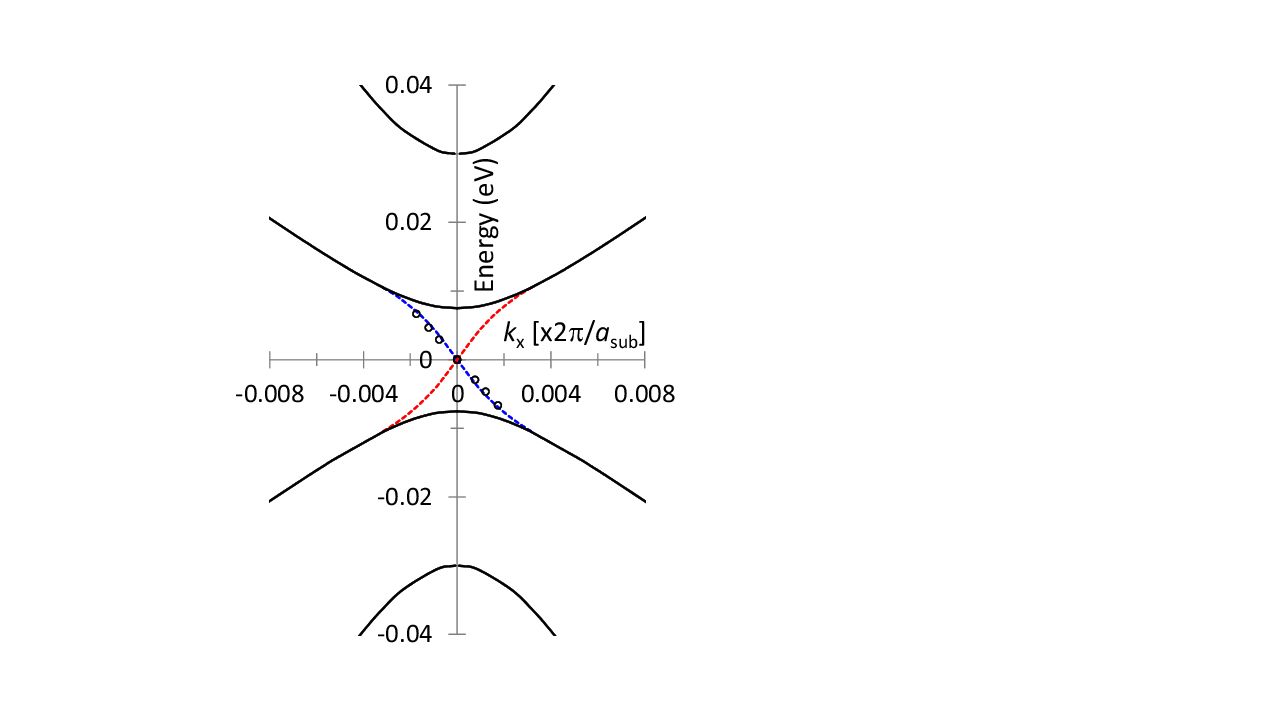} & \includegraphics[viewport=150bp 50bp 535bp 490bp,clip,scale=0.38]{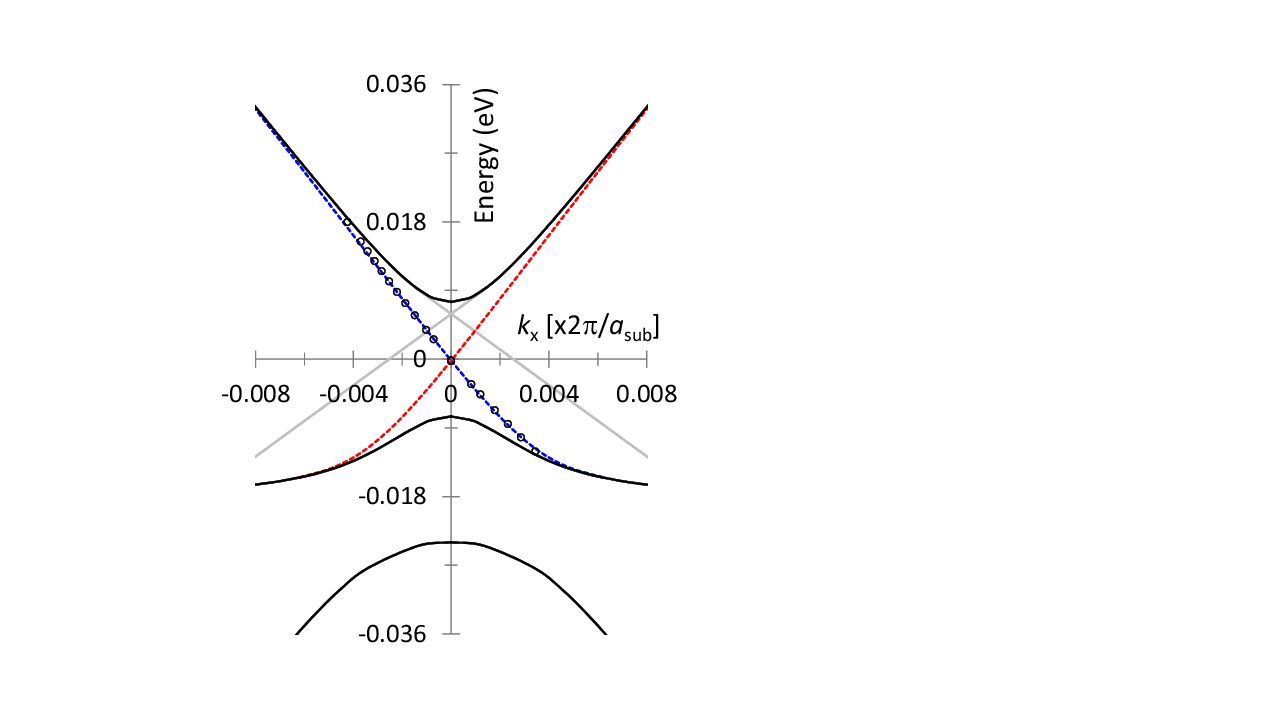}\tabularnewline
\label{5(a)}(a) & \label{5(b)}(b) & \label{5(c)}(c)\tabularnewline
\end{tabular}

\caption{{\footnotesize{}\label{fig:Figure 5}(Color online) Edge state dispersions
for the 4 band spin up (down) Hamiltonian are shown as a blue (red)
dashed curves superimposed on the nearby bulk bands (black curves).
In each case, the wall has parameters $M_{0}=20\,\mathrm{eV}$ , $A_{0}=11\,\mathrm{eV\,\mathring{A}}$,
and $Q_{10}=Q_{20}=2\,\mathrm{eV\,\mathring{A}}$. The semiconductor
parameters in (a) are the same as in Table \ref{tab:Table 2}. In
(b), the outer bands of the semiconductor are closer to the inner
bands than in (a), with $\Delta_{1}=\Delta_{2}=0.03\,\mathrm{eV}$.
In (c) the asymmetric semiconductor band parameters are those listed
in the left hand column of Table \ref{tab:Table 3}. Open circles
depict the edge dispersions of the 2 band spin up Hamiltonian (Eq.~\eqref{eq:11})
with a Dirac point energy of zero in (a) and (b) and $-0.00020\,\mathrm{eV}$
in (c). For comparison, the OBC dispersion is also shown in (c) as
gray lines.}}
\end{figure*}

The blue dashed curves in Fig.~\ref{fig:Figure 5} show three examples
of the spin up edge state dispersions calculated for symmetric or
asymmetric semiconductor band structures. In each case, the wall parameters
are $M_{0}$= 20 eV, $A_{0}$ = 11 eV\,Å, and $Q_{10}=Q_{20}=$ 2
eV\,Å, where $M_{0}$ and $A_{0}$ have the same order of magnitude
as the values estimated for a passivation material in Sec.~\ref{sec:5-WALL}.
In all three cases the dispersions are insensitive to a variation
in the value of the wall band gap by a factor of greater than 0.05
(corresponding to $M_{0}>1\,\mathrm{eV}$). In Fig.~5(a), the semiconductor
parameters are the same as in Table \ref{tab:Table 2} for a symmetric
TI band structure. The open circles depict a variation of the form:
$E=-Ak_{x}$, and there is virtually no difference between this variation
and the blue dashed curve, showing that the dispersion calculated
from the four band spin up model has the same linear variation as
for the two band case. The red dashed curve is for the spin down edge
state and is obtained from the blue curve by time reversal. 
\begin{figure}
\begin{tabular}{cc}
\includegraphics[viewport=150bp 0bp 700bp 500bp,clip,scale=0.22]{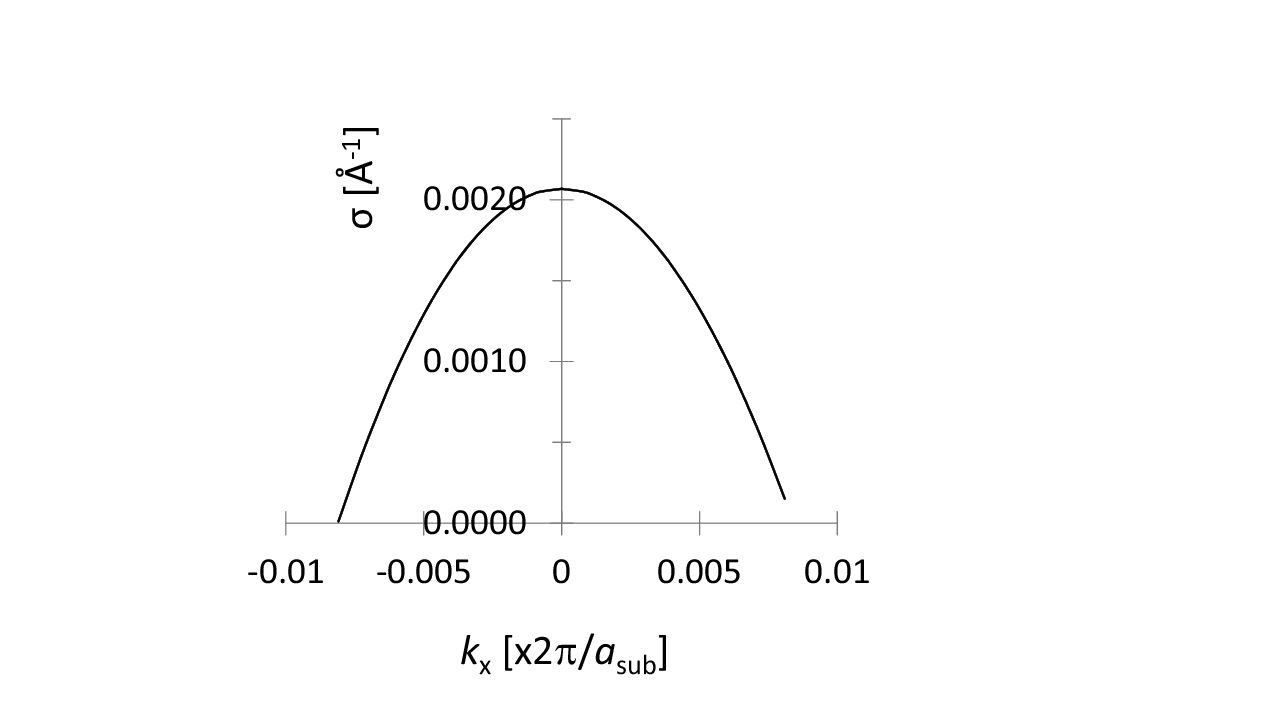} & \includegraphics[viewport=150bp 10bp 690bp 510bp,clip,scale=0.23]{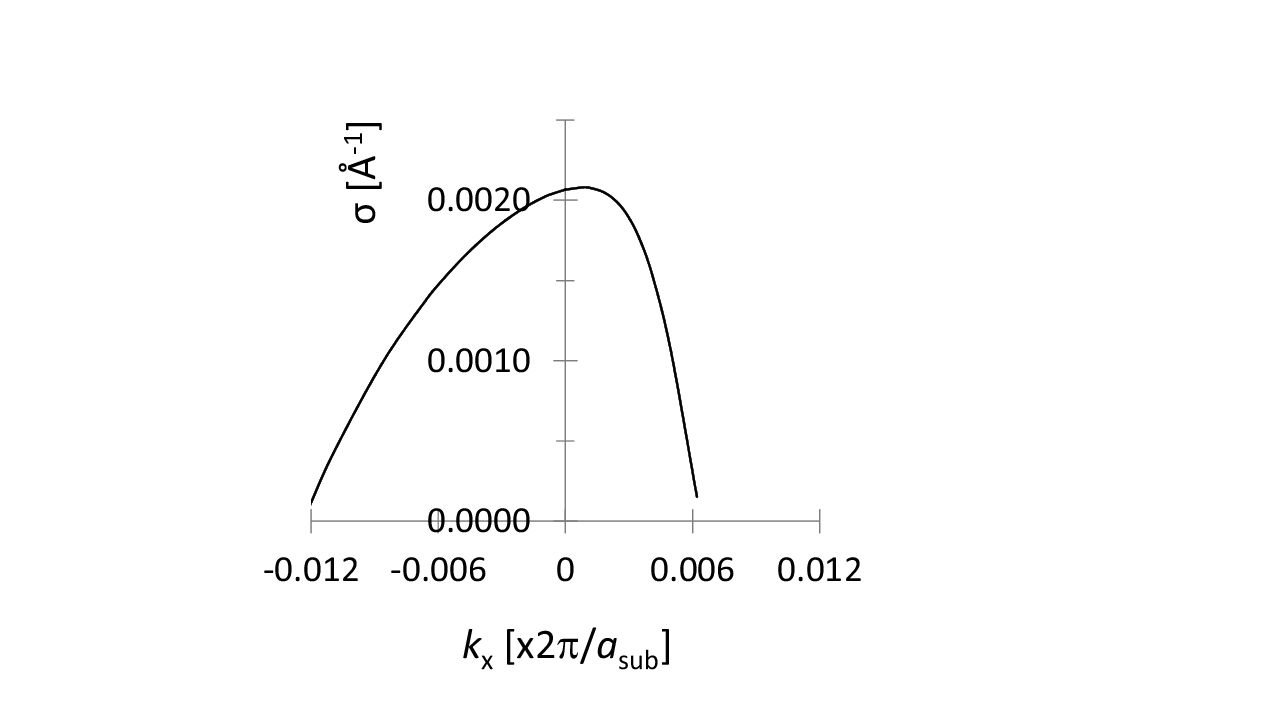}\tabularnewline
\label{6(a)}(a) & \label{6(b)}(b)\tabularnewline
\end{tabular}

\caption{{\footnotesize{}\label{fig: Figure 6}(Color online) Decay parameters
for the spin up edge states shown by the blue dashed curves in Figs.~5(a)
and 5(c), respectively.}}
\end{figure}

In Fig.~5(b), the outer band gap in the semiconductor is reduced
to $\Delta=\Delta_{1}=\Delta_{2}=0.03\,\mathrm{eV}$. Based on Eq.~\eqref{Eq. 4},
this is equivalent to $B=650\,\mathrm{eV\,\mathring{A}^{2}}$ for
the two band case. All other parameters in the wall and semiconductor
are the same as for Fig.~5(a), and except near the band edges, the
multiband result is still very close to the circles which depict the
two band variation, $E=-Ak_{x}$. Fig.~5(c) shows an example with
asymmetric semiconductor band parameters, as listed in the left hand
column of Table \ref{tab:Table 3}. As already discussed for this
case in Sec.~\ref{sec:4-MULTI-BAND-A0}, the Dirac point is very
close to mid gap, with an energy of $E_{4\times4\uparrow}^{\mathrm{D}}=-0.00020\,\mathrm{eV}$.
Using this value as an estimate for the two band Dirac point energy,
the open circles in Fig.~5(c) depict the dispersion based on the
two band spin up result, $E=-Ak_{x}+Dk_{x}^{2}+E_{2\times2\uparrow}^{\mathrm{D}}$,
where the square root in Eq.~\eqref{eq:11} is very close to unity.
The two band spin up dispersion again agrees quite well with the calculated
multiband result. 

The OBC dispersion, given by Eq.~\eqref{eq:9}, may be compared with
the SBC results in Fig.~\ref{fig:Figure 5}. It is the same as the
open circles for the symmetric cases in Figs.~5(a) and 5(b) but
very different for the asymmetric case in Fig.~5(c), where it is
shown as gray lines. For the asymmetric case, the Dirac point is strongly
shifted toward the conduction band and the edge state dispersion is
linear with a much reduced phase velocity.

The wave vector dependence of the spin up decay parameters in the
multiband treatment is shown in Fig.~\ref{fig: Figure 6}, where
it is compared for the symmetric and asymmetric cases shown in Figs.~5(a)
and 5(c), respectively. The wave vectors of the merging points can
clearly be identified where $\sigma\rightarrow0$. In both cases,
the decay parameter at the Dirac point agrees with Eq.~\eqref{eq:12}
which has a value close to $\sigma_{\mathrm{D}}\simeq M/A=0.0020~\mathrm{\mathring{A}}^{-1}$
when $A^{2}>>4MB$ as in these examples. For the symmetric case, the
two band model corresponds fairly well with the multiband model over
the whole wave vector range, with merging points in the two band model
at $\pm0.083\times\frac{2\pi}{a_{\mathrm{sub}}}$. For the asymmetric
case, however, the two band model gives merging points at $\pm0.072\times\frac{2\pi}{a_{\mathrm{sub}}}$,
while the magnitudes of the merging wave vectors in Fig.~6(b) are
similar to these values but unequal. This can be attributed to stronger
band non-parabolicities in the asymmetric multiband model. 

All of the edge dispersions in Figs.~\ref{fig:Figure 5} and \ref{fig: Figure 6}
are based on middle wave vector solutions in the wall and semiconductor.
As mentioned in Sec.~\ref{sec:3-TWO-BAND}, when the wing solutions
are used in the asymmetric two band spin up model, they only yield
an edge state when \textit{D} < \textit{B}, which then has an anomalous
dispersion, as shown for example with $A_{0}=A$ in Fig. 2(a) of Ref.
\onlinecite{Klipstein2015}. This is also true in the four band spin
up model. Using parameters corresponding to Fig.~5(c) but with $A_{0}=A$,
the edge dispersion fails to cross or merge with the bulk bands, and
no exponential edge state can be found for $A_{0}=11\,\mathrm{eV\mathring{A}}$.
Therefore, this edge state can be considered unphysical even though,
in principle, the four band spin up Hamiltonian gives physical wing
solutions, as shown, for example, in Fig.~2(b). This behavior is
related to the large magnitude of the imaginary wing wave vector in
the wall Hamiltonian, which tends to infinity as $\Delta_{10},\Delta_{20}\rightarrow-\infty$.
Even for large but finite values of these parameters, its magnitude
is well beyond the boundary of the Brillouin zone, so a hard wall
edge state based on the wing solutions cannot be described physically,
even with four bands.

Finally, it should be noted that the band gap ratios in Eq.~\eqref{Eq. A2}
that correspond to exponential edge state solutions are \textit{k}-dependent,
so the outer bands have a significant dispersion even when the inner
bands do not (i.e. constant $M_{0}$). Since all bands in the wall
are effectively at infinity, and the results in Fig.~\ref{fig:Figure 5}
are totally insensitive to the positions of the inner bands for $M_{0}>1\,\mathrm{eV}$,
it is anticipated that the effect of the outer band positions on the
energy dispersions should not be too significant. Nevertheless, in
order to test this assumption a full numerical treatment is needed
for constant values of all the wall band gaps, when the wave function
will generally exhibit a non-exponential decay. 

\section{\label{Appendix B}Topologically trivial edge solutions}

It has recently been pointed out that edge states can be found for
the BHZ Hamiltonian in the topologically trivial phase, with $M>0$.\citep{Candido2018}
In this Appendix, such behavior is confirmed for both OBC and SBC
boundary conditions. However, all such states include the wing solution
and are therefore unlikely to exist in any real physical system. No
edge state can be found using SBCs based only on the physical middle
solution.

\begin{figure}
\includegraphics[viewport=900bp 300bp 5000bp 2000bp,scale=0.115]{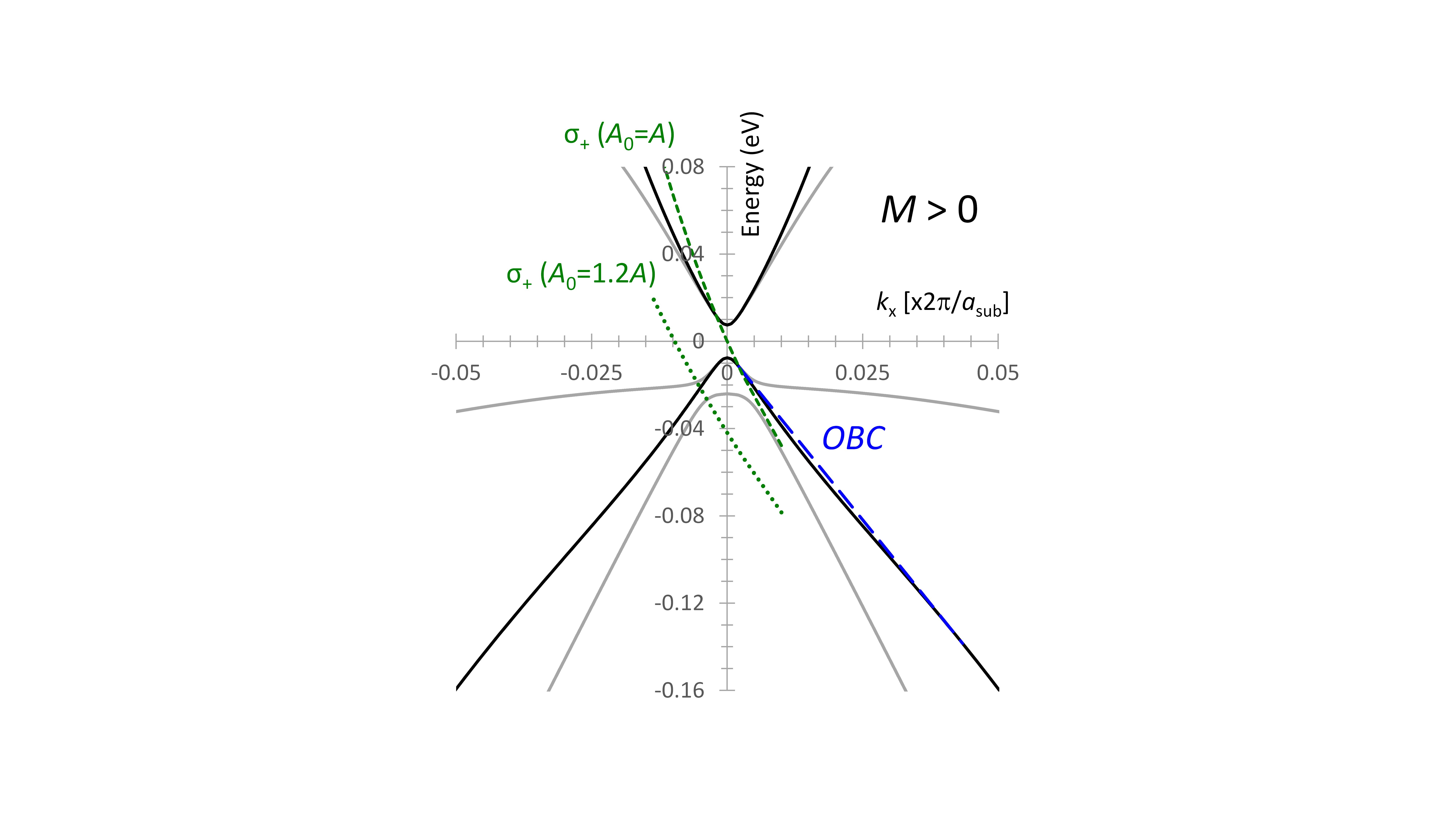}

\caption{{\footnotesize{}\label{fig:Figure 7}(Color on line) BHZ conduction
and valence bands for a topologically trivial insulator with $M=0.0075\,\mathrm{eV},$
$A=4\,\mathrm{eV\,\mathring{A}}$, $B=76.2\,\mathrm{eV\,\mathring{A}^{2}}$,
and $D=50.8\,\mathrm{eV\,\mathring{A}^{2}}$(black), superimposed
on the corresponding 8 band dispersion with $\Delta_{1}=0.15\,\mathrm{eV}$,
$\Delta_{2}=0.024\,\mathrm{eV}$ and $Q_{1}=Q_{2}=2\,\mathrm{eV\,\mathring{A}}$
(gray). There is good correspondence for $|k_{x}|<k_{\mathrm{max}}$
(defined in Sec.~\ref{sec:1-INTRODUCTION}). A spin up OBC edge solution
extends to wave vectors well beyond $k_{\mathrm{max}}$ (blue, long
dash). Spin up SBC edge states with decay parameter, $\sigma_{+}$,
are also plotted for wall hybridization parameters of $4$ and $5\,\mathrm{eV\,\mathring{A}}$
(green short-dash and dot, respectively, and only shown near the zone
center). All of these solutions do not represent real physical edge
states because they involve the spurious wing solution. There is no
SBC edge state with physical decay parameter, $\sigma_{-}$.}}
\end{figure}

The band structure near the bulk band gap of a topologically trivial
insulator is shown in gray in Fig.~\ref{fig:Figure 7}, for an eight
band Hamiltonian with $M=0.0075\,\mathrm{eV},$ $A=2Q_{1}=2Q_{2}=4\,\mathrm{eV\,\mathring{A}}$,
$\Delta_{1}=0.15\,\mathrm{eV}$, and $\Delta_{2}=0.024\,\mathrm{eV}$.
The conduction and valence bands of the BHZ Hamiltonian are superimposed
in black, with quadratic coefficients, $B=76.2\,\mathrm{eV\,\mathring{A}^{2}}$
and $D=50.8\,\mathrm{eV\,\mathring{A}^{2}}$, calculated from Eq.~\eqref{Eq. 4}.
It can be seen that for wave vectors beyond the range allowed by perturbation
theory, $k_{\mathrm{max}}\simeq0.0075\times\frac{2\pi}{a_{\mathrm{sub}}}$,
the conduction and valence bands of the BHZ Hamiltonian deviate strongly
from those of the full eight band Hamiltonian (see Sec~\ref{sec:1-INTRODUCTION}).
Therefore results of the BHZ Hamiltonian are only reliable in this
small wave vector range. 

Exponential BHZ edge states can be determined using the OBC and SBC
characteristic equations, \eqref{eq:8} and \eqref{eq:10}, respectively.
For OBCs, a topologically trivial edge solution exists with real decay
parameters when $0.0021\times\frac{2\pi}{a_{\mathrm{sub}}}<k_{x}<0.043\times\frac{2\pi}{a_{\mathrm{sub}}}$.
The dispersion, shown as a blue dashed line in Fig.~\ref{fig:Figure 7},
reproduces very well the form of the dispersion shown in Fig.~1(a)
of Ref.~\onlinecite{Candido2018}.\footnote{Note that the BHZ Hamiltonian used in Ref.~\onlinecite{Candido2018}
corresponds to spin down in this work} A topologically trivial SBC edge solution can also be found, for
the wing decay parameter, $\sigma_{+}$, in Eq.~\eqref{eq:10}. Its
dispersion is plotted near the zone center, as green dashed and dotted
lines, respectively, for wall hybridization parameters of $4$ and
$5\,\mathrm{eV\,\mathring{A}}$. The solution for $A_{0}=A$ is in
fact very similar to the spurious $\sigma_{+}$ solution shown for
the TI phase in Fig.\,2(a) of Ref.~\onlinecite{Klipstein2015}.
This is because $B_{-}\sigma_{+}^{2}>>|M|$ in Eq.~\eqref{eq:10},
so the sign of \textit{M} becomes unimportant. Noting that Eq.~\eqref{eq:10}
corresponds to a wall with negative $B_{0}\rightarrow0$, a finite
wall with positive $B_{0}$ can be intercalated next to the TI, so
that the edge state localizes on the interface between wall materials
for which $\Delta N_{C}=1$, separating from the sample edge for which
$\Delta N_{C}=0$ (analogous to Fig.~\ref{Figure 3} and Ref.~\onlinecite{Klipstein2018}).
This confirms that the $\sigma_{+}$ edge state is not a physical
solution. In addition to a wave function that includes the unphysical
wing solution, both the OBC and SBC edge states extend to wave vectors
far beyond the $k_{\mathrm{max}}-$limit. In contrast, when SBCs are
used in Eq.~\eqref{eq:10} with the physical, $\sigma_{-}$ decay
parameter, there is no topologically trivial solution. This is consistent
with $\Delta N_{C}=1$, allowing only a single edge state, which is
the unphysical, $\sigma_{+}$ solution.\footnote{This is also consistent with the mutiband model. For positive $M_{0}$,
$\frac{\Delta_{i0}}{M_{0}}=\gamma\frac{\Delta_{i}}{M}$ with $\gamma=\pm1$,
semiconductor parameters as in Fig.~\ref{fig:Figure 7}, and $A_{0}=A$,
the only eigenvectors at zero energy and wave vector that are nearly
similar on each side of the boundary are those for the wing solution,
and only for $\gamma=-1$, when the outer band gap parameters change
sign (consistent with negative $B_{0}$ and $\Delta N_{C}=1$ in the
four band model). Since $\gamma$ is negative there is no exponential
solution (see Sec.~\ref{sec:2-MULTI-BAND}C). However, given that
the eigenvectors are very close, namely $\left[i0.321,\,0.174,\,-0.421,\,i0.830\right]$
in the wall and $\left[i0.343,\,0.159,\,-0.410,\,i0.830\right]$ in
the semiconductor, this suggests that a nearby non-exponential edge
state exists in the multiband model, corresponding to the wing solution
of the topologically trivial phase. } 

In Ref.~\onlinecite{Candido2018}, edge states appear in the topologically
trivial phase for any value, apart from $\frac{\pi}{2}$, of a $k_{x}-$independent,
phenomenological boundary condition parameter, $\theta$, which is
conserved across the phase boundary, and whose value depends on the
boundary conditions used for the envelope function. For OBC edge states
as in Fig.~\ref{fig:Figure 7}, $\theta$ is a function of $\frac{D}{B}$
and is indeed independent of wave vector and the sign of the band
gap parameter, \textit{M}. For SBCs in the TI phase, with negative
\textit{M,} $A_{0}=A$, and other parameters as in Fig.~\ref{fig:Figure 7},
the physical $\sigma_{-}$ edge state behaves as $\theta=\chi\frac{\pi}{2}$,
with $\chi=1.018$ at $k_{x}=0$, and $\chi\rightarrow1$ at the merging
points. The value of $\chi$ is quite insensitive to an increase in
the wall hybridization parameter and always tends to one at the merging
points. When there is charge conjugation symmetry with $D=0$ and
a linear edge dispersion, $\chi=1$ for all $k_{x}$. Thus, even in
the presence of strong band asymmetry, the behavior of the physical
SBC solution is very close to the special case of $\theta=\frac{\pi}{2}$,
where no edge state exists in the topologically trivial phase. The
paradoxical appearance of edge states in the topologically trivial
phase may therefore be a mathematical artifact associated with the
spurious wing solution. 

\bibliographystyle{apsrev4-1}
\bibliography{Ref4bandRT4}

\begin{thebibliography}{75}%
\makeatletter
\providecommand \@ifxundefined [1]{%
 \@ifx{#1\undefined}
}%
\providecommand \@ifnum [1]{%
 \ifnum #1\expandafter \@firstoftwo
 \else \expandafter \@secondoftwo
 \fi
}%
\providecommand \@ifx [1]{%
 \ifx #1\expandafter \@firstoftwo
 \else \expandafter \@secondoftwo
 \fi
}%
\providecommand \natexlab [1]{#1}%
\providecommand \enquote  [1]{``#1''}%
\providecommand \bibnamefont  [1]{#1}%
\providecommand \bibfnamefont [1]{#1}%
\providecommand \citenamefont [1]{#1}%
\providecommand \href@noop [0]{\@secondoftwo}%
\providecommand \href [0]{\begingroup \@sanitize@url \@href}%
\providecommand \@href[1]{\@@startlink{#1}\@@href}%
\providecommand \@@href[1]{\endgroup#1\@@endlink}%
\providecommand \@sanitize@url [0]{\catcode `\\12\catcode `\$12\catcode
  `\&12\catcode `\#12\catcode `\^12\catcode `\_12\catcode `\%12\relax}%
\providecommand \@@startlink[1]{}%
\providecommand \@@endlink[0]{}%
\providecommand \url  [0]{\begingroup\@sanitize@url \@url }%
\providecommand \@url [1]{\endgroup\@href {#1}{\urlprefix }}%
\providecommand \urlprefix  [0]{URL }%
\providecommand \Eprint [0]{\href }%
\providecommand \doibase [0]{http://dx.doi.org/}%
\providecommand \selectlanguage [0]{\@gobble}%
\providecommand \bibinfo  [0]{\@secondoftwo}%
\providecommand \bibfield  [0]{\@secondoftwo}%
\providecommand \translation [1]{[#1]}%
\providecommand \BibitemOpen [0]{}%
\providecommand \bibitemStop [0]{}%
\providecommand \bibitemNoStop [0]{.\EOS\space}%
\providecommand \EOS [0]{\spacefactor3000\relax}%
\providecommand \BibitemShut  [1]{\csname bibitem#1\endcsname}%
\let\auto@bib@innerbib\@empty
\bibitem [{\citenamefont {Yu}\ and\ \citenamefont
  {Cardona}(1996)}]{YuCardona1996}%
  \BibitemOpen
  \bibfield  {author} {\bibinfo {author} {\bibfnamefont {P.}~\bibnamefont
  {Yu}}\ and\ \bibinfo {author} {\bibfnamefont {M.}~\bibnamefont {Cardona}},\
  }\href@noop {} {\emph {\bibinfo {title} {Fundamentals of Semiconductors,
  Physics and Material Properties}}}\ (\bibinfo  {publisher} {Springer Verlag,
  Berlin},\ \bibinfo {year} {1996})\BibitemShut {NoStop}%
\bibitem [{\citenamefont {Ritchie}(1957)}]{Ritchie1957}%
  \BibitemOpen
  \bibfield  {author} {\bibinfo {author} {\bibfnamefont {R.~H.}\ \bibnamefont
  {Ritchie}},\ }\href@noop {} {\bibfield  {journal} {\bibinfo  {journal} {Phys.
  Rev.}\ }\textbf {\bibinfo {volume} {106}},\ \bibinfo {pages} {874} (\bibinfo
  {year} {1957})}\BibitemShut {NoStop}%
\bibitem [{\citenamefont {Pitarke}\ \emph {et~al.}(2007)\citenamefont
  {Pitarke}, \citenamefont {Silkin}, \citenamefont {Chulkov},\ and\
  \citenamefont {Echenique}}]{Pitarke2007}%
  \BibitemOpen
  \bibfield  {author} {\bibinfo {author} {\bibfnamefont {J.~M.}\ \bibnamefont
  {Pitarke}}, \bibinfo {author} {\bibfnamefont {V.~M.}\ \bibnamefont {Silkin}},
  \bibinfo {author} {\bibfnamefont {E.~V.}\ \bibnamefont {Chulkov}}, \ and\
  \bibinfo {author} {\bibfnamefont {P.~M.}\ \bibnamefont {Echenique}},\
  }\href@noop {} {\bibfield  {journal} {\bibinfo  {journal} {Rep. Prog. Phys.}\
  }\textbf {\bibinfo {volume} {70}},\ \bibinfo {pages} {1} (\bibinfo {year}
  {2007})}\BibitemShut {NoStop}%
\bibitem [{\citenamefont {Fuchs}\ and\ \citenamefont
  {Kliewer}(1965)}]{Fuchs1965}%
  \BibitemOpen
  \bibfield  {author} {\bibinfo {author} {\bibfnamefont {R.}~\bibnamefont
  {Fuchs}}\ and\ \bibinfo {author} {\bibfnamefont {K.~L.}\ \bibnamefont
  {Kliewer}},\ }\href@noop {} {\bibfield  {journal} {\bibinfo  {journal} {Phys.
  Rev.}\ }\textbf {\bibinfo {volume} {140}},\ \bibinfo {pages} {A2076}
  (\bibinfo {year} {1965})}\BibitemShut {NoStop}%
\bibitem [{\citenamefont {Sood}\ \emph {et~al.}(1985)\citenamefont {Sood},
  \citenamefont {Menendez}, \citenamefont {Cardona},\ and\ \citenamefont
  {Ploog}}]{Sood1985}%
  \BibitemOpen
  \bibfield  {author} {\bibinfo {author} {\bibfnamefont {A.~K.}\ \bibnamefont
  {Sood}}, \bibinfo {author} {\bibfnamefont {J.}~\bibnamefont {Menendez}},
  \bibinfo {author} {\bibfnamefont {M.}~\bibnamefont {Cardona}}, \ and\
  \bibinfo {author} {\bibfnamefont {K.}~\bibnamefont {Ploog}},\ }\href@noop {}
  {\bibfield  {journal} {\bibinfo  {journal} {Phys. Rev. Lett.}\ }\textbf
  {\bibinfo {volume} {54}},\ \bibinfo {pages} {2115} (\bibinfo {year}
  {1985})}\BibitemShut {NoStop}%
\bibitem [{\citenamefont {Qi}\ and\ \citenamefont {Zhang}(2010)}]{Qi2010}%
  \BibitemOpen
  \bibfield  {author} {\bibinfo {author} {\bibfnamefont {X.-L.}\ \bibnamefont
  {Qi}}\ and\ \bibinfo {author} {\bibfnamefont {S.-C.}\ \bibnamefont {Zhang}},\
  }\href@noop {} {\bibfield  {journal} {\bibinfo  {journal} {Physics Today}\
  }\textbf {\bibinfo {volume} {63}},\ \bibinfo {pages} {33} (\bibinfo {year}
  {2010})}\BibitemShut {NoStop}%
\bibitem [{\citenamefont {Bernevig}\ \emph {et~al.}(2006)\citenamefont
  {Bernevig}, \citenamefont {Hughes},\ and\ \citenamefont
  {Zhang}}]{Bernevig2006}%
  \BibitemOpen
  \bibfield  {author} {\bibinfo {author} {\bibfnamefont {B.~A.}\ \bibnamefont
  {Bernevig}}, \bibinfo {author} {\bibfnamefont {T.~L.}\ \bibnamefont
  {Hughes}}, \ and\ \bibinfo {author} {\bibfnamefont {S.-C.}\ \bibnamefont
  {Zhang}},\ }\href@noop {} {\bibfield  {journal} {\bibinfo  {journal}
  {Science}\ }\textbf {\bibinfo {volume} {314}},\ \bibinfo {pages} {1757}
  (\bibinfo {year} {2006})}\BibitemShut {NoStop}%
\bibitem [{\citenamefont {K{\"{o}}nig}\ \emph {et~al.}(2008)\citenamefont
  {K{\"{o}}nig}, \citenamefont {Buhmann}, \citenamefont {Molenkamp},
  \citenamefont {Hughes}, \citenamefont {Liu}, \citenamefont {Qi},\ and\
  \citenamefont {Zhang}}]{Konig2008}%
  \BibitemOpen
  \bibfield  {author} {\bibinfo {author} {\bibfnamefont {M.}~\bibnamefont
  {K{\"{o}}nig}}, \bibinfo {author} {\bibfnamefont {H.}~\bibnamefont
  {Buhmann}}, \bibinfo {author} {\bibfnamefont {L.~W.}\ \bibnamefont
  {Molenkamp}}, \bibinfo {author} {\bibfnamefont {T.~L.}\ \bibnamefont
  {Hughes}}, \bibinfo {author} {\bibfnamefont {C.-X.}\ \bibnamefont {Liu}},
  \bibinfo {author} {\bibfnamefont {X.-L.}\ \bibnamefont {Qi}}, \ and\ \bibinfo
  {author} {\bibfnamefont {S.-C.}\ \bibnamefont {Zhang}},\ }\href@noop {}
  {\bibfield  {journal} {\bibinfo  {journal} {J. Phys. Soc. Jpn.}\ }\textbf
  {\bibinfo {volume} {77}},\ \bibinfo {pages} {031007} (\bibinfo {year}
  {2008})}\BibitemShut {NoStop}%
\bibitem [{\citenamefont {Nichele}\ \emph {et~al.}(2016)\citenamefont
  {Nichele}, \citenamefont {Suominen}, \citenamefont {Kjaergaard},
  \citenamefont {Marcus}, \citenamefont {Sajadi}, \citenamefont {Folk},
  \citenamefont {Qu}, \citenamefont {Beukman}, \citenamefont {de~Vries},
  \citenamefont {van Veen}, \citenamefont {Nadj-Perge}, \citenamefont
  {Kouwenhouven}, \citenamefont {Nguyen}, \citenamefont {Kisilev},
  \citenamefont {Yi}, \citenamefont {Sokolovich}, \citenamefont {Manfra},
  \citenamefont {Spanton},\ and\ \citenamefont {Moler}}]{Nichelel2016}%
  \BibitemOpen
  \bibfield  {author} {\bibinfo {author} {\bibfnamefont {F.}~\bibnamefont
  {Nichele}}, \bibinfo {author} {\bibfnamefont {H.~J.}\ \bibnamefont
  {Suominen}}, \bibinfo {author} {\bibfnamefont {M.}~\bibnamefont
  {Kjaergaard}}, \bibinfo {author} {\bibfnamefont {C.~M.}\ \bibnamefont
  {Marcus}}, \bibinfo {author} {\bibfnamefont {E.}~\bibnamefont {Sajadi}},
  \bibinfo {author} {\bibfnamefont {J.~A.}\ \bibnamefont {Folk}}, \bibinfo
  {author} {\bibfnamefont {F.}~\bibnamefont {Qu}}, \bibinfo {author}
  {\bibfnamefont {A.~J.~A.}\ \bibnamefont {Beukman}}, \bibinfo {author}
  {\bibfnamefont {F.~K.}\ \bibnamefont {de~Vries}}, \bibinfo {author}
  {\bibfnamefont {J.}~\bibnamefont {van Veen}}, \bibinfo {author}
  {\bibfnamefont {S.}~\bibnamefont {Nadj-Perge}}, \bibinfo {author}
  {\bibfnamefont {L.~P.}\ \bibnamefont {Kouwenhouven}}, \bibinfo {author}
  {\bibfnamefont {B.-M.}\ \bibnamefont {Nguyen}}, \bibinfo {author}
  {\bibfnamefont {A.~A.}\ \bibnamefont {Kisilev}}, \bibinfo {author}
  {\bibfnamefont {W.}~\bibnamefont {Yi}}, \bibinfo {author} {\bibfnamefont
  {M.}~\bibnamefont {Sokolovich}}, \bibinfo {author} {\bibfnamefont {M.~J.}\
  \bibnamefont {Manfra}}, \bibinfo {author} {\bibfnamefont {E.~M.}\
  \bibnamefont {Spanton}}, \ and\ \bibinfo {author} {\bibfnamefont {K.~A.}\
  \bibnamefont {Moler}},\ }\href@noop {} {\bibfield  {journal} {\bibinfo
  {journal} {New. Journ. Phys.}\ }\textbf {\bibinfo {volume} {18}},\ \bibinfo
  {pages} {083005} (\bibinfo {year} {2016})}\BibitemShut {NoStop}%
\bibitem [{\citenamefont {Klipstein}(2015)}]{Klipstein2015}%
  \BibitemOpen
  \bibfield  {author} {\bibinfo {author} {\bibfnamefont {P.~C.}\ \bibnamefont
  {Klipstein}},\ }\href@noop {} {\bibfield  {journal} {\bibinfo  {journal}
  {Phys. Rev. B}\ }\textbf {\bibinfo {volume} {91}},\ \bibinfo {pages} {035310}
  (\bibinfo {year} {2015})}\BibitemShut {NoStop}%
\bibitem [{\citenamefont {(erratum)}(2016)}]{KlipErratum2016}%
  \BibitemOpen
  \bibfield  {author} {\bibinfo {author} {\bibnamefont {(erratum)}},\
  }\href@noop {} {\bibfield  {journal} {\bibinfo  {journal} {Phys. Rev. B}\
  }\textbf {\bibinfo {volume} {93}},\ \bibinfo {pages} {199905} (\bibinfo
  {year} {2016})}\BibitemShut {NoStop}%
\bibitem [{\citenamefont {Tkachov}\ and\ \citenamefont
  {Hankiewicz}(2010)}]{Tkachov2010}%
  \BibitemOpen
  \bibfield  {author} {\bibinfo {author} {\bibfnamefont {G.}~\bibnamefont
  {Tkachov}}\ and\ \bibinfo {author} {\bibfnamefont {E.~M.}\ \bibnamefont
  {Hankiewicz}},\ }\href@noop {} {\bibfield  {journal} {\bibinfo  {journal}
  {Phs. Rev. Lett.}\ }\textbf {\bibinfo {volume} {104}},\ \bibinfo {pages}
  {166803} (\bibinfo {year} {2010})}\BibitemShut {NoStop}%
\bibitem [{\citenamefont {Tkachov}\ and\ \citenamefont
  {Hankiewicz}(2013)}]{Tkachov2013}%
  \BibitemOpen
  \bibfield  {author} {\bibinfo {author} {\bibfnamefont {G.}~\bibnamefont
  {Tkachov}}\ and\ \bibinfo {author} {\bibfnamefont {E.~M.}\ \bibnamefont
  {Hankiewicz}},\ }\href@noop {} {\bibfield  {journal} {\bibinfo  {journal}
  {Phys. Staus. Solidi B}\ }\textbf {\bibinfo {volume} {250}},\ \bibinfo
  {pages} {215} (\bibinfo {year} {2013})}\BibitemShut {NoStop}%
\bibitem [{\citenamefont {Medhi}\ and\ \citenamefont
  {Shenoy}(2012)}]{Medhi2012}%
  \BibitemOpen
  \bibfield  {author} {\bibinfo {author} {\bibfnamefont {A.}~\bibnamefont
  {Medhi}}\ and\ \bibinfo {author} {\bibfnamefont {V.~B.}\ \bibnamefont
  {Shenoy}},\ }\href@noop {} {\bibfield  {journal} {\bibinfo  {journal} {J.
  Phys.: Condens. Matter}\ }\textbf {\bibinfo {volume} {24}},\ \bibinfo {pages}
  {355001} (\bibinfo {year} {2012})}\BibitemShut {NoStop}%
\bibitem [{\citenamefont {Zhou}\ \emph {et~al.}(2008)\citenamefont {Zhou},
  \citenamefont {Li}, \citenamefont {Chu}, \citenamefont {Shen},\ and\
  \citenamefont {Niu}}]{Zhou2008}%
  \BibitemOpen
  \bibfield  {author} {\bibinfo {author} {\bibfnamefont {B.}~\bibnamefont
  {Zhou}}, \bibinfo {author} {\bibfnamefont {H.-Z.}\ \bibnamefont {Li}},
  \bibinfo {author} {\bibfnamefont {R.-L.}\ \bibnamefont {Chu}}, \bibinfo
  {author} {\bibfnamefont {S.-Q.}\ \bibnamefont {Shen}}, \ and\ \bibinfo
  {author} {\bibfnamefont {Q.}~\bibnamefont {Niu}},\ }\href@noop {} {\bibfield
  {journal} {\bibinfo  {journal} {Phys. Rev. Lett.}\ }\textbf {\bibinfo
  {volume} {101}},\ \bibinfo {pages} {246807} (\bibinfo {year}
  {2008})}\BibitemShut {NoStop}%
\bibitem [{\citenamefont {Essert}(2015)}]{Essert2015}%
  \BibitemOpen
  \bibfield  {author} {\bibinfo {author} {\bibfnamefont {S.}~\bibnamefont
  {Essert}},\ }\href@noop {} {Ph.D. thesis},\ \bibinfo  {school} {Univ. of
  Regensberg} (\bibinfo {year} {2015})\BibitemShut {NoStop}%
\bibitem [{\citenamefont {Qi}\ \emph {et~al.}(2006)\citenamefont {Qi},
  \citenamefont {Wu},\ and\ \citenamefont {Zhang}}]{Qi2006}%
  \BibitemOpen
  \bibfield  {author} {\bibinfo {author} {\bibfnamefont {X.-L.}\ \bibnamefont
  {Qi}}, \bibinfo {author} {\bibfnamefont {Y.-S.}\ \bibnamefont {Wu}}, \ and\
  \bibinfo {author} {\bibfnamefont {S.-C.}\ \bibnamefont {Zhang}},\ }\href@noop
  {} {\bibfield  {journal} {\bibinfo  {journal} {Phys. Rev. B}\ }\textbf
  {\bibinfo {volume} {74}},\ \bibinfo {pages} {085308} (\bibinfo {year}
  {2006})}\BibitemShut {NoStop}%
\bibitem [{\citenamefont {K{\"{o}}nig}\ \emph {et~al.}(2007)\citenamefont
  {K{\"{o}}nig}, \citenamefont {Wiedmann}, \citenamefont {Br{\"{u}}ne},
  \citenamefont {Roth}, \citenamefont {Buhmann}, \citenamefont {Molenkamp},
  \citenamefont {Qi},\ and\ \citenamefont {Zhang}}]{Konig2007}%
  \BibitemOpen
  \bibfield  {author} {\bibinfo {author} {\bibfnamefont {M.}~\bibnamefont
  {K{\"{o}}nig}}, \bibinfo {author} {\bibfnamefont {S.}~\bibnamefont
  {Wiedmann}}, \bibinfo {author} {\bibfnamefont {C.}~\bibnamefont
  {Br{\"{u}}ne}}, \bibinfo {author} {\bibfnamefont {A.}~\bibnamefont {Roth}},
  \bibinfo {author} {\bibfnamefont {H.}~\bibnamefont {Buhmann}}, \bibinfo
  {author} {\bibfnamefont {L.~W.}\ \bibnamefont {Molenkamp}}, \bibinfo {author}
  {\bibfnamefont {X.-L.}\ \bibnamefont {Qi}}, \ and\ \bibinfo {author}
  {\bibfnamefont {S.-C.}\ \bibnamefont {Zhang}},\ }\href@noop {} {\bibfield
  {journal} {\bibinfo  {journal} {Science}\ }\textbf {\bibinfo {volume}
  {318}},\ \bibinfo {pages} {766} (\bibinfo {year} {2007})}\BibitemShut
  {NoStop}%
\bibitem [{\citenamefont {Dai}\ \emph {et~al.}(2008)\citenamefont {Dai},
  \citenamefont {Hughes}, \citenamefont {Qi}, \citenamefont {Fang},\ and\
  \citenamefont {Zhang}}]{Dai2008}%
  \BibitemOpen
  \bibfield  {author} {\bibinfo {author} {\bibfnamefont {X.}~\bibnamefont
  {Dai}}, \bibinfo {author} {\bibfnamefont {T.~L.}\ \bibnamefont {Hughes}},
  \bibinfo {author} {\bibfnamefont {X.-L.}\ \bibnamefont {Qi}}, \bibinfo
  {author} {\bibfnamefont {Z.}~\bibnamefont {Fang}}, \ and\ \bibinfo {author}
  {\bibfnamefont {S.-C.}\ \bibnamefont {Zhang}},\ }\href@noop {} {\bibfield
  {journal} {\bibinfo  {journal} {Phys. Rev. B}\ }\textbf {\bibinfo {volume}
  {77}},\ \bibinfo {pages} {125319} (\bibinfo {year} {2008})}\BibitemShut
  {NoStop}%
\bibitem [{\citenamefont {Liu}\ \emph {et~al.}(2008)\citenamefont {Liu},
  \citenamefont {Hughes}, \citenamefont {Qi}, \citenamefont {Wang},\ and\
  \citenamefont {Zhang}}]{Liu2008}%
  \BibitemOpen
  \bibfield  {author} {\bibinfo {author} {\bibfnamefont {C.-X.}\ \bibnamefont
  {Liu}}, \bibinfo {author} {\bibfnamefont {T.~L.}\ \bibnamefont {Hughes}},
  \bibinfo {author} {\bibfnamefont {X.-L.}\ \bibnamefont {Qi}}, \bibinfo
  {author} {\bibfnamefont {K.}~\bibnamefont {Wang}}, \ and\ \bibinfo {author}
  {\bibfnamefont {S.-C.}\ \bibnamefont {Zhang}},\ }\href@noop {} {\bibfield
  {journal} {\bibinfo  {journal} {Phys. Rev. Lett.}\ }\textbf {\bibinfo
  {volume} {100}},\ \bibinfo {pages} {236601} (\bibinfo {year}
  {2008})}\BibitemShut {NoStop}%
\bibitem [{\citenamefont {Linder}\ \emph {et~al.}(2009)\citenamefont {Linder},
  \citenamefont {Yokoyama},\ and\ \citenamefont {Sudb{\o}}}]{Linder2009}%
  \BibitemOpen
  \bibfield  {author} {\bibinfo {author} {\bibfnamefont {J.}~\bibnamefont
  {Linder}}, \bibinfo {author} {\bibfnamefont {T.}~\bibnamefont {Yokoyama}}, \
  and\ \bibinfo {author} {\bibfnamefont {A.}~\bibnamefont {Sudb{\o}}},\
  }\href@noop {} {\bibfield  {journal} {\bibinfo  {journal} {Phys. Rev. B}\
  }\textbf {\bibinfo {volume} {80}},\ \bibinfo {pages} {205401} (\bibinfo
  {year} {2009})}\BibitemShut {NoStop}%
\bibitem [{\citenamefont {Liu}\ \emph {et~al.}(2010)\citenamefont {Liu},
  \citenamefont {Qi}, \citenamefont {Zhang}, \citenamefont {Dai}, \citenamefont
  {Fang},\ and\ \citenamefont {Zhang}}]{Liu2010}%
  \BibitemOpen
  \bibfield  {author} {\bibinfo {author} {\bibfnamefont {C.-X.}\ \bibnamefont
  {Liu}}, \bibinfo {author} {\bibfnamefont {X.-L.}\ \bibnamefont {Qi}},
  \bibinfo {author} {\bibfnamefont {H.-J.}\ \bibnamefont {Zhang}}, \bibinfo
  {author} {\bibfnamefont {X.}~\bibnamefont {Dai}}, \bibinfo {author}
  {\bibfnamefont {Z.}~\bibnamefont {Fang}}, \ and\ \bibinfo {author}
  {\bibfnamefont {S.-C.}\ \bibnamefont {Zhang}},\ }\href@noop {} {\bibfield
  {journal} {\bibinfo  {journal} {Phys. Rev. B}\ }\textbf {\bibinfo {volume}
  {82}},\ \bibinfo {pages} {045122} (\bibinfo {year} {2010})}\BibitemShut
  {NoStop}%
\bibitem [{\citenamefont {Lu}\ \emph {et~al.}(2010)\citenamefont {Lu},
  \citenamefont {Shan}, \citenamefont {Yao}, \citenamefont {Niu},\ and\
  \citenamefont {Shen}}]{Lu2010}%
  \BibitemOpen
  \bibfield  {author} {\bibinfo {author} {\bibfnamefont {H.~Z.}\ \bibnamefont
  {Lu}}, \bibinfo {author} {\bibfnamefont {W.~Y.}\ \bibnamefont {Shan}},
  \bibinfo {author} {\bibfnamefont {W.}~\bibnamefont {Yao}}, \bibinfo {author}
  {\bibfnamefont {Q.}~\bibnamefont {Niu}}, \ and\ \bibinfo {author}
  {\bibfnamefont {S.~Q.}\ \bibnamefont {Shen}},\ }\href@noop {} {\bibfield
  {journal} {\bibinfo  {journal} {Phys. Rev. B}\ }\textbf {\bibinfo {volume}
  {81}},\ \bibinfo {pages} {115407} (\bibinfo {year} {2010})}\BibitemShut
  {NoStop}%
\bibitem [{\citenamefont {Shan}\ \emph {et~al.}(2010)\citenamefont {Shan},
  \citenamefont {Lu},\ and\ \citenamefont {Shen}}]{Shan2010}%
  \BibitemOpen
  \bibfield  {author} {\bibinfo {author} {\bibfnamefont {W.~Y.}\ \bibnamefont
  {Shan}}, \bibinfo {author} {\bibfnamefont {H.~Z.}\ \bibnamefont {Lu}}, \ and\
  \bibinfo {author} {\bibfnamefont {S.~Q.}\ \bibnamefont {Shen}},\ }\href@noop
  {} {\bibfield  {journal} {\bibinfo  {journal} {New. Journal of Physics}\
  }\textbf {\bibinfo {volume} {12}},\ \bibinfo {pages} {043048} (\bibinfo
  {year} {2010})}\BibitemShut {NoStop}%
\bibitem [{\citenamefont {Sonin}(2010)}]{Sonin2010}%
  \BibitemOpen
  \bibfield  {author} {\bibinfo {author} {\bibfnamefont {E.~B.}\ \bibnamefont
  {Sonin}},\ }\href@noop {} {\bibfield  {journal} {\bibinfo  {journal} {Phys.
  Rev. B.}\ }\textbf {\bibinfo {volume} {82}},\ \bibinfo {pages} {113307}
  (\bibinfo {year} {2010})}\BibitemShut {NoStop}%
\bibitem [{\citenamefont {Murakami}(2011)}]{Murakami2011}%
  \BibitemOpen
  \bibfield  {author} {\bibinfo {author} {\bibfnamefont {S.}~\bibnamefont
  {Murakami}},\ }\href@noop {} {\bibfield  {journal} {\bibinfo  {journal} {J.
  Phys.: Conference Series}\ }\textbf {\bibinfo {volume} {302}},\ \bibinfo
  {pages} {012019} (\bibinfo {year} {2011})}\BibitemShut {NoStop}%
\bibitem [{\citenamefont {Shen}\ \emph {et~al.}(2011)\citenamefont {Shen},
  \citenamefont {Shan},\ and\ \citenamefont {Lu}}]{Shen2011}%
  \BibitemOpen
  \bibfield  {author} {\bibinfo {author} {\bibfnamefont {S.~Q.}\ \bibnamefont
  {Shen}}, \bibinfo {author} {\bibfnamefont {W.~Y.}\ \bibnamefont {Shan}}, \
  and\ \bibinfo {author} {\bibfnamefont {H.}~\bibnamefont {Lu}},\ }\href@noop
  {} {\bibfield  {journal} {\bibinfo  {journal} {Spin}\ }\textbf {\bibinfo
  {volume} {1}},\ \bibinfo {pages} {33} (\bibinfo {year} {2011})}\BibitemShut
  {NoStop}%
\bibitem [{\citenamefont {Wada}\ \emph {et~al.}(2011)\citenamefont {Wada},
  \citenamefont {Murakami}, \citenamefont {Freimuth},\ and\ \citenamefont
  {Bihlmayer}}]{Wada2011}%
  \BibitemOpen
  \bibfield  {author} {\bibinfo {author} {\bibfnamefont {M.}~\bibnamefont
  {Wada}}, \bibinfo {author} {\bibfnamefont {S.}~\bibnamefont {Murakami}},
  \bibinfo {author} {\bibfnamefont {F.}~\bibnamefont {Freimuth}}, \ and\
  \bibinfo {author} {\bibfnamefont {G.}~\bibnamefont {Bihlmayer}},\ }\href@noop
  {} {\bibfield  {journal} {\bibinfo  {journal} {Phys. Rev. B}\ }\textbf
  {\bibinfo {volume} {83}},\ \bibinfo {pages} {121301} (\bibinfo {year}
  {2011})}\BibitemShut {NoStop}%
\bibitem [{\citenamefont {Michetti}\ \emph
  {et~al.}(2012{\natexlab{a}})\citenamefont {Michetti}, \citenamefont {Budich},
  \citenamefont {Novik},\ and\ \citenamefont {Recher}}]{Michetti2012a}%
  \BibitemOpen
  \bibfield  {author} {\bibinfo {author} {\bibfnamefont {P.}~\bibnamefont
  {Michetti}}, \bibinfo {author} {\bibfnamefont {J.~C.}\ \bibnamefont
  {Budich}}, \bibinfo {author} {\bibfnamefont {E.~G.}\ \bibnamefont {Novik}}, \
  and\ \bibinfo {author} {\bibfnamefont {P.}~\bibnamefont {Recher}},\
  }\href@noop {} {\bibfield  {journal} {\bibinfo  {journal} {Phys. Rev. B}\
  }\textbf {\bibinfo {volume} {85}},\ \bibinfo {pages} {125309} (\bibinfo
  {year} {2012}{\natexlab{a}})}\BibitemShut {NoStop}%
\bibitem [{\citenamefont {Michetti}\ \emph
  {et~al.}(2012{\natexlab{b}})\citenamefont {Michetti}, \citenamefont
  {Penteado}, \citenamefont {Egues},\ and\ \citenamefont
  {Recher}}]{Michetti2012b}%
  \BibitemOpen
  \bibfield  {author} {\bibinfo {author} {\bibfnamefont {P.}~\bibnamefont
  {Michetti}}, \bibinfo {author} {\bibfnamefont {P.~H.}\ \bibnamefont
  {Penteado}}, \bibinfo {author} {\bibfnamefont {J.~C.}\ \bibnamefont {Egues}},
  \ and\ \bibinfo {author} {\bibfnamefont {P.}~\bibnamefont {Recher}},\
  }\href@noop {} {\bibfield  {journal} {\bibinfo  {journal} {Semicond. Sci.
  Technol.}\ }\textbf {\bibinfo {volume} {27}},\ \bibinfo {pages} {124007}
  (\bibinfo {year} {2012}{\natexlab{b}})}\BibitemShut {NoStop}%
\bibitem [{\citenamefont {Takagaki}(2012)}]{Takagaki2012}%
  \BibitemOpen
  \bibfield  {author} {\bibinfo {author} {\bibfnamefont {Y.}~\bibnamefont
  {Takagaki}},\ }\href@noop {} {\bibfield  {journal} {\bibinfo  {journal} {J.
  Phys.: Condens Matter}\ }\textbf {\bibinfo {volume} {24}},\ \bibinfo {pages}
  {435301} (\bibinfo {year} {2012})}\BibitemShut {NoStop}%
\bibitem [{\citenamefont {Cano-Corte's}\ \emph {et~al.}(2013)\citenamefont
  {Cano-Corte's}, \citenamefont {Ortix},\ and\ \citenamefont {van~den
  Brink}}]{CanoCortes2013}%
  \BibitemOpen
  \bibfield  {author} {\bibinfo {author} {\bibfnamefont {L.}~\bibnamefont
  {Cano-Corte's}}, \bibinfo {author} {\bibfnamefont {C.}~\bibnamefont {Ortix}},
  \ and\ \bibinfo {author} {\bibfnamefont {J.}~\bibnamefont {van~den Brink}},\
  }\href@noop {} {\bibfield  {journal} {\bibinfo  {journal} {Phys. Rev. Lett.}\
  }\textbf {\bibinfo {volume} {111}},\ \bibinfo {pages} {146801} (\bibinfo
  {year} {2013})}\BibitemShut {NoStop}%
\bibitem [{\citenamefont {Hohenadler}\ and\ \citenamefont
  {Assaad}(2013)}]{Hohenadler2013}%
  \BibitemOpen
  \bibfield  {author} {\bibinfo {author} {\bibfnamefont {M.}~\bibnamefont
  {Hohenadler}}\ and\ \bibinfo {author} {\bibfnamefont {F.~F.}\ \bibnamefont
  {Assaad}},\ }\href@noop {} {\bibfield  {journal} {\bibinfo  {journal} {J.
  Phys.: Condens. Matter}\ }\textbf {\bibinfo {volume} {25}},\ \bibinfo {pages}
  {143201} (\bibinfo {year} {2013})}\BibitemShut {NoStop}%
\bibitem [{\citenamefont {Sengupta}\ \emph {et~al.}(2013)\citenamefont
  {Sengupta}, \citenamefont {Tillmann}, \citenamefont {Yaohua}, \citenamefont
  {Povolotskyi},\ and\ \citenamefont {Klimeck}}]{Sengupta2013}%
  \BibitemOpen
  \bibfield  {author} {\bibinfo {author} {\bibfnamefont {P.}~\bibnamefont
  {Sengupta}}, \bibinfo {author} {\bibfnamefont {K.}~\bibnamefont {Tillmann}},
  \bibinfo {author} {\bibfnamefont {T.}~\bibnamefont {Yaohua}}, \bibinfo
  {author} {\bibfnamefont {M.}~\bibnamefont {Povolotskyi}}, \ and\ \bibinfo
  {author} {\bibfnamefont {G.}~\bibnamefont {Klimeck}},\ }\href@noop {}
  {\bibfield  {journal} {\bibinfo  {journal} {Journ. of Appl. Phys.}\ }\textbf
  {\bibinfo {volume} {114}},\ \bibinfo {pages} {043702} (\bibinfo {year}
  {2013})}\BibitemShut {NoStop}%
\bibitem [{\citenamefont {Wang}\ \emph {et~al.}(2014)\citenamefont {Wang},
  \citenamefont {Xu},\ and\ \citenamefont {Zhang}}]{Wang2014}%
  \BibitemOpen
  \bibfield  {author} {\bibinfo {author} {\bibfnamefont {J.}~\bibnamefont
  {Wang}}, \bibinfo {author} {\bibfnamefont {Y.}~\bibnamefont {Xu}}, \ and\
  \bibinfo {author} {\bibfnamefont {S.-C.}\ \bibnamefont {Zhang}},\ }\href@noop
  {} {\bibfield  {journal} {\bibinfo  {journal} {Physical Review B}\ }\textbf
  {\bibinfo {volume} {90}},\ \bibinfo {pages} {054503} (\bibinfo {year}
  {2014})}\BibitemShut {NoStop}%
\bibitem [{\citenamefont {Xu}\ \emph {et~al.}(2014)\citenamefont {Xu},
  \citenamefont {Gao}, \citenamefont {Liu}, \citenamefont {Sun}, \citenamefont
  {Zhang},\ and\ \citenamefont {Zhou}}]{Xu2014}%
  \BibitemOpen
  \bibfield  {author} {\bibinfo {author} {\bibfnamefont {D.-H.}\ \bibnamefont
  {Xu}}, \bibinfo {author} {\bibfnamefont {J.-H.}\ \bibnamefont {Gao}},
  \bibinfo {author} {\bibfnamefont {C.-X.}\ \bibnamefont {Liu}}, \bibinfo
  {author} {\bibfnamefont {J.-H.}\ \bibnamefont {Sun}}, \bibinfo {author}
  {\bibfnamefont {F.-C.}\ \bibnamefont {Zhang}}, \ and\ \bibinfo {author}
  {\bibfnamefont {Y.}~\bibnamefont {Zhou}},\ }\href@noop {} {\bibfield
  {journal} {\bibinfo  {journal} {Phys. Rev. B}\ }\textbf {\bibinfo {volume}
  {89}},\ \bibinfo {pages} {195104} (\bibinfo {year} {2014})}\BibitemShut
  {NoStop}%
\bibitem [{\citenamefont {Enaldiev}\ \emph {et~al.}(2015)\citenamefont
  {Enaldiev}, \citenamefont {Zagorodnev},\ and\ \citenamefont
  {Volkov}}]{Enaldiev2015}%
  \BibitemOpen
  \bibfield  {author} {\bibinfo {author} {\bibfnamefont {V.~V.}\ \bibnamefont
  {Enaldiev}}, \bibinfo {author} {\bibfnamefont {I.~V.}\ \bibnamefont
  {Zagorodnev}}, \ and\ \bibinfo {author} {\bibfnamefont {V.~A.}\ \bibnamefont
  {Volkov}},\ }\href@noop {} {\bibfield  {journal} {\bibinfo  {journal} {JETP
  Letters}\ }\textbf {\bibinfo {volume} {101}},\ \bibinfo {pages} {89}
  (\bibinfo {year} {2015})}\BibitemShut {NoStop}%
\bibitem [{\citenamefont {Durnev}\ and\ \citenamefont
  {Tarasenko}(2016)}]{Durnev2016}%
  \BibitemOpen
  \bibfield  {author} {\bibinfo {author} {\bibfnamefont {M.~V.}\ \bibnamefont
  {Durnev}}\ and\ \bibinfo {author} {\bibfnamefont {S.~A.}\ \bibnamefont
  {Tarasenko}},\ }\href@noop {} {\bibfield  {journal} {\bibinfo  {journal}
  {Phys. Rev. B}\ }\textbf {\bibinfo {volume} {93}},\ \bibinfo {pages} {075434}
  (\bibinfo {year} {2016})}\BibitemShut {NoStop}%
\bibitem [{\citenamefont {Entin}\ \emph {et~al.}(2017)\citenamefont {Entin},
  \citenamefont {Magarill},\ and\ \citenamefont {Mahmoodian}}]{Entin2017}%
  \BibitemOpen
  \bibfield  {author} {\bibinfo {author} {\bibfnamefont {M.~V.}\ \bibnamefont
  {Entin}}, \bibinfo {author} {\bibfnamefont {L.~I.}\ \bibnamefont {Magarill}},
  \ and\ \bibinfo {author} {\bibfnamefont {M.}~\bibnamefont {Mahmoodian}},\
  }\href@noop {} {\bibfield  {journal} {\bibinfo  {journal} {Europhysics
  Letters}\ }\textbf {\bibinfo {volume} {118}},\ \bibinfo {pages} {57002}
  (\bibinfo {year} {2017})}\BibitemShut {NoStop}%
\bibitem [{\citenamefont {Candido}\ \emph {et~al.}(2020)\citenamefont
  {Candido}, \citenamefont {Kharitonov}, \citenamefont {Egues},\ and\
  \citenamefont {Hankiewicz}}]{Candido2018}%
  \BibitemOpen
  \bibfield  {author} {\bibinfo {author} {\bibfnamefont {D.~R.}\ \bibnamefont
  {Candido}}, \bibinfo {author} {\bibfnamefont {M.}~\bibnamefont {Kharitonov}},
  \bibinfo {author} {\bibfnamefont {J.~C.}\ \bibnamefont {Egues}}, \ and\
  \bibinfo {author} {\bibfnamefont {E.~M.}\ \bibnamefont {Hankiewicz}},\
  }\href@noop {} {\bibfield  {journal} {\bibinfo  {journal} {Phys. Rev. B.}\
  }\textbf {\bibinfo {volume} {98}},\ \bibinfo {pages} {161111} (\bibinfo
  {year} {2020})}\BibitemShut {NoStop}%
\bibitem [{\citenamefont {Gioia}\ \emph {et~al.}(2018)\citenamefont {Gioia},
  \citenamefont {Z{\"{u}}licke}, \citenamefont {Governale},\ and\ \citenamefont
  {Winkler}}]{Gioia2018}%
  \BibitemOpen
  \bibfield  {author} {\bibinfo {author} {\bibfnamefont {L.}~\bibnamefont
  {Gioia}}, \bibinfo {author} {\bibfnamefont {U.}~\bibnamefont
  {Z{\"{u}}licke}}, \bibinfo {author} {\bibfnamefont {M.}~\bibnamefont
  {Governale}}, \ and\ \bibinfo {author} {\bibfnamefont {R.}~\bibnamefont
  {Winkler}},\ }\href@noop {} {\bibfield  {journal} {\bibinfo  {journal} {Phys.
  Rev. B}\ }\textbf {\bibinfo {volume} {97}},\ \bibinfo {pages} {205421}
  (\bibinfo {year} {2018})}\BibitemShut {NoStop}%
\bibitem [{\citenamefont {Durnev}\ and\ \citenamefont
  {Tarasenko}(2019)}]{Durnev2019}%
  \BibitemOpen
  \bibfield  {author} {\bibinfo {author} {\bibfnamefont {M.~V.}\ \bibnamefont
  {Durnev}}\ and\ \bibinfo {author} {\bibfnamefont {S.~A.}\ \bibnamefont
  {Tarasenko}},\ }\href@noop {} {\bibfield  {journal} {\bibinfo  {journal}
  {Ann. Phys.}\ }\textbf {\bibinfo {volume} {531}},\ \bibinfo {pages} {1800418}
  (\bibinfo {year} {2019})}\BibitemShut {NoStop}%
\bibitem [{\citenamefont {Gioia}\ \emph {et~al.}(2019)\citenamefont {Gioia},
  \citenamefont {Christie}, \citenamefont {Z{\"{u}}licke}, \citenamefont
  {Governale},\ and\ \citenamefont {Sneyd}}]{Gioia2019}%
  \BibitemOpen
  \bibfield  {author} {\bibinfo {author} {\bibfnamefont {L.}~\bibnamefont
  {Gioia}}, \bibinfo {author} {\bibfnamefont {M.~G.}\ \bibnamefont {Christie}},
  \bibinfo {author} {\bibfnamefont {U.}~\bibnamefont {Z{\"{u}}licke}}, \bibinfo
  {author} {\bibfnamefont {M.}~\bibnamefont {Governale}}, \ and\ \bibinfo
  {author} {\bibfnamefont {A.~J.}\ \bibnamefont {Sneyd}},\ }\href {\doibase
  10.1103/physrevb.100.205417} {\bibfield  {journal} {\bibinfo  {journal}
  {Phys. Rev. B}\ }\textbf {\bibinfo {volume} {100}},\ \bibinfo {pages}
  {205417} (\bibinfo {year} {2019})}\BibitemShut {NoStop}%
\bibitem [{\citenamefont {B{\"{o}}ttcher}\ \emph {et~al.}(2020)\citenamefont
  {B{\"{o}}ttcher}, \citenamefont {Tutschku},\ and\ \citenamefont
  {Hankiewicz}}]{Bottcher2020}%
  \BibitemOpen
  \bibfield  {author} {\bibinfo {author} {\bibfnamefont {J.}~\bibnamefont
  {B{\"{o}}ttcher}}, \bibinfo {author} {\bibfnamefont {C.}~\bibnamefont
  {Tutschku}}, \ and\ \bibinfo {author} {\bibfnamefont {E.~M.}\ \bibnamefont
  {Hankiewicz}},\ }\href@noop {} {\bibfield  {journal} {\bibinfo  {journal}
  {Phys. Rev. B}\ }\textbf {\bibinfo {volume} {101}},\ \bibinfo {pages}
  {195433} (\bibinfo {year} {2020})}\BibitemShut {NoStop}%
\bibitem [{\citenamefont {Durnev}(2020)}]{Durnev2020}%
  \BibitemOpen
  \bibfield  {author} {\bibinfo {author} {\bibfnamefont {M.~V.}\ \bibnamefont
  {Durnev}},\ }\href@noop {} {\bibfield  {journal} {\bibinfo  {journal} {Phys.
  Solid State}\ }\textbf {\bibinfo {volume} {62}},\ \bibinfo {pages} {504}
  (\bibinfo {year} {2020})}\BibitemShut {NoStop}%
\bibitem [{\citenamefont {Kane}(2013)}]{Kane2013}%
  \BibitemOpen
  \bibfield  {author} {\bibinfo {author} {\bibfnamefont {C.~H.}\ \bibnamefont
  {Kane}},\ }in\ \href@noop {} {\emph {\bibinfo {booktitle} {Contemporary
  Concepts of Condensed Matter Science}}},\ Vol.~\bibinfo {volume} {6},\
  \bibinfo {editor} {edited by\ \bibinfo {editor} {\bibfnamefont
  {M.}~\bibnamefont {Franz}}\ and\ \bibinfo {editor} {\bibfnamefont
  {L.}~\bibnamefont {Molenkamp}}}\ (\bibinfo  {publisher} {Elsevier,
  Amsterdam},\ \bibinfo {year} {2013})\ Chap.~\bibinfo {chapter}
  {1}\BibitemShut {NoStop}%
\bibitem [{\citenamefont {Klipstein}(2010)}]{Klipstein2010}%
  \BibitemOpen
  \bibfield  {author} {\bibinfo {author} {\bibfnamefont {P.~C.}\ \bibnamefont
  {Klipstein}},\ }\href@noop {} {\bibfield  {journal} {\bibinfo  {journal}
  {Phys. Rev. B}\ }\textbf {\bibinfo {volume} {81}},\ \bibinfo {pages} {235314}
  (\bibinfo {year} {2010})}\BibitemShut {NoStop}%
\bibitem [{\citenamefont {White}\ and\ \citenamefont {Sham}(1981)}]{White1981}%
  \BibitemOpen
  \bibfield  {author} {\bibinfo {author} {\bibfnamefont {S.~R.}\ \bibnamefont
  {White}}\ and\ \bibinfo {author} {\bibfnamefont {L.~J.}\ \bibnamefont
  {Sham}},\ }\href@noop {} {\bibfield  {journal} {\bibinfo  {journal} {Phys.
  Rev. Lett.}\ }\textbf {\bibinfo {volume} {47}},\ \bibinfo {pages} {879}
  (\bibinfo {year} {1981})}\BibitemShut {NoStop}%
\bibitem [{\citenamefont {Schuurmans}\ and\ \citenamefont
  {t'Hooft}(1985)}]{Schuurmans1985}%
  \BibitemOpen
  \bibfield  {author} {\bibinfo {author} {\bibfnamefont {M.~F.~H.}\
  \bibnamefont {Schuurmans}}\ and\ \bibinfo {author} {\bibfnamefont {G.~W.}\
  \bibnamefont {t'Hooft}},\ }\href@noop {} {\bibfield  {journal} {\bibinfo
  {journal} {Phys. Rev. B}\ }\textbf {\bibinfo {volume} {31}},\ \bibinfo
  {pages} {8041} (\bibinfo {year} {1985})}\BibitemShut {NoStop}%
\bibitem [{Note1()}]{Note1}%
  \BibitemOpen
  \bibinfo {note} {A physical dispersion must be evanescent in the band gap,
  starting and finishing at a band edge.}\BibitemShut {Stop}%
\bibitem [{\citenamefont {Klipstein}(2018)}]{Klipstein2018}%
  \BibitemOpen
  \bibfield  {author} {\bibinfo {author} {\bibfnamefont {P.~C.}\ \bibnamefont
  {Klipstein}},\ }\href@noop {} {\bibfield  {journal} {\bibinfo  {journal}
  {J.Phys.: Condens. Matter}\ }\textbf {\bibinfo {volume} {30}},\ \bibinfo
  {pages} {275302} (\bibinfo {year} {2018})}\BibitemShut {NoStop}%
\bibitem [{\citenamefont {Luttinger}\ and\ \citenamefont
  {Kohn}(1955)}]{LuttKohn1955}%
  \BibitemOpen
  \bibfield  {author} {\bibinfo {author} {\bibfnamefont {J.~M.}\ \bibnamefont
  {Luttinger}}\ and\ \bibinfo {author} {\bibfnamefont {W.}~\bibnamefont
  {Kohn}},\ }\href@noop {} {\bibfield  {journal} {\bibinfo  {journal} {Phys.
  Rev.}\ }\textbf {\bibinfo {volume} {97}},\ \bibinfo {pages} {869} (\bibinfo
  {year} {1955})}\BibitemShut {NoStop}%
\bibitem [{\citenamefont {Volkov}\ and\ \citenamefont
  {Takhtamirov}(1997)}]{Volkov1997}%
  \BibitemOpen
  \bibfield  {author} {\bibinfo {author} {\bibfnamefont {V.~A.}\ \bibnamefont
  {Volkov}}\ and\ \bibinfo {author} {\bibfnamefont {{\'{E}}.~E.}\ \bibnamefont
  {Takhtamirov}},\ }\href@noop {} {\bibfield  {journal} {\bibinfo  {journal}
  {Physics-Uspekhi}\ }\textbf {\bibinfo {volume} {40}},\ \bibinfo {pages}
  {1071} (\bibinfo {year} {1997})}\BibitemShut {NoStop}%
\bibitem [{\citenamefont {Takhtamirov}\ and\ \citenamefont
  {Volkov}(1997)}]{Takhtamirov1997}%
  \BibitemOpen
  \bibfield  {author} {\bibinfo {author} {\bibfnamefont {{\'{E}}.~E.}\
  \bibnamefont {Takhtamirov}}\ and\ \bibinfo {author} {\bibfnamefont {V.~A.}\
  \bibnamefont {Volkov}},\ }\href@noop {} {\bibfield  {journal} {\bibinfo
  {journal} {Semicond. Sci. and Technol.}\ }\textbf {\bibinfo {volume} {12}},\
  \bibinfo {pages} {77} (\bibinfo {year} {1997})}\BibitemShut {NoStop}%
\bibitem [{\citenamefont {Takhtamirov}\ and\ \citenamefont
  {Volkov}(1999)}]{Takhtamirov1999}%
  \BibitemOpen
  \bibfield  {author} {\bibinfo {author} {\bibfnamefont {{\'{E}}.~E.}\
  \bibnamefont {Takhtamirov}}\ and\ \bibinfo {author} {\bibfnamefont {V.~A.}\
  \bibnamefont {Volkov}},\ }\href@noop {} {\bibfield  {journal} {\bibinfo
  {journal} {JETP}\ }\textbf {\bibinfo {volume} {89}},\ \bibinfo {pages} {1000}
  (\bibinfo {year} {1999})}\BibitemShut {NoStop}%
\bibitem [{\citenamefont {Cardona}\ and\ \citenamefont
  {Pollak}(1966)}]{Cardona1966}%
  \BibitemOpen
  \bibfield  {author} {\bibinfo {author} {\bibfnamefont {M.}~\bibnamefont
  {Cardona}}\ and\ \bibinfo {author} {\bibfnamefont {F.~H.}\ \bibnamefont
  {Pollak}},\ }\href {\doibase 10.1103/physrev.142.530} {\bibfield  {journal}
  {\bibinfo  {journal} {Phys. Rev.}\ }\textbf {\bibinfo {volume} {142}},\
  \bibinfo {pages} {530} (\bibinfo {year} {1966})}\BibitemShut {NoStop}%
\bibitem [{\citenamefont {Bir}\ and\ \citenamefont
  {Pikus}(1974)}]{BirPikus1974}%
  \BibitemOpen
  \bibfield  {author} {\bibinfo {author} {\bibfnamefont {G.~L.}\ \bibnamefont
  {Bir}}\ and\ \bibinfo {author} {\bibfnamefont {G.~E.}\ \bibnamefont
  {Pikus}},\ }\href@noop {} {\emph {\bibinfo {title} {Symmetry and Strain
  induced effects in semiconductors}}}\ (\bibinfo  {publisher} {Wiley, New
  York},\ \bibinfo {year} {1974})\BibitemShut {NoStop}%
\bibitem [{Note2()}]{Note2}%
  \BibitemOpen
  \bibinfo {note} {If $k_{+}$ and $k_{-}$ are interchanged in the \protect
  \textit {Q}-terms of Eq.~\protect \textup {\hbox {\mathsurround \z@ \protect
  \normalfont (\ignorespaces \ref {eq:3}\unskip \@@italiccorr )}},
  corresponding to a change in symmetry of the remote states, there is no
  change to the energy of the Dirac point based on the present multiband model,
  and the edge state dispersions are essentially the same as those derived in
  Appendix A.}\BibitemShut {Stop}%
\bibitem [{\citenamefont {Lawaetz}(1971)}]{Lawaetz1971}%
  \BibitemOpen
  \bibfield  {author} {\bibinfo {author} {\bibfnamefont {P.}~\bibnamefont
  {Lawaetz}},\ }\href@noop {} {\bibfield  {journal} {\bibinfo  {journal} {Phys.
  Rev. B.}\ }\textbf {\bibinfo {volume} {4}},\ \bibinfo {pages} {3460}
  (\bibinfo {year} {1971})}\BibitemShut {NoStop}%
\bibitem [{\citenamefont {Klipstein}(2016)}]{Klipstein2016}%
  \BibitemOpen
  \bibfield  {author} {\bibinfo {author} {\bibfnamefont {P.~C.}\ \bibnamefont
  {Klipstein}},\ }\href@noop {} {\bibfield  {journal} {\bibinfo  {journal} {J.
  Phys.: Condens. Matter}\ }\textbf {\bibinfo {volume} {28}},\ \bibinfo {pages}
  {375801} (\bibinfo {year} {2016})}\BibitemShut {NoStop}%
\bibitem [{Note3()}]{Note3}%
  \BibitemOpen
  \bibinfo {note} {Even though the wing solution of the four band spin up
  Hamiltonian is not necessarily spurious, the reason it gives an unphysical
  edge state is discussed at the end of Appendix A.}\BibitemShut {Stop}%
\bibitem [{\citenamefont {Knez}\ \emph
  {et~al.}(2014{\natexlab{a}})\citenamefont {Knez}, \citenamefont {Rettner},
  \citenamefont {Yang}, \citenamefont {Parkin}, \citenamefont {Du},
  \citenamefont {Du},\ and\ \citenamefont {Sullivan}}]{KnezParkin2014}%
  \BibitemOpen
  \bibfield  {author} {\bibinfo {author} {\bibfnamefont {I.}~\bibnamefont
  {Knez}}, \bibinfo {author} {\bibfnamefont {C.~I.}\ \bibnamefont {Rettner}},
  \bibinfo {author} {\bibfnamefont {S.-H.}\ \bibnamefont {Yang}}, \bibinfo
  {author} {\bibfnamefont {S.~S.~P.}\ \bibnamefont {Parkin}}, \bibinfo {author}
  {\bibfnamefont {L.}~\bibnamefont {Du}}, \bibinfo {author} {\bibfnamefont
  {R.-R.}\ \bibnamefont {Du}}, \ and\ \bibinfo {author} {\bibfnamefont
  {G.}~\bibnamefont {Sullivan}},\ }\href@noop {} {\bibfield  {journal}
  {\bibinfo  {journal} {Phys. Rev. Lett.}\ }\textbf {\bibinfo {volume} {112}},\
  \bibinfo {pages} {026602} (\bibinfo {year} {2014}{\natexlab{a}})}\BibitemShut
  {NoStop}%
\bibitem [{\citenamefont {Harrison}(1980)}]{Harrison1980}%
  \BibitemOpen
  \bibfield  {author} {\bibinfo {author} {\bibfnamefont {W.~A.}\ \bibnamefont
  {Harrison}},\ }\enquote {\bibinfo {title} {Electronic structure and the
  properties of solids: the physics of the chemical bond},}\ \ (\bibinfo
  {publisher} {W. H. Freeman and Co., San Francisco},\ \bibinfo {year} {1980})\
  Chap.\ \bibinfo {chapter} {11-D}, pp.\ \bibinfo {pages}
  {267--270}\BibitemShut {NoStop}%
\bibitem [{\citenamefont {Coh}\ and\ \citenamefont
  {Vanderbilt}(2008)}]{Coh2008}%
  \BibitemOpen
  \bibfield  {author} {\bibinfo {author} {\bibfnamefont {S.}~\bibnamefont
  {Coh}}\ and\ \bibinfo {author} {\bibfnamefont {D.}~\bibnamefont
  {Vanderbilt}},\ }\href@noop {} {\bibfield  {journal} {\bibinfo  {journal}
  {Phys Rev. B}\ }\textbf {\bibinfo {volume} {78}},\ \bibinfo {pages} {054117}
  (\bibinfo {year} {2008})}\BibitemShut {NoStop}%
\bibitem [{\citenamefont {Tashmukhamedova}\ and\ \citenamefont
  {Yusupjanova}(2016)}]{Tash2016}%
  \BibitemOpen
  \bibfield  {author} {\bibinfo {author} {\bibfnamefont {D.~A.}\ \bibnamefont
  {Tashmukhamedova}}\ and\ \bibinfo {author} {\bibfnamefont {M.~B.}\
  \bibnamefont {Yusupjanova}},\ }\href@noop {} {\bibfield  {journal} {\bibinfo
  {journal} {Journ. Surface Investigation. X-ray, Synchrotron and Neutron
  Techniques}\ }\textbf {\bibinfo {volume} {10}},\ \bibinfo {pages} {1273}
  (\bibinfo {year} {2016})}\BibitemShut {NoStop}%
\bibitem [{\citenamefont {Voitsekhovskii}\ \emph {et~al.}(2010)\citenamefont
  {Voitsekhovskii}, \citenamefont {Gorn}, \citenamefont {Nesmelov},\ and\
  \citenamefont {Kokhanenko}}]{Voitsekhovskii2010}%
  \BibitemOpen
  \bibfield  {author} {\bibinfo {author} {\bibfnamefont {A.~V.}\ \bibnamefont
  {Voitsekhovskii}}, \bibinfo {author} {\bibfnamefont {D.~I.}\ \bibnamefont
  {Gorn}}, \bibinfo {author} {\bibfnamefont {S.~N.}\ \bibnamefont {Nesmelov}},
  \ and\ \bibinfo {author} {\bibfnamefont {A.~P.}\ \bibnamefont {Kokhanenko}},\
  }\href@noop {} {\bibfield  {journal} {\bibinfo  {journal} {Opto-electronics
  Review}\ }\textbf {\bibinfo {volume} {18}},\ \bibinfo {pages} {241} (\bibinfo
  {year} {2010})}\BibitemShut {NoStop}%
\bibitem [{\citenamefont {Hellwege}\ and\ \citenamefont
  {Madelung}(1982)}]{Madelung1982}%
  \BibitemOpen
  \bibinfo {editor} {\bibfnamefont {K.~H.}\ \bibnamefont {Hellwege}}\ and\
  \bibinfo {editor} {\bibfnamefont {O.}~\bibnamefont {Madelung}},\ eds.,\
  \href@noop {} {\emph {\bibinfo {title} {Physics of Group IV elements and
  III-V Compounds}}},\ \bibinfo {series} {Landolt-B\"{o}rnstein New Series,
  Group III}, Vol.\ \bibinfo {volume} {17a}\ (\bibinfo  {publisher} {Springer
  Verlag, Berlin},\ \bibinfo {year} {1982})\BibitemShut {NoStop}%
\bibitem [{\citenamefont {Knez}\ \emph
  {et~al.}(2014{\natexlab{b}})\citenamefont {Knez}, \citenamefont {Du},\ and\
  \citenamefont {Sullivan}}]{Knez2014}%
  \BibitemOpen
  \bibfield  {author} {\bibinfo {author} {\bibfnamefont {I.}~\bibnamefont
  {Knez}}, \bibinfo {author} {\bibfnamefont {R.-R.}\ \bibnamefont {Du}}, \ and\
  \bibinfo {author} {\bibfnamefont {G.}~\bibnamefont {Sullivan}},\ }\href@noop
  {} {\bibfield  {journal} {\bibinfo  {journal} {Phys. Rev. B}\ }\textbf
  {\bibinfo {volume} {81}},\ \bibinfo {pages} {201301} (\bibinfo {year}
  {2014}{\natexlab{b}})}\BibitemShut {NoStop}%
\bibitem [{\citenamefont {Foreman}(1998)}]{Foreman1998}%
  \BibitemOpen
  \bibfield  {author} {\bibinfo {author} {\bibfnamefont {B.~A.}\ \bibnamefont
  {Foreman}},\ }\href@noop {} {\bibfield  {journal} {\bibinfo  {journal} {Phys.
  Rev. Lett.}\ }\textbf {\bibinfo {volume} {81}},\ \bibinfo {pages} {425}
  (\bibinfo {year} {1998})}\BibitemShut {NoStop}%
\bibitem [{\citenamefont {Livneh}\ \emph {et~al.}(2012)\citenamefont {Livneh},
  \citenamefont {Klipstein}, \citenamefont {Klin}, \citenamefont {Snapi},
  \citenamefont {Grossman}, \citenamefont {Glozman},\ and\ \citenamefont
  {Weiss}}]{Livneh2012}%
  \BibitemOpen
  \bibfield  {author} {\bibinfo {author} {\bibfnamefont {Y.}~\bibnamefont
  {Livneh}}, \bibinfo {author} {\bibfnamefont {P.~C.}\ \bibnamefont
  {Klipstein}}, \bibinfo {author} {\bibfnamefont {O.}~\bibnamefont {Klin}},
  \bibinfo {author} {\bibfnamefont {N.}~\bibnamefont {Snapi}}, \bibinfo
  {author} {\bibfnamefont {S.}~\bibnamefont {Grossman}}, \bibinfo {author}
  {\bibfnamefont {A.}~\bibnamefont {Glozman}}, \ and\ \bibinfo {author}
  {\bibfnamefont {E.}~\bibnamefont {Weiss}},\ }\href@noop {} {\bibfield
  {journal} {\bibinfo  {journal} {Phys. Rev. B}\ }\textbf {\bibinfo {volume}
  {86}},\ \bibinfo {pages} {235311} (\bibinfo {year} {2012})}\BibitemShut
  {NoStop}%
\bibitem [{\citenamefont {(erratum)}(2014)}]{LivErratum2014}%
  \BibitemOpen
  \bibfield  {author} {\bibinfo {author} {\bibnamefont {(erratum)}},\
  }\href@noop {} {\bibfield  {journal} {\bibinfo  {journal} {Phys. Rev. B}\
  }\textbf {\bibinfo {volume} {90}},\ \bibinfo {pages} {039903} (\bibinfo
  {year} {2014})}\BibitemShut {NoStop}%
\bibitem [{\citenamefont {Pezo}\ \emph {et~al.}(2020)\citenamefont {Pezo},
  \citenamefont {Focassio}, \citenamefont {Schleder}, \citenamefont {Costa},
  \citenamefont {Lewenkopf},\ and\ \citenamefont {Fazzio}}]{Pezo2020}%
  \BibitemOpen
  \bibfield  {author} {\bibinfo {author} {\bibfnamefont {A.}~\bibnamefont
  {Pezo}}, \bibinfo {author} {\bibfnamefont {B.}~\bibnamefont {Focassio}},
  \bibinfo {author} {\bibfnamefont {G.~R.}\ \bibnamefont {Schleder}}, \bibinfo
  {author} {\bibfnamefont {M.}~\bibnamefont {Costa}}, \bibinfo {author}
  {\bibfnamefont {C.}~\bibnamefont {Lewenkopf}}, \ and\ \bibinfo {author}
  {\bibfnamefont {A.}~\bibnamefont {Fazzio}},\ }\href@noop {} {\bibfield
  {journal} {\bibinfo  {journal} {arXiv:2010.11693}\ } (\bibinfo {year}
  {2020})}\BibitemShut {NoStop}%
\bibitem [{Note4()}]{Note4}%
  \BibitemOpen
  \bibinfo {note} {This is also true for solutions derived with $k_{+}$ and
  $k_{-}$ interchanged in the \protect \textit {Q}-terms of Eq.~\protect
  \textup {\hbox {\mathsurround \z@ \protect \normalfont (\ignorespaces \ref
  {eq:3}\unskip \@@italiccorr )}}, corresponding to a change in symmetry of the
  remote states. Note that the interchange leads to similar but modified
  expressions for the eigenvectors in Table~\ref {tab:Table 4} and the band gap
  ratios in Eq.~\protect \textup {\hbox {\mathsurround \z@ \protect \normalfont
  (\ignorespaces \ref {Eq. A2}\unskip \@@italiccorr )}}.}\BibitemShut {Stop}%
\bibitem [{Note5()}]{Note5}%
  \BibitemOpen
  \bibinfo {note} {Note that the BHZ Hamiltonian used in Ref.~\protect
  \rev@citealpnum {Candido2018} corresponds to spin down in this
  work}\BibitemShut {NoStop}%
\bibitem [{Note6()}]{Note6}%
  \BibitemOpen
  \bibinfo {note} {This is also consistent with the mutiband model. For
  positive $M_{0}$, $\protect \frac {\Delta _{i0}}{M_{0}}=\gamma \protect \frac
  {\Delta _{i}}{M}$ with $\gamma =\pm 1$, semiconductor parameters as in
  Fig.~\ref {fig:Figure 7}, and $A_{0}=A$, the only eigenvectors at zero energy
  and wave vector that are nearly similar on each side of the boundary are
  those for the wing solution, and only for $\gamma =-1$, when the outer band
  gap parameters change sign (consistent with negative $B_{0}$ and $\Delta
  N_{C}=1$ in the four band model). Since $\gamma $ is negative there is no
  exponential solution (see Sec.~\ref {sec:2-MULTI-BAND}C). However, given that
  the eigenvectors are very close, namely $\left [i0.321,\protect \tmspace
  +\thinmuskip {.1667em}0.174,\protect \tmspace +\thinmuskip
  {.1667em}-0.421,\protect \tmspace +\thinmuskip {.1667em}i0.830\right ]$ in
  the wall and $\left [i0.343,\protect \tmspace +\thinmuskip
  {.1667em}0.159,\protect \tmspace +\thinmuskip {.1667em}-0.410,\protect
  \tmspace +\thinmuskip {.1667em}i0.830\right ]$ in the semiconductor, this
  suggests that a nearby non-exponential edge state exists in the multiband
  model, corresponding to the wing solution of the topologically trivial
  phase.}\BibitemShut {Stop}%
\end{thebibliography}%

\end{document}